\documentclass[useAMS,usenatbib]{mn2e}
\usepackage{float}
\usepackage{natbib}
\usepackage{lipsum}
\usepackage{tikz}
\usetikzlibrary{spy}
\usepackage{subcaption}
\usepackage{epstopdf}

\begin{document}
\title[A method of complex background estimation in astronomical images]{A method of complex background estimation in astronomical images}
\author[A. Popowicz and B. Smolka]{A. Popowicz\thanks{E-mail:apopowicz@polsl.pl} and B. Smolka \\
Silesian University of Technology, Institute of Automatic Control, Poland, 44-100 Gliwice, Akademicka 16}

\date{~}
\pagerange{\pageref{firstpage}--\pageref{lastpage}} \pubyear{2015}
\maketitle
\label{firstpage}

\begin{abstract}
In this paper,  we present a novel approach to the estimation of strongly varying backgrounds in astronomical images by means of small objects removal and subsequent missing pixels interpolation. The method is based on the analysis of a pixel local neighborhood and utilizes the morphological distance transform. In contrast to popular background estimation techniques, our algorithm allows for accurate extraction of complex structures, like galaxies or nebulae. Moreover, it does not require multiple tuning parameters, since it relies on physical properties of CCD image sensors - the gain and the read-out noise characteristics. The comparison with other widely used background estimators revealed higher accuracy of the proposed technique. The superiority of the novel method is especially significant for the most challenging fluctuating backgrounds. The size of filtered out objects is tunable, therefore the algorithm may eliminate a wide range of foreground structures, including the dark current impulses, cosmic rays or even entire galaxies in deep field images.
\end{abstract}


\section{Introduction}

\begin{figure}
        \centering
        \begin{subfigure}[b]{\linewidth}
         \centering
                \includegraphics[width=6cm]{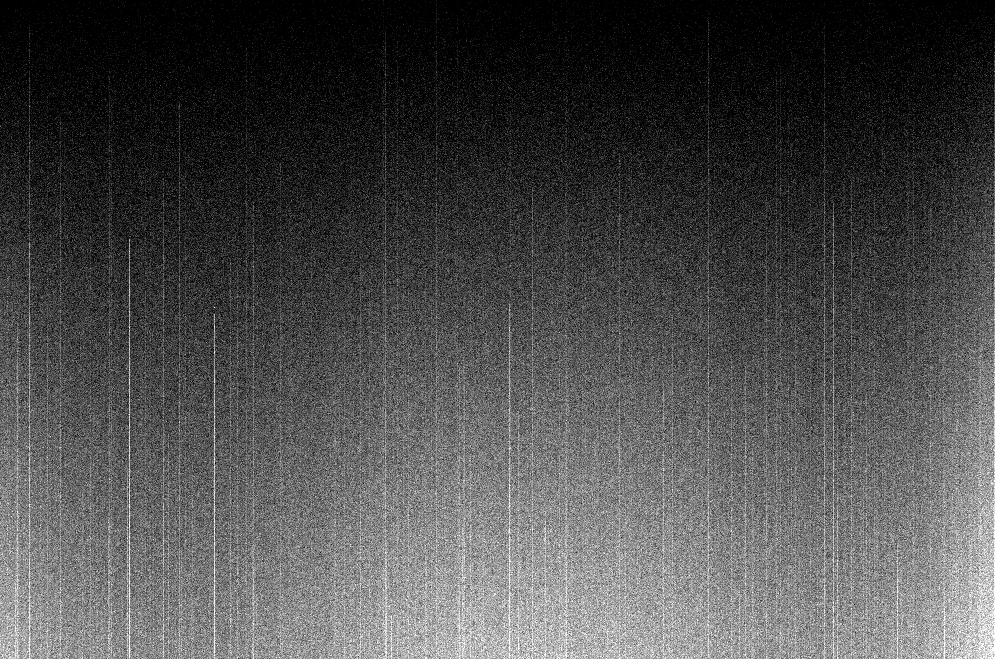}
		   \captionsetup{justification=centering}
                \caption{Gradient in bias frame due to the thermal charge accumulation during the readout phase (Kodak CCD KAI11002).\vspace{2mm}}
                \label{}
        \end{subfigure}
                \begin{subfigure}[b]{\linewidth}
                         \centering
                \includegraphics[width=6cm]{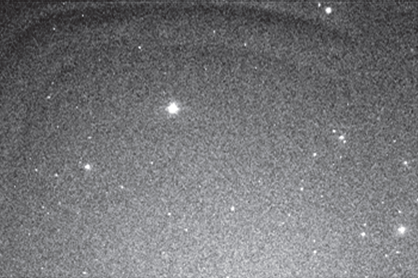}
		   \captionsetup{justification=centering}
                \caption{Condensation of moisture at the CCD image sensor edge (Kodak CCD KAF8300, \cite{ThermalEffectsPopowicz}).\vspace{2mm}}
                \label{}
        \end{subfigure}
                        \begin{subfigure}[b]{\linewidth}
                         \centering
                \includegraphics[width=6cm]{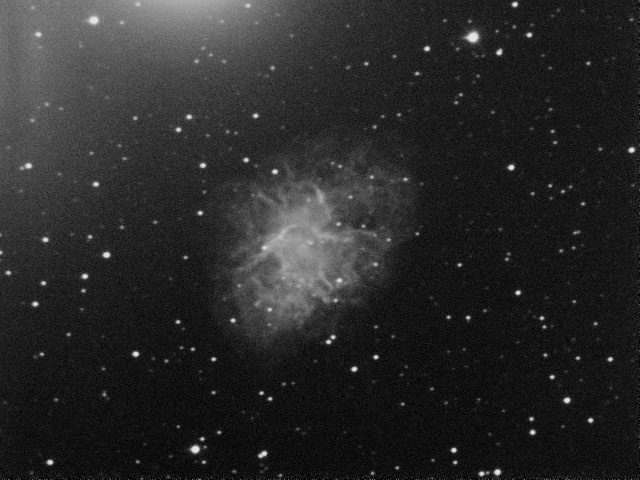}
		   \captionsetup{justification=centering}
                \caption{Amplifier electro-luminescence glow in upper left corner (CCD camera TouCAM PCVC740).\vspace{2mm}}
        \end{subfigure}
                        \begin{subfigure}[b]{\linewidth}
                         \centering
                \includegraphics[width=6cm]{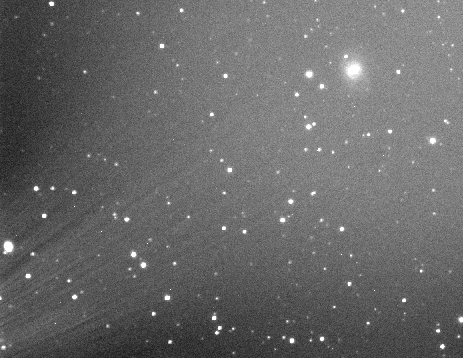}
		   \captionsetup{justification=centering}
                \caption{Straylight from Mars.}
        \end{subfigure}
        \caption{Exemplary background variations due to the CCD operational problems (a-c) and a straylight (d).}
        \label{ccdvariations}
\end{figure}

The background estimation is a fundamental step in image reduction techniques. It is performed to diminish the impact of both instrumental and non-instrumental offsets, which may lead to imprecise object detection and erroneous flux estimation. The instrumental sources of background variations are mainly related to the problems caused by CCD image detectors. They may be due to the nonuniformity of CCD bias frame (e.g. due to the additional dark charge created during the readout cycle), the temperature variations over the detector, which result in local condensation within the CCD plane (\cite{ThermalEffectsPopowicz}), or it may result from the electro-luminescence of a CCD amplifier during the exposure (\cite{SCCDSJanesick}) (see examples presented in Fig. \ref{ccdvariations}). However, it should be noted, that the most of those problems are mitigated nowadays by novel CCD structures, proper CCD calibration and strong cooling. 

The main source of non-instrumental background variations is the sky glow, which is dependent on the altitude above the horizon and the observed wavelength range and originates from the local light pollution. The situation is much worse in the infrared, since the emission of the sky background and the radiation from the telescope itself is time- and space-dependent and is also much higher in this wavelength range. In such a case, the techniques, like chopping or nodding, have to be employed to subtract the background bias frames in real-time (\cite{Bertero,Fiorucci}). However, the scattered light in the instrument, resulting from e.g. nearby very bright object, is still difficult to remove (see Fig. \ref{ccdvariations}d).

\begin{figure}
        \centering
        \begin{subfigure}[b]{0.49\linewidth}
         \centering
                \includegraphics[height=6cm]{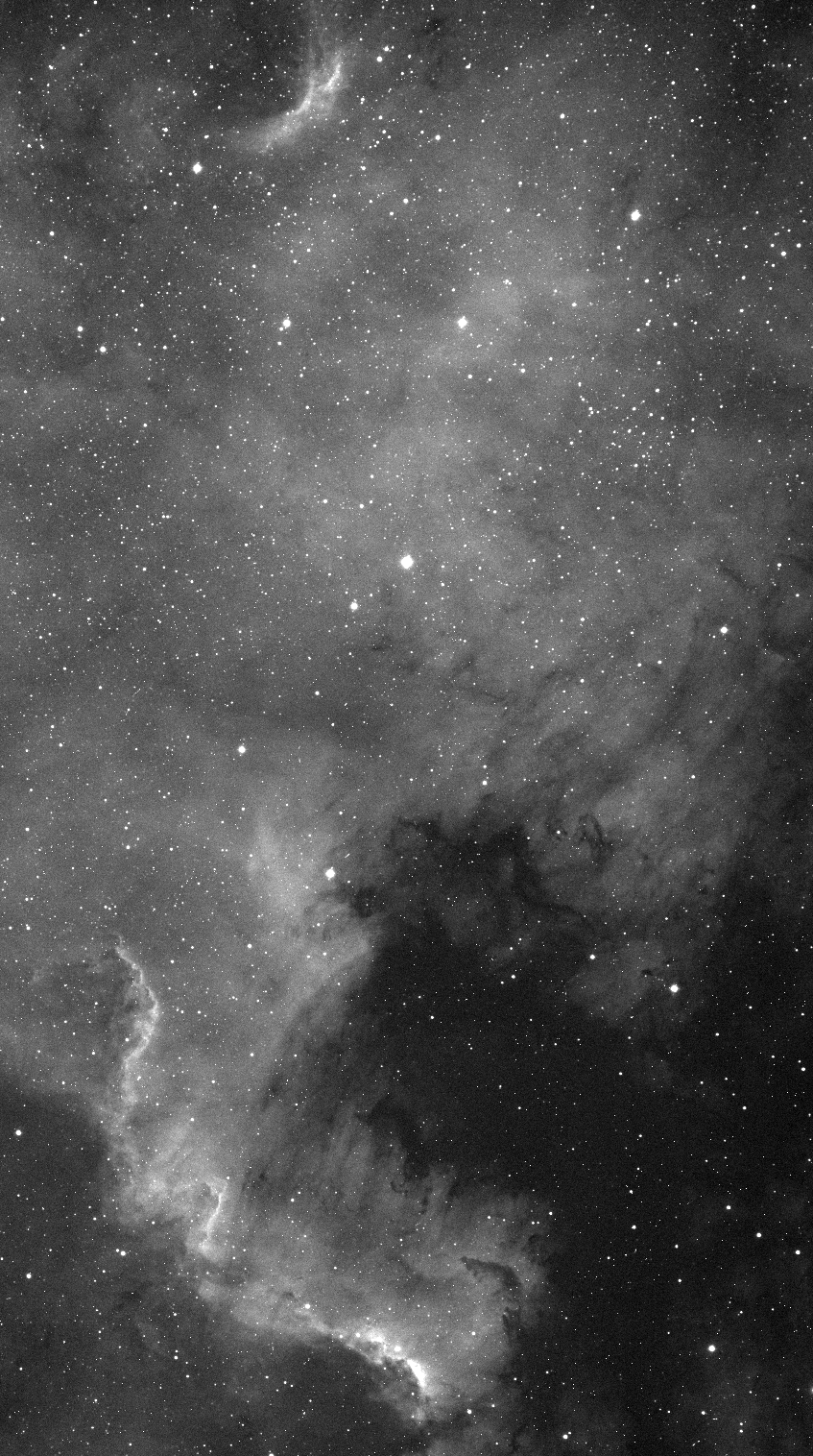}
		   \captionsetup{justification=centering}
                \caption{}
                \label{}
        \end{subfigure}
                \begin{subfigure}[b]{0.49\linewidth}
                         \centering
                \includegraphics[height=6cm]{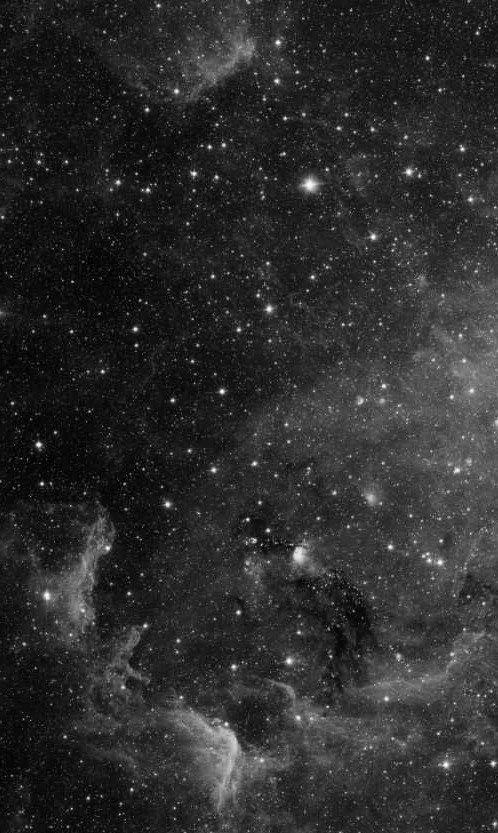}
		   \captionsetup{justification=centering}
                \caption{}
                \label{}
        \end{subfigure}
        \caption{NGC7000 North America nebula: (a) - image obtained by the author in the visible, (b) - the same region in the infrared, (IRAC camera, \protect\cite{IRAC}).}
        \label{vis-ir}
\end{figure}

Additionally, the challenging backgrounds are produced also by extended astronomical objects, like the nebulae. They may have very complex structures including varying intensity gradients, which may be erroneously detected as the objects of interest - the stars. Hence, it is very difficult to perform automatic analysis, including the detection and the photometry of the regions covered by complex background structures. It is also to be noted that, depending on the observed wavelength range, the degree of background intensity variations may differ as depicted in Fig. \ref{vis-ir}. Finally, in some cases, the background subtle structures, like the  nebulae or galaxies filaments, may provide viable information (\cite{galaxy1}, \cite{galaxy2}), thus the background has to be extracted.

In this paper, we present a novel technique of astronomical background estimation. In contrast to most of background determination methods, which are designed mainly for the star detection purposes, we focus on the background as an object of interest. The method aims to estimate the complex background structures by eliminating smaller foreground objects like stars, traces of cosmic rays or impulsive noise artifacts.

The article is organized as follows. In Section 2 we enumerate and characterize current background estimation techniques. Next, in Section 3, we describe the distance transformation, since it constitutes the key element of our algorithm. In Section 4, we consider the impact of CCD noise on the background estimation results. Some suggestions for the adjustment of our method for the analysis of images obtained in different wavelength ranges are also given. The details and the results of the comparison of our method with other background estimators are provided in Section 5 and 6. In Section 7 we consider the over-resolved sources in images and the presence of close multiple objects. Finally, in Section 8, we present the applications of our background estimator using various astronomical images. We conclude the paper and summarize the main results in Section 9.

%----------------------------------------------------------------------------------------------------------------------------------
\section{Background estimation algorithms}

The most popular techniques of background estimation are implemented in well-know astronomical object extraction packages dedicated to automatic image analysis: \emph{DAOPHOT} (\cite{DAOPHOT}) and \emph{SExtractor} (SE) (\cite{SExtractor}). They are iterative methods, wherein the image is divided into small patches to perform the analysis of local intensity histograms. The authors iteratively apply $\sigma$-clipping in each image patch, until the intensities belong to the $\pm3\sigma$ range around the local intensity median. If the final standard deviation $\sigma$ is larger than 20$\%$ of its initial value, the field is considered to be uncrowded (i.e. there are few stars within the region) and, in such a case, the authors suggest to take the mean of the clipped histogram as an intensity of the background. Otherwise, the field is considered to be crowded by the objects, hence the following formula is used to obtain the mode of background intensity:
\begin{equation}
\centering
\textrm{mode} = 2.5 \! \times \! \textrm{median} - 1.5 \! \times \! \textrm{mean}.
\label{mode1}
\end{equation}
\cite{SExtractor} claim that such an estimation is more accurate than the one suggested by \cite{Kendall}:
\begin{equation}
\centering
\textrm{mode} = 3 \! \times \! \textrm{median} - 2 \! \times \! \textrm{mean}.
\label{mode1}
\end{equation}
This method was observed to be less affected by crowding than simple trimmed mean, utilized in FOCAS photometric package (Faint Object Classification), included in meaningful astronomical image processing environment - IRAF (\cite{iraf}), and also employed by other authors (\cite{Vikhlinin}, \cite{Lazzati} and \cite{Perret}). According to the most recent astronomical image processing review presented by \cite{Masias2,Masias1}, such a range of techniques, involving $\sigma$-clipping, is currently the most widespread, due to their simplicity and low computational burden. 

In several publications (\cite{Irwin}, \cite{Slezak} and \cite{LeFevre}), the authors utilized the algorithm proposed by \cite{Bijaoui}. The approach involves iterative background level estimation based on Bayesian scheme. Although the technique was originally developed for photographic plates, it was also successfully adopted to digital images. In contrast to $\sigma$-clipping based approach, the computational effort of the method is significantly higher and it is not useful for strongly varying backgrounds. It is due to the low number of pixels in very small patches required to follow high background fluctuations. In such applications, the artifacts in form of rings or crosses can be observed.

Since the foreground objects may be interpreted as the outliers within the background structures, the median filtering of the image may be also suitable for astronomical purposes (\cite{Remazeilles}). In such a technique, each pixel intensity is replaced by a median (or a trimmed median, as presented by \cite{Bickel}, \cite{Davies} and \cite{Smolka}) of the pixel intensities in the local neighborhood.

%----------------------------------------------------------------------------------------------------------------------------------
\section{Modified distance transformation}

\begin{figure}
\centering
\begin{subfigure}[b]{0.4\linewidth}
\centering
\begin{tikzpicture}[scale=0.75]
\filldraw[fill=white, draw=white] (1,1) rectangle +(1,1);
\filldraw[fill=white] (2,2) rectangle +(1,1); \node at (2.5,2.5)  {$P_0$};
\filldraw[fill=gray!30] (1,2) rectangle + (1,1);  \node at (1.5,2.5)  {$P_1$};
\filldraw[fill=gray!30] (1,3) rectangle + (1,1); \node at  (1.5,3.5)  {$P_2$};
\filldraw[fill=gray!30] (2,3) rectangle + (1,1); \node at  (2.5,3.5)  {$P_3$};
\filldraw[fill=gray!30] (3,3) rectangle + (1,1); \node at  (3.5,3.5)  {$P_4$};
\end{tikzpicture}
\caption{First scan mask.}
\end{subfigure}~~~~
\begin{subfigure}[b]{0.4\linewidth}
\centering
\begin{tikzpicture}[scale=0.75]
\filldraw[fill=white] (2,2) rectangle +(1,1); \node at (2.5,2.5)  {$P_0$};
\filldraw[fill=gray!30] (3,2) rectangle + (1,1);  \node at (3.5,2.5)  {$P_1$};
\filldraw[fill=gray!30] (1,1) rectangle + (1,1); \node at  (1.5,1.5)  {$P_2$};
\filldraw[fill=gray!30] (2,1) rectangle + (1,1); \node at  (2.5,1.5)  {$P_3$};
\filldraw[fill=gray!30] (3,1) rectangle + (1,1); \node at  (3.5,1.5)  {$P_4$};
\end{tikzpicture}
\caption{Second scan mask.}
\end{subfigure}

\caption{Processing masks utilized in double-scan algorithm.}
\label{masks}
\end{figure}
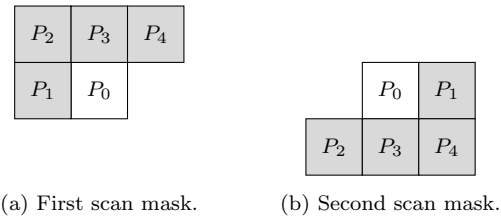

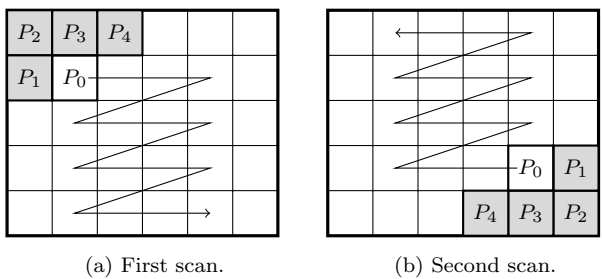
\begin{figure}
        \centering
\begin{subfigure}[b]{0.47\linewidth}
\begin{tikzpicture}[scale=0.6]
\draw[black,very thick] (1,1) rectangle (7,6);
\draw[step=1,black] (1,1) grid (7,6);

\draw[thick,fill=gray!30] (1,6) rectangle (2,5);   \draw[black] (1.5,5.5) node{\footnotesize $P_2$};
\draw[thick,fill=gray!30] (3,6) rectangle (2,5);   \draw[black] (2.5,5.5) node{\footnotesize $P_3$};
\draw[thick,fill=gray!30] (2,5) rectangle (1,4);   \draw[black] (1.5,4.5) node{\footnotesize $P_1$};
\draw[thick,fill=gray!30] (4,6) rectangle (3,5);   \draw[black] (3.5,5.5) node{\footnotesize $P_4$};
\draw[thick] (2,4) rectangle (3,5);   \draw[black] (2.5,4.5) node{\footnotesize $P_0$};
\draw[->] (2.8,4.5) -- (5.5,4.5) -- (2.5,3.5) -- (5.5,3.5) -- (2.5,2.5) -- (5.5,2.5) -- (2.5,1.5) -- (5.5,1.5);

\end{tikzpicture}
\caption{First scan.}
\end{subfigure}~~
\begin{subfigure}[b]{0.47\linewidth}
\begin{tikzpicture}[scale=0.6]
\draw[black,very thick] (1,1) rectangle (7,6);
\draw[step=1,black] (1,1) grid (7,6);

\draw[thick,fill=gray!30] (6,1) rectangle (7,2);   \draw[black] (6.5,1.5) node{\footnotesize $P_2$};
\draw[thick,fill=gray!30] (6,2) rectangle (5,1);   \draw[black] (5.5,1.5) node{\footnotesize $P_3$};
\draw[thick,fill=gray!30] (7,3) rectangle (6,2);   \draw[black] (6.5,2.5) node{\footnotesize $P_1$};
\draw[thick,fill=gray!30] (5,2) rectangle (4,1);   \draw[black] (4.5,1.5) node{\footnotesize $P_4$};
\draw[thick] (5,2) rectangle (6,3);   \draw[black] (5.5,2.5) node{\footnotesize $P_0$};
\draw[->] (5.2,2.5) -- (2.5,2.5) -- (5.5,3.5) -- (2.5,3.5) -- (5.5,4.5) -- (2.5,4.5) -- (5.5,5.5) -- (2.5,5.5);
\end{tikzpicture}
\caption{Second scan.}
\end{subfigure}

\caption{Visualization of double-scan algorithm with indicated direction of scanning in each run.}
\label{2-scan}
\end{figure}

In its original formulation, the distance transformation provides the estimation of metric distance between given object and other image pixels \cite{Klette}. The popular, fast implementation is based on  so-called double-scan algorithm (\cite{Rosenfeld}, \cite{Borgefors}), which involves simple replacement operations between pixels $P_0$, $P_1$, $P_2$ and $P_3$ in a local sliding window (see Fig. \ref{masks}). The algorithm passes the distance array twice: from left to right, up to bottom (first scan) and then, it returns - right to left, bottom to the top (second scan). The double-scan technique employing 8-neighborhood scanning mask is explained in Fig. \ref{2-scan}. 

\begin{figure}
        \centering
\begin{subfigure}[b]{0.47\linewidth}
\begin{tikzpicture}[scale=0.6]
\draw[black,very thick] (1,1) rectangle (7,6);
\draw[step=1,black] (1,1) grid (7,6);

\draw[fill=gray!30] (4,4) rectangle (3,3);   \draw[white] (3.5,3.5) node{};
\draw[fill=gray!30] (5,4) rectangle (4,3);   \draw[white] (4.5,3.5) node{};
\draw[fill=gray!30] (4,5) rectangle (3,4);   \draw[white] (4.5,3.5) node{};

%^\draw[fill=blue] (6,4) rectangle (7,3);   \draw[white] (6.5,3.5) node{6};

\end{tikzpicture}
\caption{Marked object (gray).\\~}
\end{subfigure}
\begin{subfigure}[b]{0.47\linewidth}
\begin{tikzpicture}[scale=0.6]
\draw[black,very thick] (1,1) rectangle (7,6);
\draw[step=1,black] (1,1) grid (7,6);
\draw[fill=gray!30] (4,5) rectangle (3,4);   \draw[white] (4.5,3.5) node{};

\draw[black] (1.5,5.5) node {\footnotesize $\infty$};
\draw[black] (2.5,5.5) node {\footnotesize  $\infty$};
\draw[black] (3.5,5.5) node {\footnotesize $\infty$};
\draw[black] (4.5,5.5) node {\footnotesize  $\infty$};
\draw[black] (5.5,5.5) node {\footnotesize  $\infty$};
\draw[black] (6.5,5.5) node {\footnotesize  $\infty$};

\draw[black] (1.5,4.5) node {\footnotesize $\infty$};
\draw[black] (2.5,4.5) node {\footnotesize $\infty$};
\draw[black] (3.5,4.5) node {\footnotesize $0$};
\draw[black] (4.5,4.5) node {\footnotesize  $\infty$};
\draw[black] (5.5,4.5) node {\footnotesize  $\infty$};
\draw[black] (6.5,4.5) node {\footnotesize  $\infty$};

\draw[black] (1.5,3.5) node {\footnotesize $\infty$};
\draw[black] (2.5,3.5) node {\footnotesize $\infty$};
\draw[fill=gray!30] (4,4) rectangle (3,3);   \draw[black] (3.5,3.5) node{\footnotesize 0};
\draw[fill=gray!30] (5,4) rectangle (4,3);   \draw[black] (4.5,3.5) node{\footnotesize 0};
\draw[black] (5.5,3.5) node {\footnotesize  $\infty$};
\draw[black] (6.5,3.5) node {\footnotesize  $\infty$};

\draw[black] (1.5,2.5) node {\footnotesize $\infty$};
\draw[black] (2.5,2.5) node {\footnotesize  $\infty$};
\draw[black] (3.5,2.5) node {\footnotesize  $\infty$};
\draw[black] (4.5,2.5) node {\footnotesize  $\infty$};
\draw[black] (5.5,2.5) node {\footnotesize  $\infty$};
\draw[black] (6.5,2.5) node {\footnotesize  $\infty$};

\draw[black] (1.5,1.5) node {\footnotesize $\infty$};
\draw[black] (2.5,1.5) node {\footnotesize  $\infty$};
\draw[black] (3.5,1.5) node {\footnotesize $\infty$};
\draw[black] (4.5,1.5) node {\footnotesize  $\infty$};
\draw[black] (5.5,1.5) node {\footnotesize  $\infty$};
\draw[black] (6.5,1.5) node {\footnotesize  $\infty$};

%^\draw[fill=blue] (6,4) rectangle (7,3);   \draw[white] (6.5,3.5) node{6};

\end{tikzpicture}
\captionsetup{justification=centering}
\caption{Distance matrix initialization.}
\end{subfigure}

\begin{subfigure}[b]{0.47\linewidth}
\begin{tikzpicture}[scale=0.6]
\draw[black,very thick] (1,1) rectangle (7,6);
\draw[step=1,black] (1,1) grid (7,6);
\draw[fill=gray!30] (4,5) rectangle (3,4);   \draw[white] (4.5,3.5) node{};

\draw[black] (1.5,5.5) node {\footnotesize $\infty$};
\draw[black] (2.5,5.5) node {\footnotesize  $\infty$};
\draw[black] (3.5,5.5) node {\footnotesize $\infty$};
\draw[black] (4.5,5.5) node {\footnotesize  $\infty$};
\draw[black] (5.5,5.5) node {\footnotesize  $\infty$};
\draw[black] (6.5,5.5) node {\footnotesize  $\infty$};

\draw[black] (1.5,4.5) node {\footnotesize $\infty$};
\draw[black] (2.5,4.5) node {\footnotesize $\infty$};
\draw[black] (3.5,4.5) node {\footnotesize $0$};
\draw[black] (4.5,4.5) node {\footnotesize  $1$};
\draw[black] (5.5,4.5) node {\footnotesize  $2$};
\draw[black] (6.5,4.5) node {\footnotesize  $2$};

\draw[black] (1.5,3.5) node {\footnotesize $\infty$};
\draw[black] (2.5,3.5) node {\footnotesize $\infty$};
\draw[fill=gray!30] (4,4) rectangle (3,3);   \draw[black] (3.5,3.5) node{\footnotesize 0};
\draw[fill=gray!30] (5,4) rectangle (4,3);   \draw[black] (4.5,3.5) node{\footnotesize 0};
\draw[black] (5.5,3.5) node {\footnotesize  $1$};
\draw[black] (6.5,3.5) node {\footnotesize  $2$};

\draw[black] (1.5,2.5) node {\footnotesize $\infty$};
\draw[black] (2.5,2.5) node {\footnotesize  $1$};
\draw[black] (3.5,2.5) node {\footnotesize  $1$};
\draw[black] (4.5,2.5) node {\footnotesize  $1$};
\draw[black] (5.5,2.5) node {\footnotesize  $1$};
\draw[black] (6.5,2.5) node {\footnotesize  $2$};

\draw[black] (1.5,1.5) node {\footnotesize $2$};
\draw[black] (2.5,1.5) node {\footnotesize  $2$};
\draw[black] (3.5,1.5) node {\footnotesize $2$};
\draw[black] (4.5,1.5) node {\footnotesize  $2$};
\draw[black] (5.5,1.5) node {\footnotesize  $2$};
\draw[black] (6.5,1.5) node {\footnotesize  $2$};

%^\draw[fill=blue] (6,4) rectangle (7,3);   \draw[white] (6.5,3.5) node{6};

\end{tikzpicture}
\caption{Result after first scan.}
\end{subfigure}
\begin{subfigure}[b]{0.47\linewidth}
\begin{tikzpicture}[scale=0.6]
\draw[black,very thick] (1,1) rectangle (7,6);
\draw[step=1,black] (1,1) grid (7,6);
\draw[fill=gray!30] (4,5) rectangle (3,4);   \draw[white] (4.5,3.5) node{};

\draw[black] (1.5,5.5) node {\footnotesize $2$};
\draw[black] (2.5,5.5) node {\footnotesize  $1$};
\draw[black] (3.5,5.5) node {\footnotesize $1$};
\draw[black] (4.5,5.5) node {\footnotesize  $1$};
\draw[black] (5.5,5.5) node {\footnotesize  $2$};
\draw[black] (6.5,5.5) node {\footnotesize  $2$};

\draw[black] (1.5,4.5) node {\footnotesize $2$};
\draw[black] (2.5,4.5) node {\footnotesize $1$};
\draw[black] (3.5,4.5) node {\footnotesize $0$};
\draw[black] (4.5,4.5) node {\footnotesize  $1$};
\draw[black] (5.5,4.5) node {\footnotesize  $1$};
\draw[black] (6.5,4.5) node {\footnotesize  $2$};

\draw[black] (1.5,3.5) node {\footnotesize $2$};
\draw[black] (2.5,3.5) node {\footnotesize $1$};
\draw[fill=gray!30] (4,4) rectangle (3,3);   \draw[black] (3.5,3.5) node{\footnotesize 0};
\draw[fill=gray!30] (5,4) rectangle (4,3);   \draw[black] (4.5,3.5) node{\footnotesize 0};
\draw[black] (5.5,3.5) node {\footnotesize  $1$};
\draw[black] (6.5,3.5) node {\footnotesize  $2$};

\draw[black] (1.5,2.5) node {\footnotesize $2$};
\draw[black] (2.5,2.5) node {\footnotesize  $1$};
\draw[black] (3.5,2.5) node {\footnotesize  $1$};
\draw[black] (4.5,2.5) node {\footnotesize  $1$};
\draw[black] (5.5,2.5) node {\footnotesize  $1$};
\draw[black] (6.5,2.5) node {\footnotesize  $2$};

\draw[black] (1.5,1.5) node {\footnotesize $2$};
\draw[black] (2.5,1.5) node {\footnotesize  $2$};
\draw[black] (3.5,1.5) node {\footnotesize $2$};
\draw[black] (4.5,1.5) node {\footnotesize  $2$};
\draw[black] (5.5,1.5) node {\footnotesize  $2$};
\draw[black] (6.5,1.5) node {\footnotesize  $2$};

%^\draw[fill=blue] (6,4) rectangle (7,3);   \draw[white] (6.5,3.5) node{6};

\end{tikzpicture}
\caption{Result after second scan.}
\end{subfigure}

\caption{An exemplary application of distance transformation. }
\label{disttransf}
\end{figure}
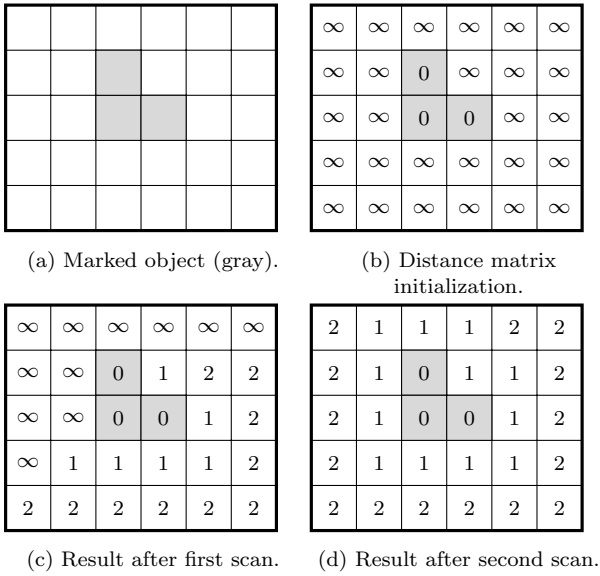

An example of distance transformation utilization for simple 3-pixel object is depicted in Fig. \ref{disttransf}. Initially, all the pixels are given infinite distance, while the distances of pixels included in the object are set to zero (see Fig. \ref{disttransf} b). At each pixel position, the following simple update of distance values is performed:
\begin{equation}
d_{0} = \textrm{min}\{ d_{0}, d_{j}+1\}, 
\label{eq1}
\end{equation}
where $j$ $\in \{1, 2, 3, 4\}$; $d_{0}$ and $d_{j}$ are the distance values at respective pixels $P_0$ and $P_j$. After the first run, only a part of image pixels is given its distance (see Fig. \ref{disttransf} c). Eventually, the second scan allows for filling the whole distance matrix.

\begin{figure}
\centering
\begin{subfigure}[b]{0.45\linewidth}
\includegraphics[width=\textwidth]{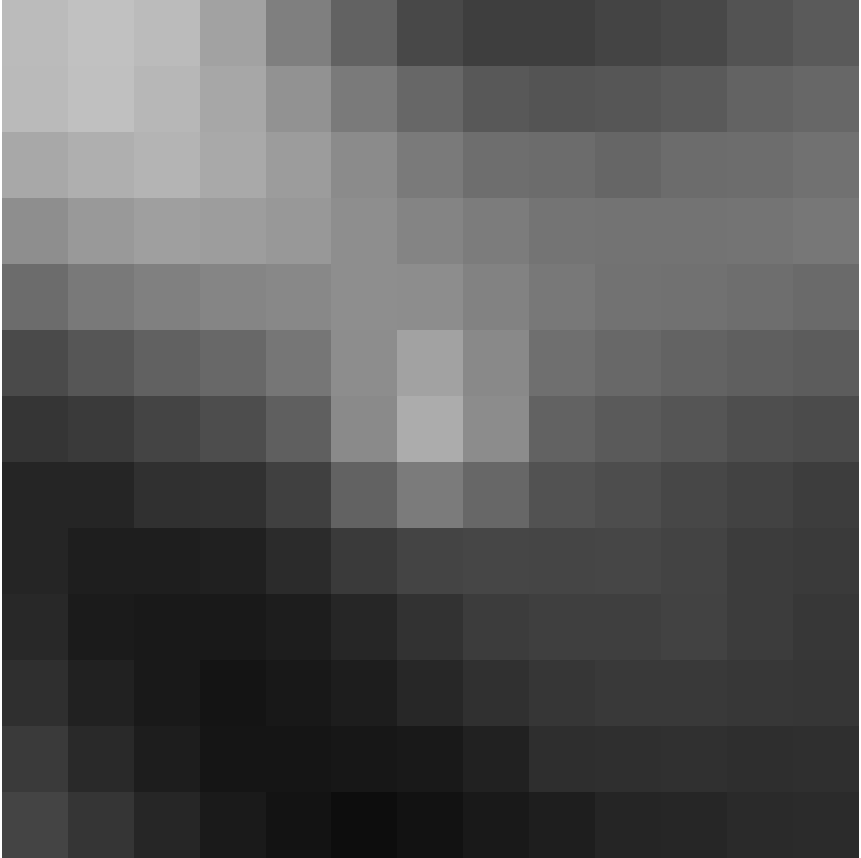}
 \captionsetup{justification=centering}
\caption{Window (13$\times$13) with a central pixel representing the object, (no ascending routes from the center).}
\end{subfigure}~~~
\begin{subfigure}[b]{0.45\linewidth}
\includegraphics[width=\textwidth]{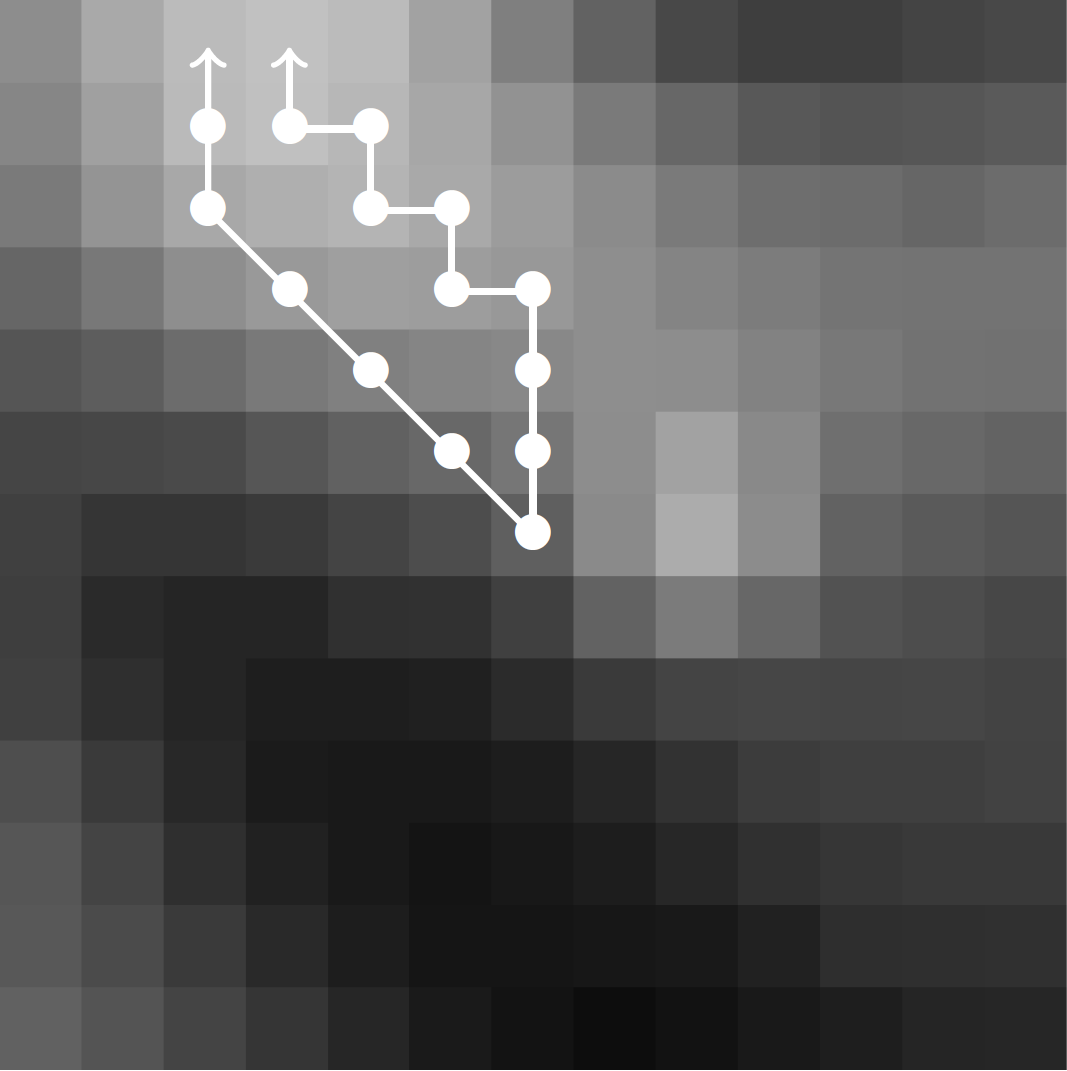}
 \captionsetup{justification=centering}
\caption{Window (13$\times$13) with a central pixel belonging to the background and indicated exemplary ascending routes.}
\end{subfigure}
\\
\begin{subfigure}[b]{0.45\linewidth}
\includegraphics[width=\textwidth]{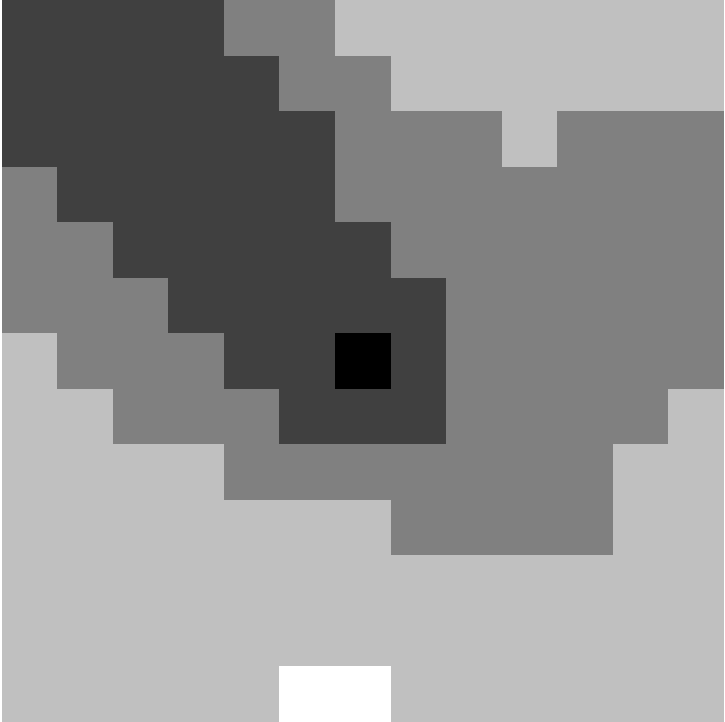}
 \captionsetup{justification=centering}
\caption{Distance array of the pixels from the window (a).}
\end{subfigure}~~~
\begin{subfigure}[b]{0.45\linewidth}
\includegraphics[width=\textwidth]{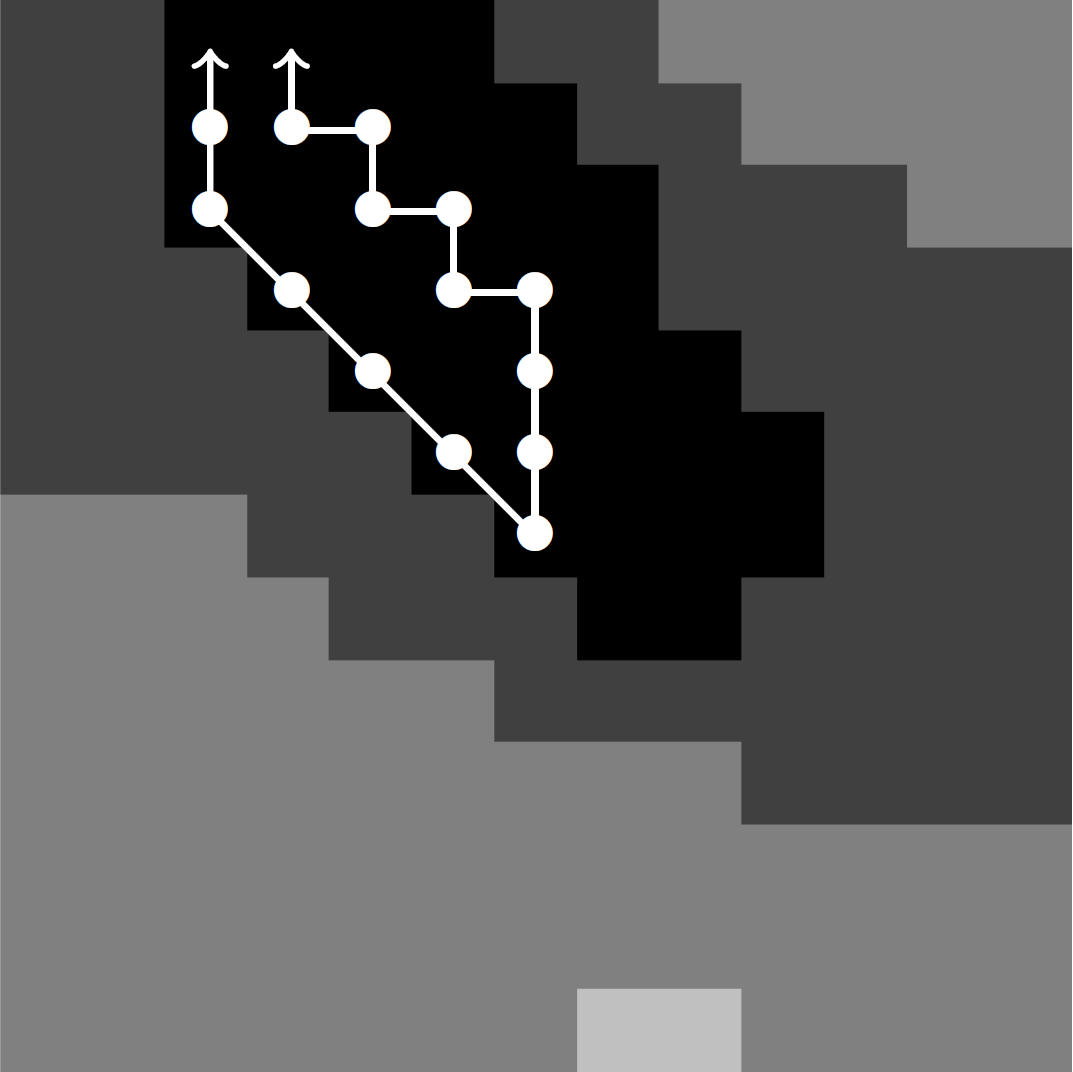}
 \captionsetup{justification=centering}
\caption{Distance array of the pixels from the window (b).}
\end{subfigure}

\caption{Exemplary distance analysis for object and background pixel. White lines highlight exemplary \emph{ascending routs} to window border. The gray scale map in (c) and (d) was chosen, so that the black color corresponds to 0 distance, while the white denotes 4.}
\label{patches}
\end{figure}

In our proposed method, we aim to determine, if a given image pixel contains counts from a foreground object. To this end, we analyze local square patches sliding over the image domain. The size of such a patch should be at least twice as big as the objects to be removed. With a use of distance transformations, we investigate the digital paths between the patch central pixel and the pixels belonging to the border. Since the astronomical objects within the image are brighter than the background, we make an assumption, that for the case of background pixel, there should be at least one digital path leading from the window center to the patch border, wherein we encounter only intensity growth. We call such a path the \emph{ascending route}. For object pixels, like within the stars, there will be no such route available. Exemplary \emph{ascending routs} for background pixel are depicted in Fig. \ref{patches} b. As can be seen, we allow for both orthogonal and diagonal steps between pixels in a path.


To find the aforementioned digital routs, we employ the distance transformation in a local processing window. Initially only the window's central pixel is given zero distance, while the others are set to infinity. Then, we run the double-scan algorithm using Eq. \ref{eq1} modified in a following way:
\begin{equation}
\begin{array}{l}
d_{0}=\textrm{min}\{ d_{0}, d_{j}+\Delta\}, \\
\Delta=1 \textrm{~~~if~~~}  I_{0} \leq I_{j},\\
\Delta=0 \textrm{~~~if~~~}  I_{0} > I_{j},\\
\end{array}
\label{eq_simple}
\end{equation}
where $I_0$ and $I_j$ is the gray scale value in respective pixels $P_0$ and $P_j$; $\Delta$ is a trigger, which controls the distance incrementation. If there exists an \emph{ascending route}, then there should be at least one zero-distance pixel at the window's border. Otherwise, the central pixel is classified as object pixel. In Fig. \ref{patches} we depict the distance transformation of two test images, in which the central pixel represents correspondingly the bright object and the background.

\begin{figure}
        \centering
         \captionsetup{justification=centering}
\begin{subfigure}[b]{0.32\linewidth}
\includegraphics[width=\textwidth]{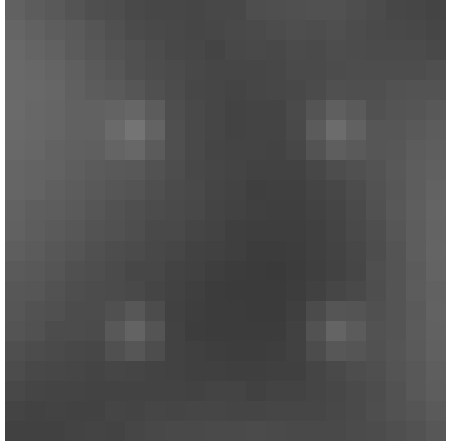}
\caption{Exemplary image patch with 4 stars.}
\end{subfigure}
\begin{subfigure}[b]{0.32\linewidth}
\includegraphics[width=\textwidth]{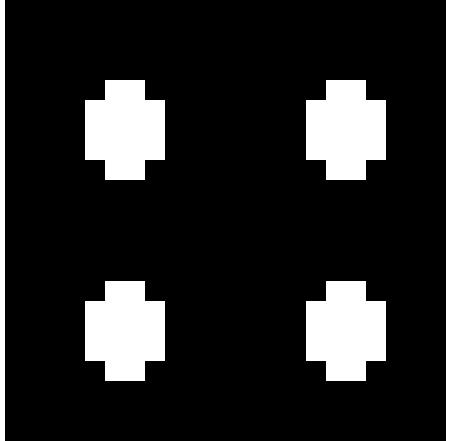}
 \captionsetup{justification=centering}
\caption{Detected object pixels\\~}
\end{subfigure}
\begin{subfigure}[b]{0.32\linewidth}
\includegraphics[width=\textwidth]{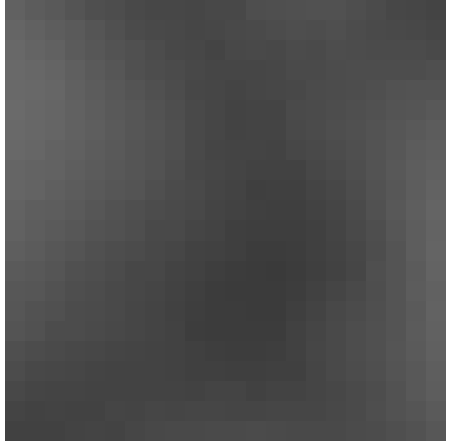}
 \captionsetup{justification=centering}
\caption{Result of cubic interpolation.\\~}
\end{subfigure}
\caption{The consecutive steps of our selective background interpolation technique: (a) - the original input image, (b) - detected object pixels using the distance transform and (c) - the result of the selective interpolation.}
\label{interp1}
\end{figure}

\vspace{0.5cm}
To estimate the background, we performed the interpolation over the pixels classified as a part of foreground objects and also over the neighboring pixels, so that the tails in objects point spread function (PSF) do not influence the interpolation. According to our previous research (\cite{BadPixelPopowicz}), there are several well-known interpolation algorithms, which are employed in the astronomical imaging: nearest neighbor, linear (used e.g. in IRAF image facility \cite{iraf}), cubic (\cite{cubic}) and biharmonic interpolation (\cite{Sandwell}). For the purpose of comparison presented in Section 5, we used all the mentioned methods and evaluated their accuracy. An illustrative result of our selective interpolation technique is presented in Fig. \ref{interp1}, where we depict respectively, an exemplary image patch, the classified object pixels with their neighborhood and the outcome of cubic interpolation. 

%-------------------------------------------------------------
\section{Image noise considerations}

In real images, one should always consider the influence of noise. There are two main noise types accompanying the CCD imaging: the Poisson noise originating in the physics of charge collection process, and the Gaussian noise resulting in the amplifier gain fluctuations (\cite{Janesick1,Janesick2}). Since the astronomical images are obtained by well calibrated detectors, both noise types are well characterized. Hence, the distribution of flux measurements in a pixel is governed by the Gaussian distribution, where the standard deviation ($\sigma_{CCD}$) is given by:
\begin{equation}
\sigma_{CCD} = \sqrt{I+\sigma_{el}^2},
\label{CCD_noise}
\end{equation}
where $I$ is the number of counts in a pixel, $\sigma_{el}$ is the standard deviation of electronic noise.

To give consideration to possible local intensity variations due to the noise impact, the inequalities in Eq. \ref{eq_simple} have to be modified:
\begin{equation}
\begin{array}{rl}
\Delta=1 \textrm{~~if~~}  I_{0} \leq I_{j}-k\sqrt{I_{j}+\sigma_{el}^2},\\
\Delta=0 \textrm{~~if~~}  I_{0} > I_{j}-k\sqrt{I_{j}+\sigma_{el}^2},\\
\end{array}
\label{formula1}
\end{equation}
where $k \geq 0$ is a sensitivity parameter (the lower the $k$, the higher the sensitivity, but also the lower the robustness). Such a formulation of the $\Delta$ trigger allows to neglect small intensity drops, which can be caused by noise, rather than by real decrease of image object intensity.

We would like to note, that the inequalities for the calculation of factor $\Delta$ may be modified to fit the detector characteristics. The proposed scheme (\ref{formula1}) was created for application in images acquired by CCD and CMOS sensors. To optimize the algorithm, so that it works for images obtained e.g. by radio-telescopes, we recommend to modify $\Delta$ parameter according to a given measurement uncertainty. This flexibility of our algorithm is an important feature, since it makes the approach beneficial in a wide range of imaging techniques.

%---------------------------------------

%---------------------------------------
\section{Background estimation comparison}

\begin{figure*}
\centering
\begin{subfigure}[b]{0.32\linewidth}
\includegraphics[width=\textwidth]{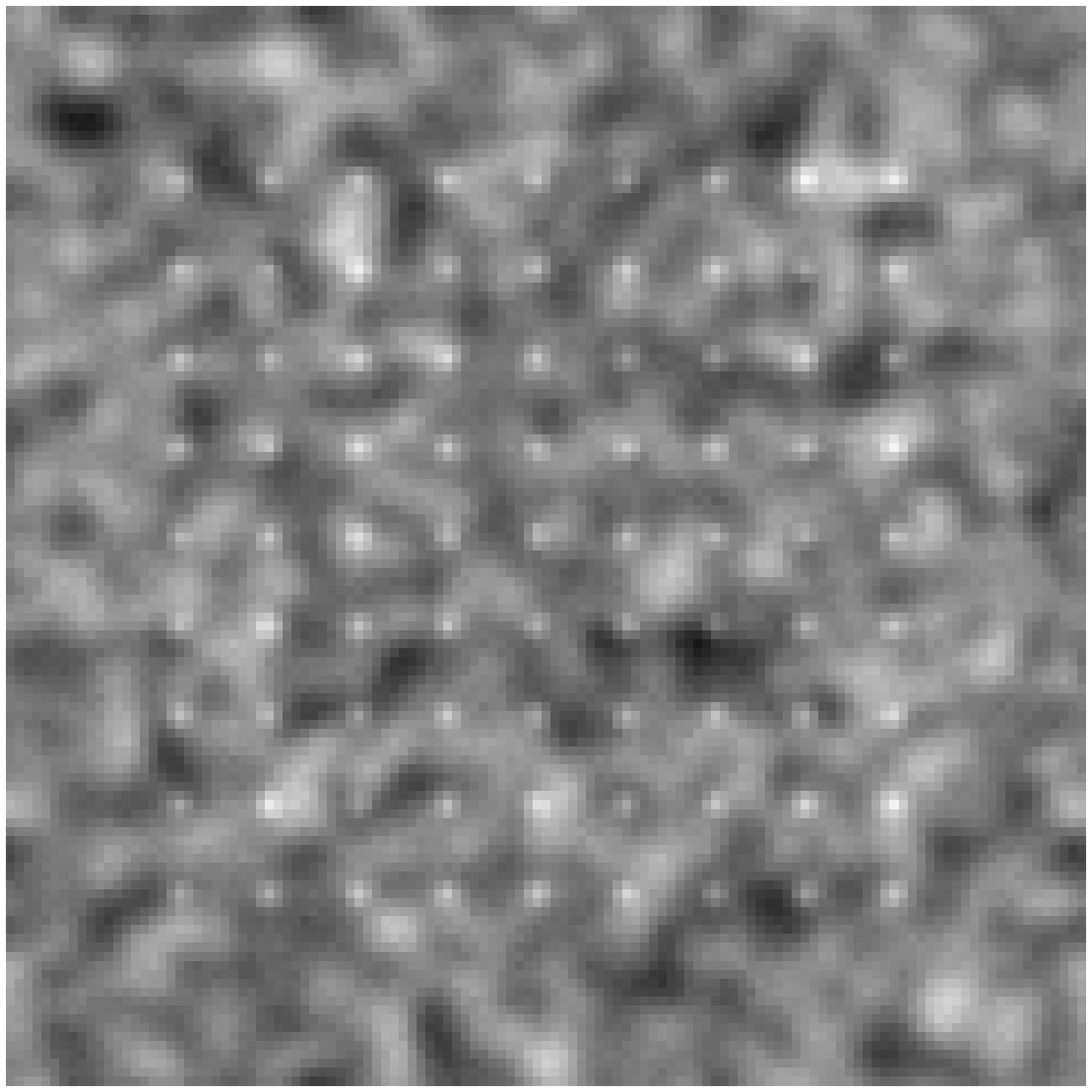}
\caption{$\sigma_s=2000~{[} e^{-} {]}$, $l=4~{[}\textrm{pix}{]}$}
\end{subfigure}
\begin{subfigure}[b]{0.32\linewidth}
\includegraphics[width=\textwidth]{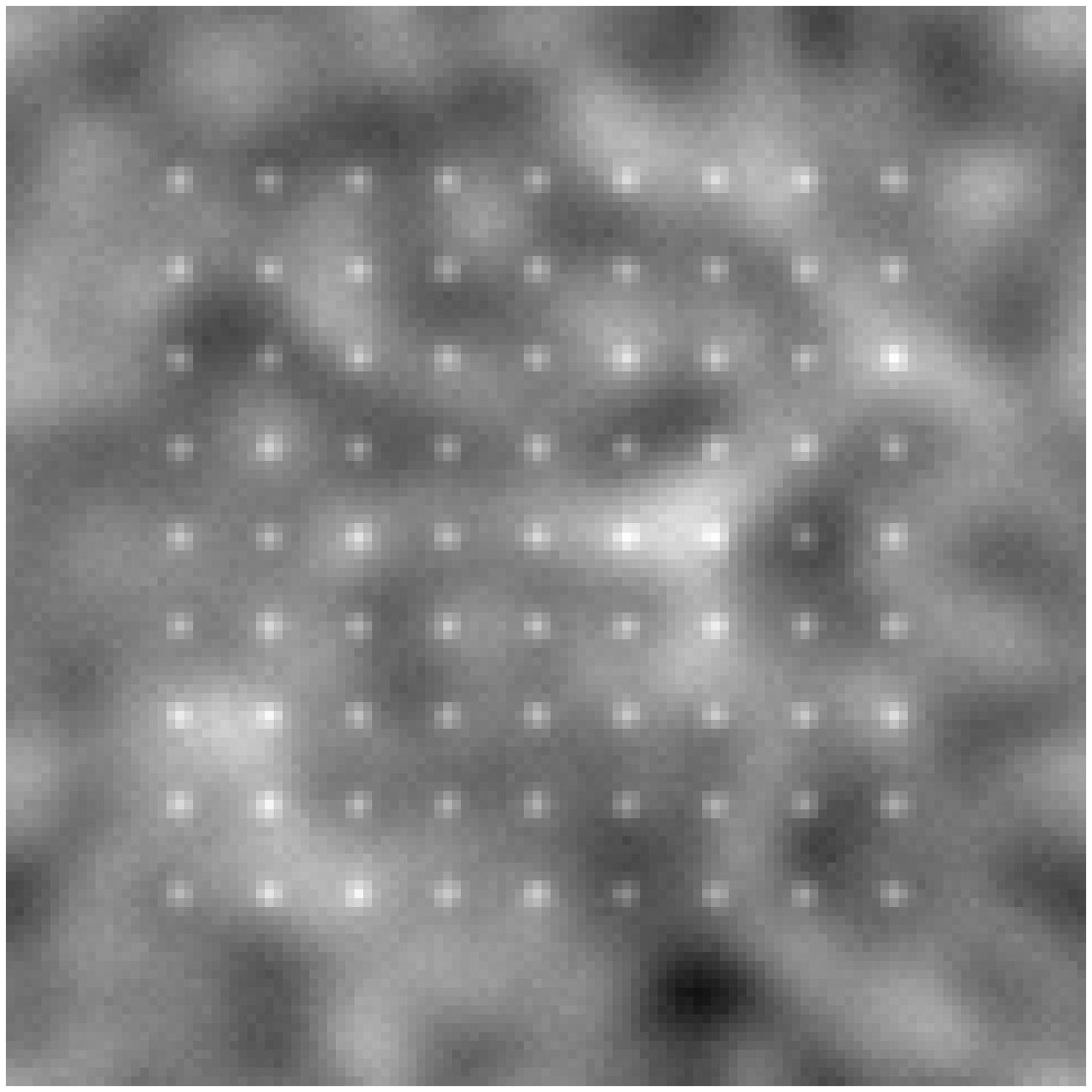}
\caption{$\sigma_s=2000~{[} e^{-} {]}$, $l=8~{[}\textrm{pix}{]}$}
\end{subfigure}
\begin{subfigure}[b]{0.32\linewidth}
\includegraphics[width=\textwidth]{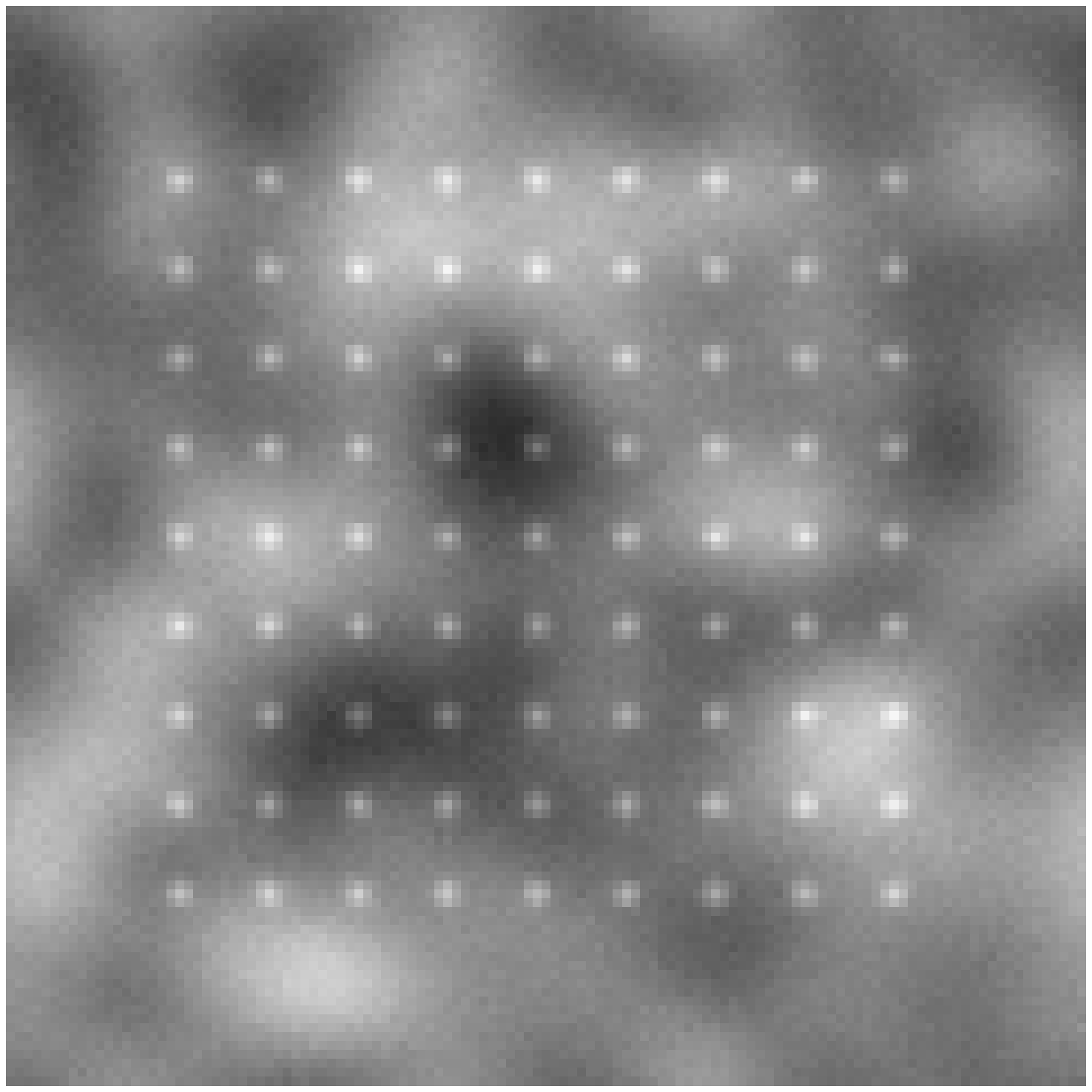}
\caption{$\sigma_s=2000~{[} e^{-} {]}$, $l=12~{[}\textrm{pix}{]}$}
\end{subfigure}
\\\vspace{0.2mm}
\begin{subfigure}[b]{0.32\linewidth}
\includegraphics[width=\textwidth]{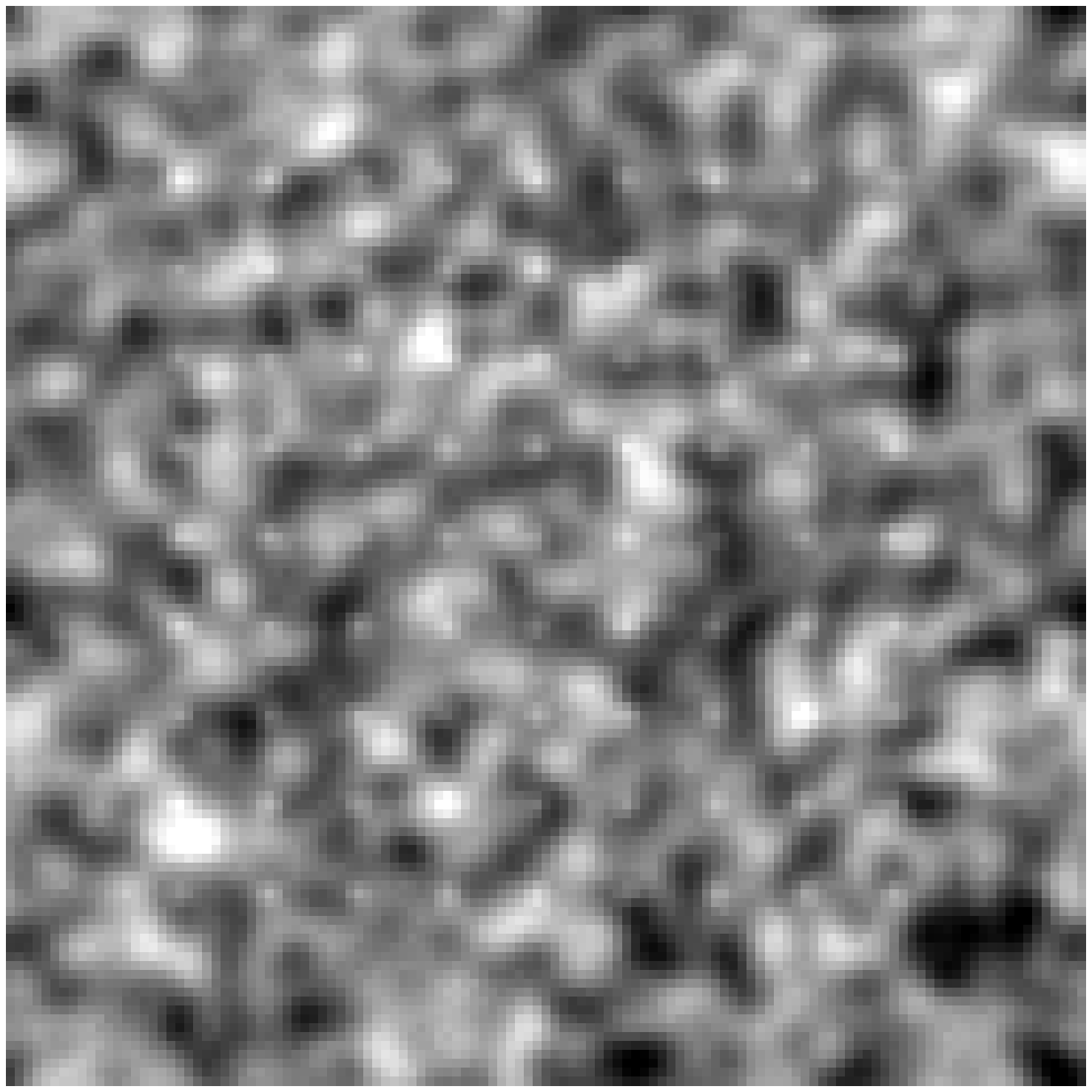}
\caption{$\sigma_s=4000~{[} e^{-} {]}$, $l=4~{[}\textrm{pix}{]}$}
\end{subfigure}
\begin{subfigure}[b]{0.32\linewidth}
\includegraphics[width=\textwidth]{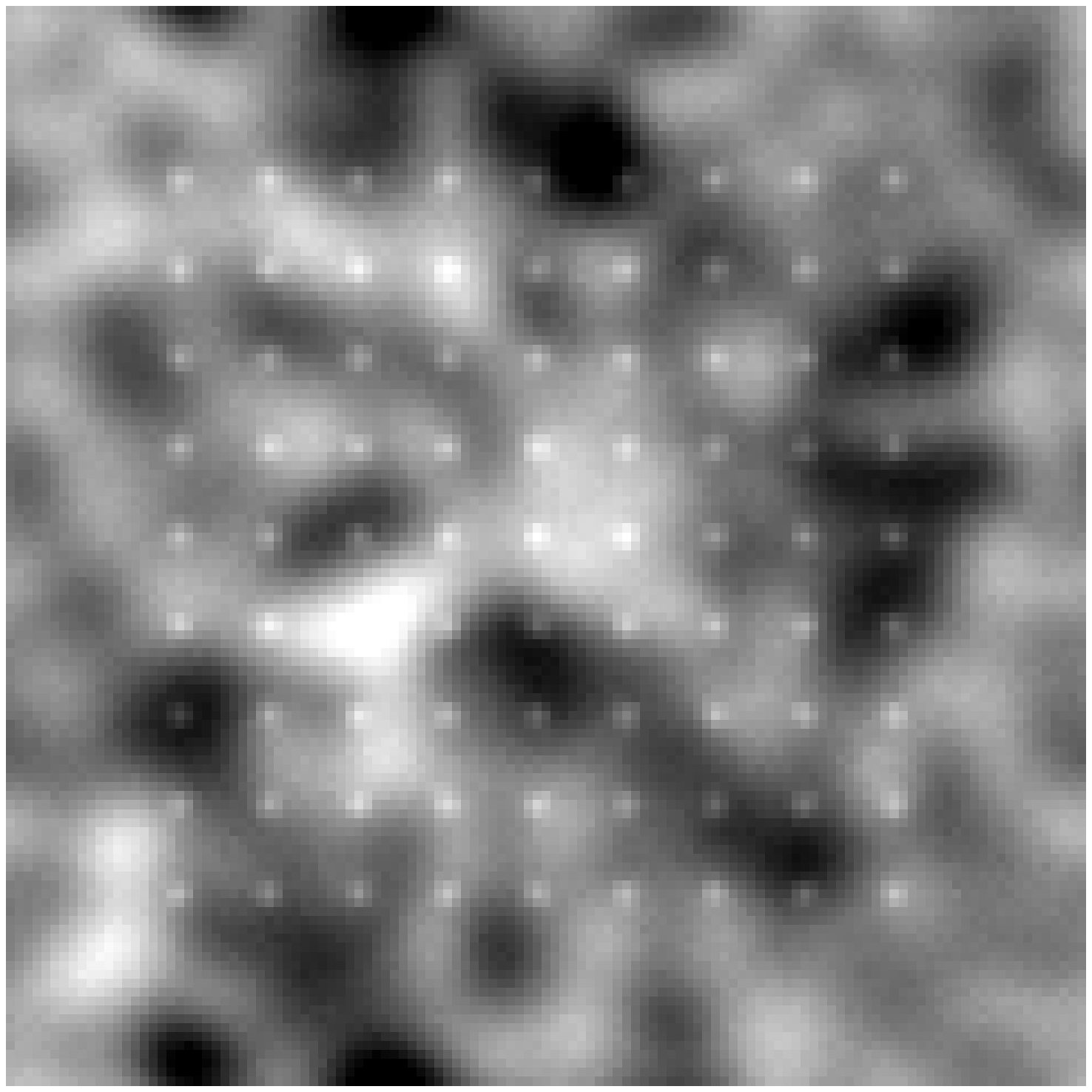}
\caption{$\sigma_s=4000~{[} e^{-} {]}$, $l=8~{[}\textrm{pix}{]}$}
\end{subfigure}
\begin{subfigure}[b]{0.32\linewidth}
\includegraphics[width=\textwidth]{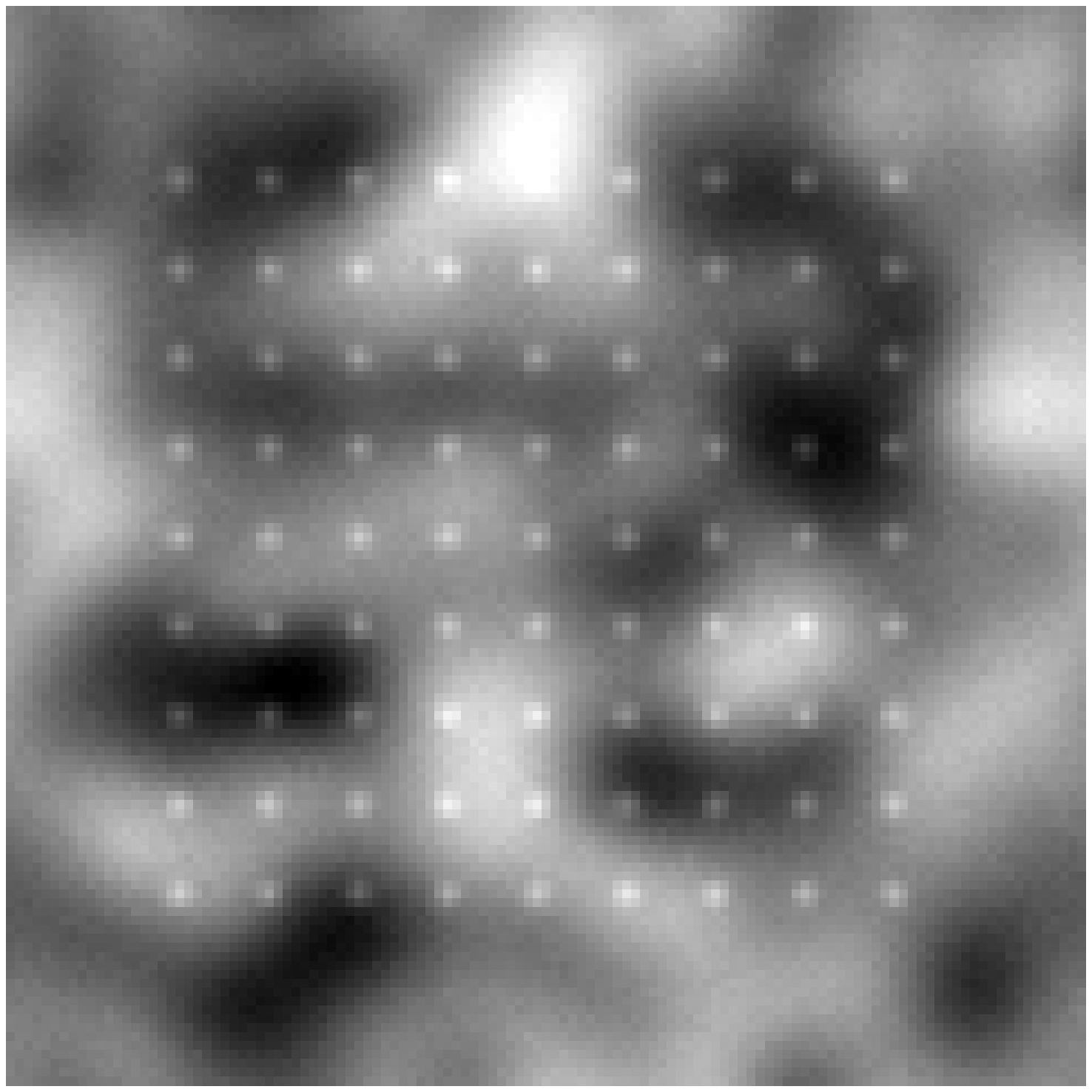}
\caption{$\sigma_s=4000~{[} e^{-} {]}$, $l=12~{[}\textrm{pix}{]}$}
\end{subfigure}
\caption{Exemplary background structures generated using \emph{random rough surfaces} method (\protect\cite{RandomRoughSurfaces}) with inserted stars.}
\label{surfexamples}
\end{figure*}

Almost all astronomical complex background structures contain unwanted objects in the foreground. Hence, we decided to produce artificial backgrounds and simulate a variety of its fluctuation properties. To this end, we utilized the idea of \emph{random rough surfaces} presented in \cite{RandomRoughSurfaces}. In this method, each surface pixel is defined by its height $z(r)$, where $r = (x,y)$ is the pixel position. Such surface has the following properties:
\begin{equation}
\langle z(r)\cdot z(r')\rangle=\sigma_s^2\textrm{ exp}\bigg(\frac{|r-r'|^2}{l^2}\bigg),
\end{equation}
where the angled brackets $\langle\cdot\rangle$ denote an average over an ensemble of realizations; $\sigma_s$ is the RMS (root mean square) of $z$ height; $l$ is the correlation length. By modifying the first parameter, one may obtain different fluctuation amplitudes. Adjusting the correlation length modifies the radius of local intensity changes. To simulate a real background of astronomical CCD images, we included the Poisson distribution of counts and added the Gaussian amplifier noise. 

Similarly to the simulations presented in \cite{BadPixelPopowicz} and \cite{Sun1,Sun2}, we assumed Gaussian PSF of stars. They were added to previously generated 120$\times$120 background structures. To avoid too many adjustable parameters and for the sake of readability of comparison results, only the background properties ($\sigma_s$ and $l$) were modified, while the electronic noise level ($\sigma_{el}$), stars amplitude ($A_{PSF}$) and the standard deviation of stars PSF ($\sigma_{PSF}$) were constant:  $\sigma_{el} = 10$ [e$^-$], $A_{PSF} = 5000$ [e$^-$], $\sigma_{PSF} = 1$ [pix]. We present the set of generated images in Fig. \ref{surfexamples}.

Our dataset consists of 3400 simulated 120$\times$120 pixel images, where the surface parameters were in the following range: $\sigma_s = 200\sim4000$ [e$^-$] (with step 200 [e$^-$]), $l = 3\sim12$ [pix] (with step 0.5 [pix]). Each parameters combination was repeated 10 times to enable the averaging of results. We did not include correlation length smaller than 4 [pix] and RMS value bigger than 4000 [e$^-$], due to the inability to visually distinguish objects from the local background fluctuations (see the example in Fig. \ref{surfexamples} d).

Three methods and their modifications were employed in the comparison: \emph{SExtractor} background estimation (\cite{SExtractor}), median filtering and our approach. The first of the methods is an example of currently most popular $\sigma$-clipping based estimator, while the second involves the simplest way to filter out the outliers by calculating the median within local sliding window. For both algorithms, we used 3 window sizes: 5$\times$5, 7$\times$7 and 9$\times$9. 

For our method, we employed different interpolation techniques changing the sensitivity parameter: $k=1,2,3$. As it was mentioned, the patch size in our distance transformation should be at least twice as big as the objects to be removed. Since the stars were simulated by Gaussian PSF with $\sigma_{PSF} = 1$ [pix], therefore in our approach we set the patch size at 7$\times$7 pixels (greater than 3$\cdot\sigma_{PSF}$). 

\begin{figure*}
\centering
\begin{subfigure}[b]{0.24\linewidth}
\includegraphics[width=1\textwidth]{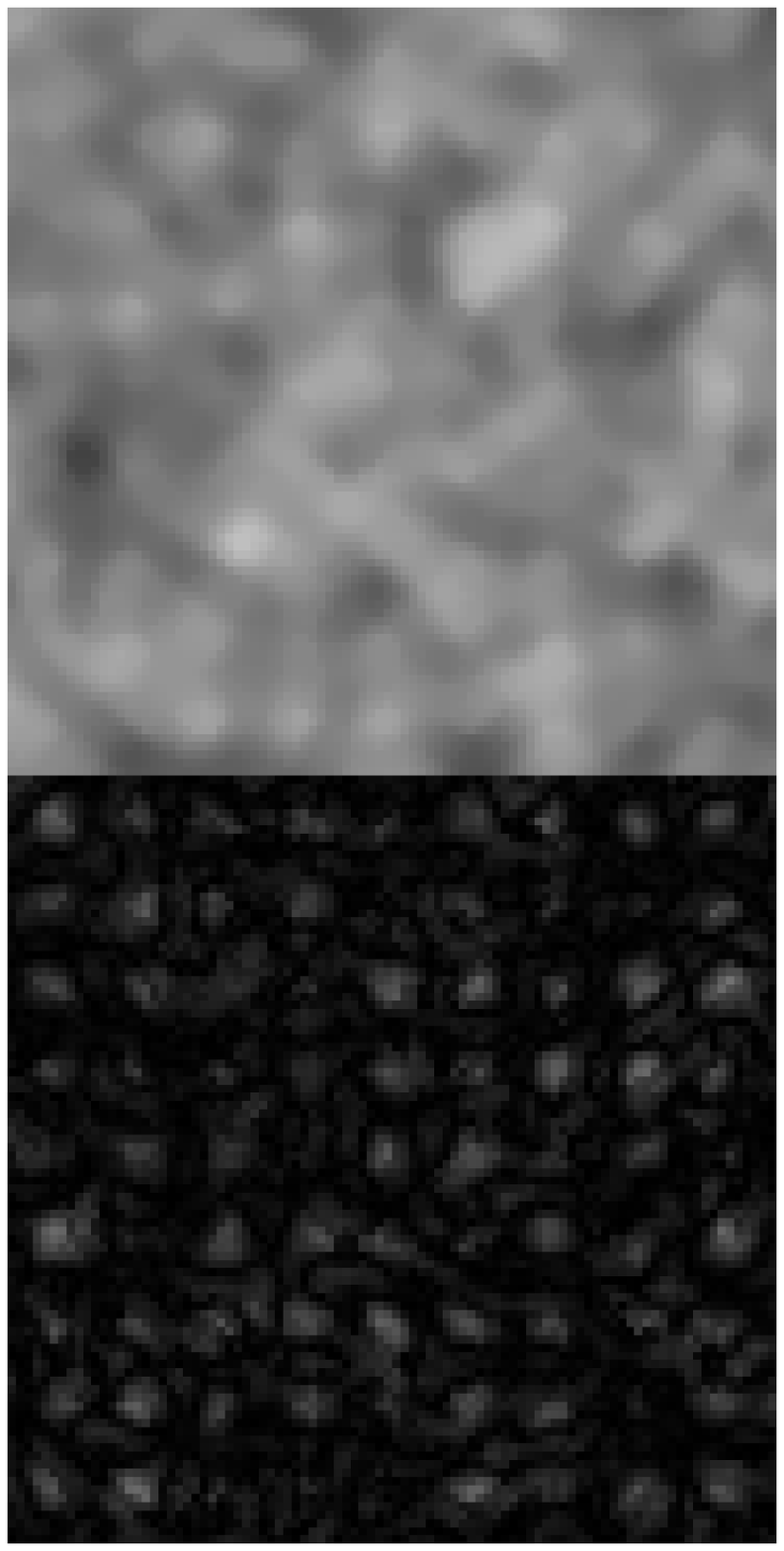}
\caption{Median, (5$\times$5)}
\end{subfigure}
\begin{subfigure}[b]{0.24\linewidth}
\includegraphics[width=1\textwidth]{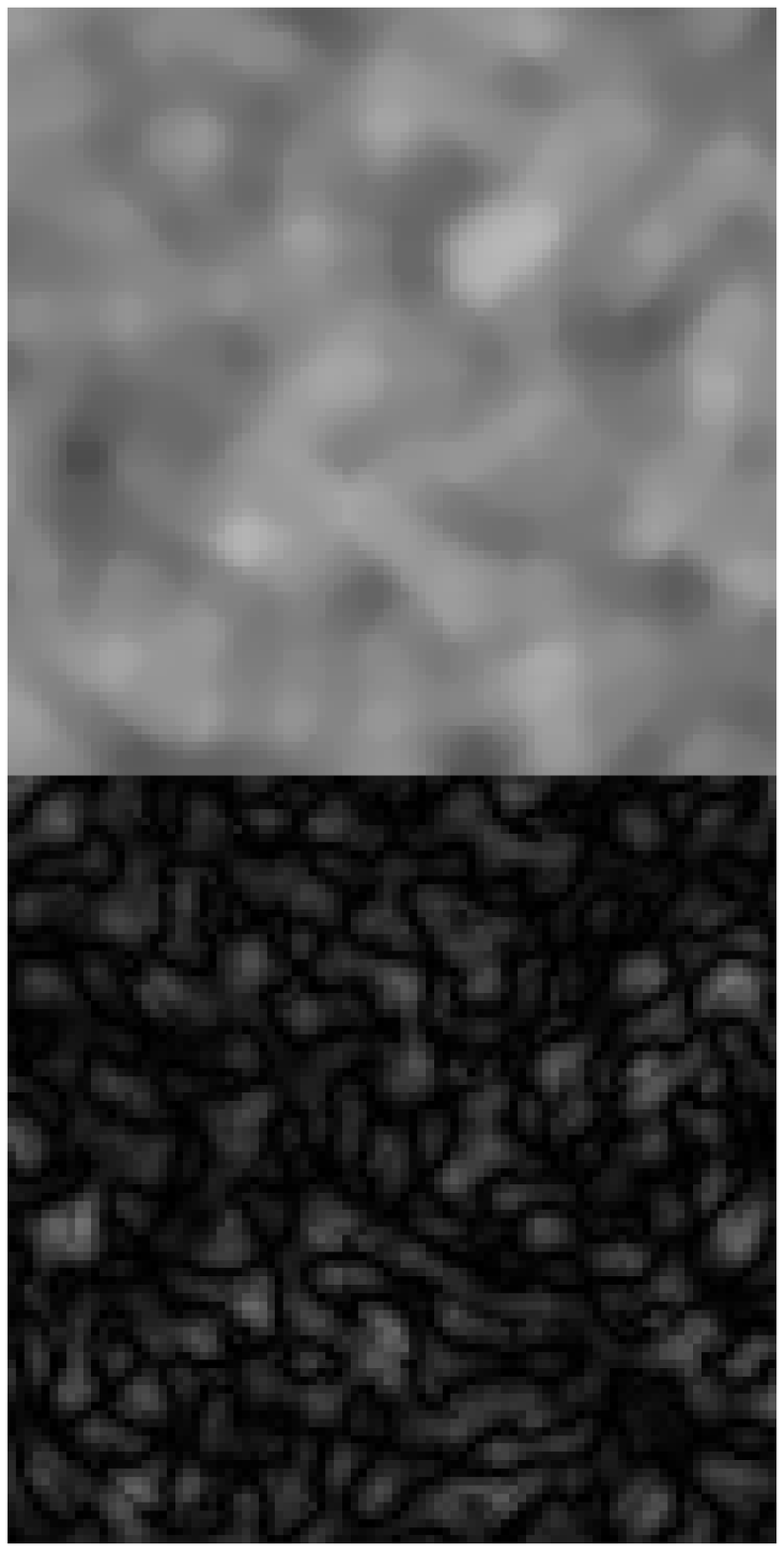}
\caption{Median, (7$\times$7)}
\end{subfigure}
\begin{subfigure}[b]{0.24\linewidth}
\includegraphics[width=1\textwidth]{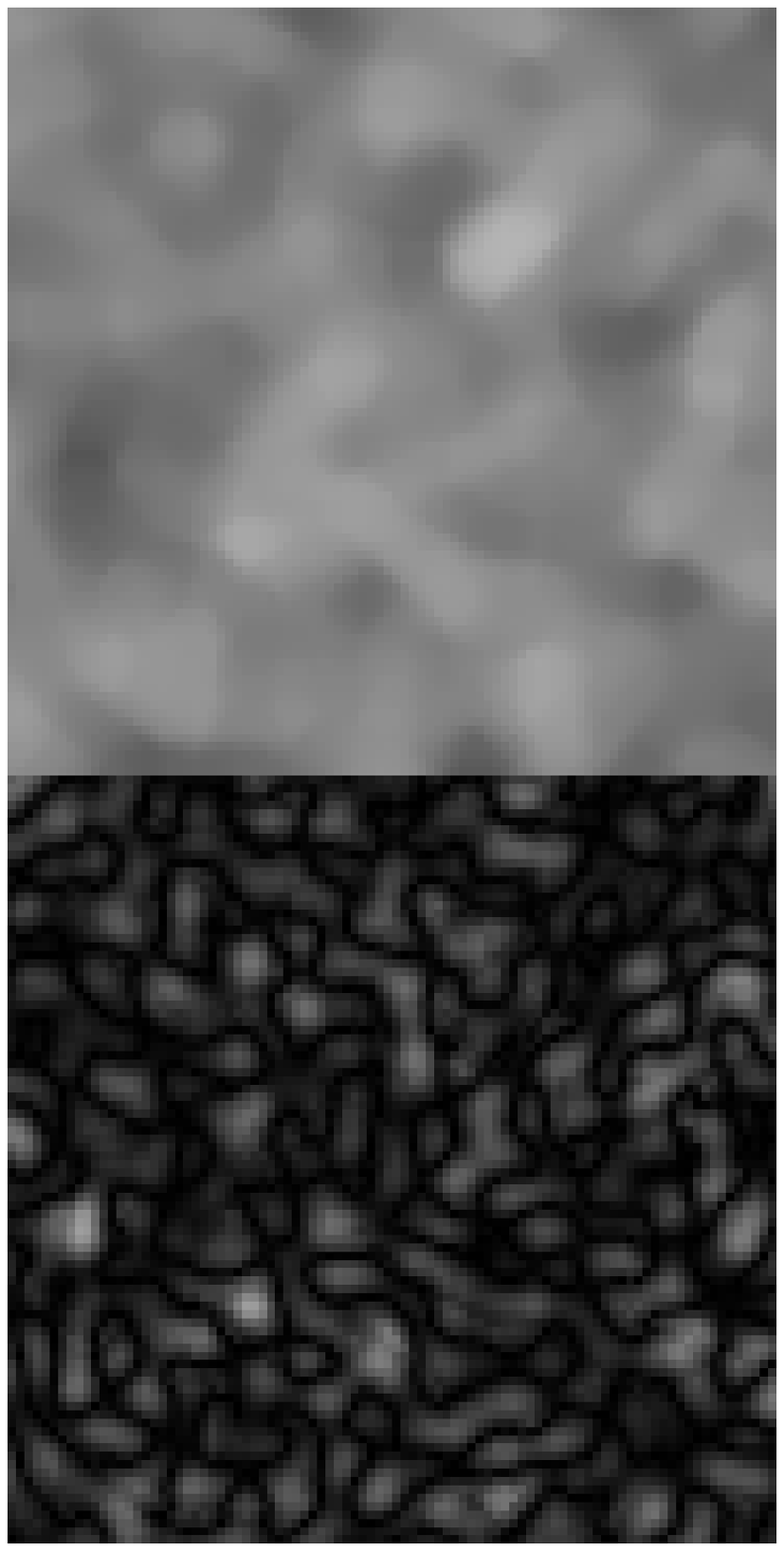}
\caption{Median, (9$\times$9)} 
\end{subfigure}\vspace{5mm}
\\
\begin{subfigure}[b]{0.24\linewidth}	
\includegraphics[width=\textwidth]{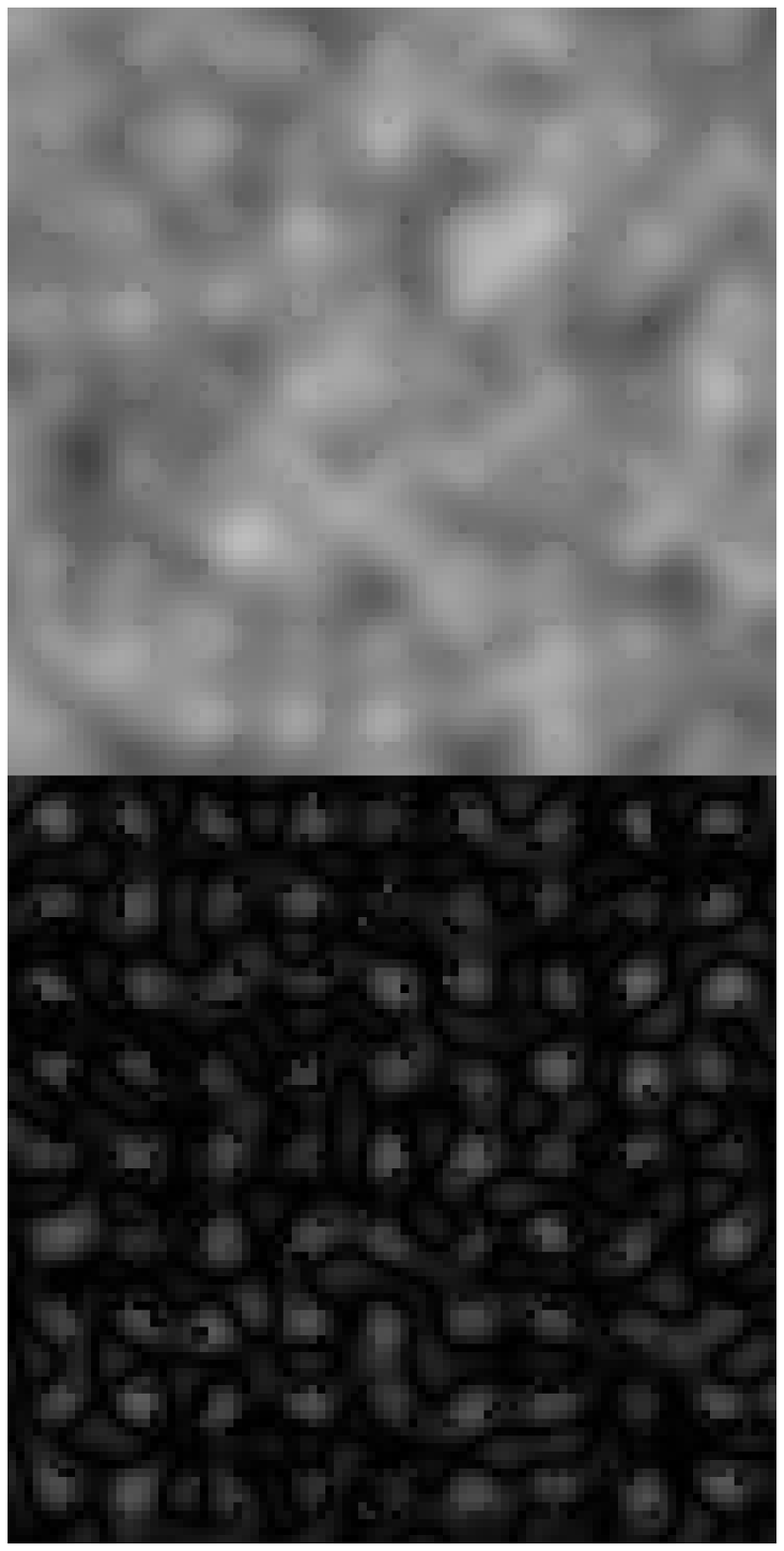}
\caption{\emph{SE}, (5$\times$5)}
\end{subfigure}
\begin{subfigure}[b]{0.24\linewidth}
\includegraphics[width=\textwidth]{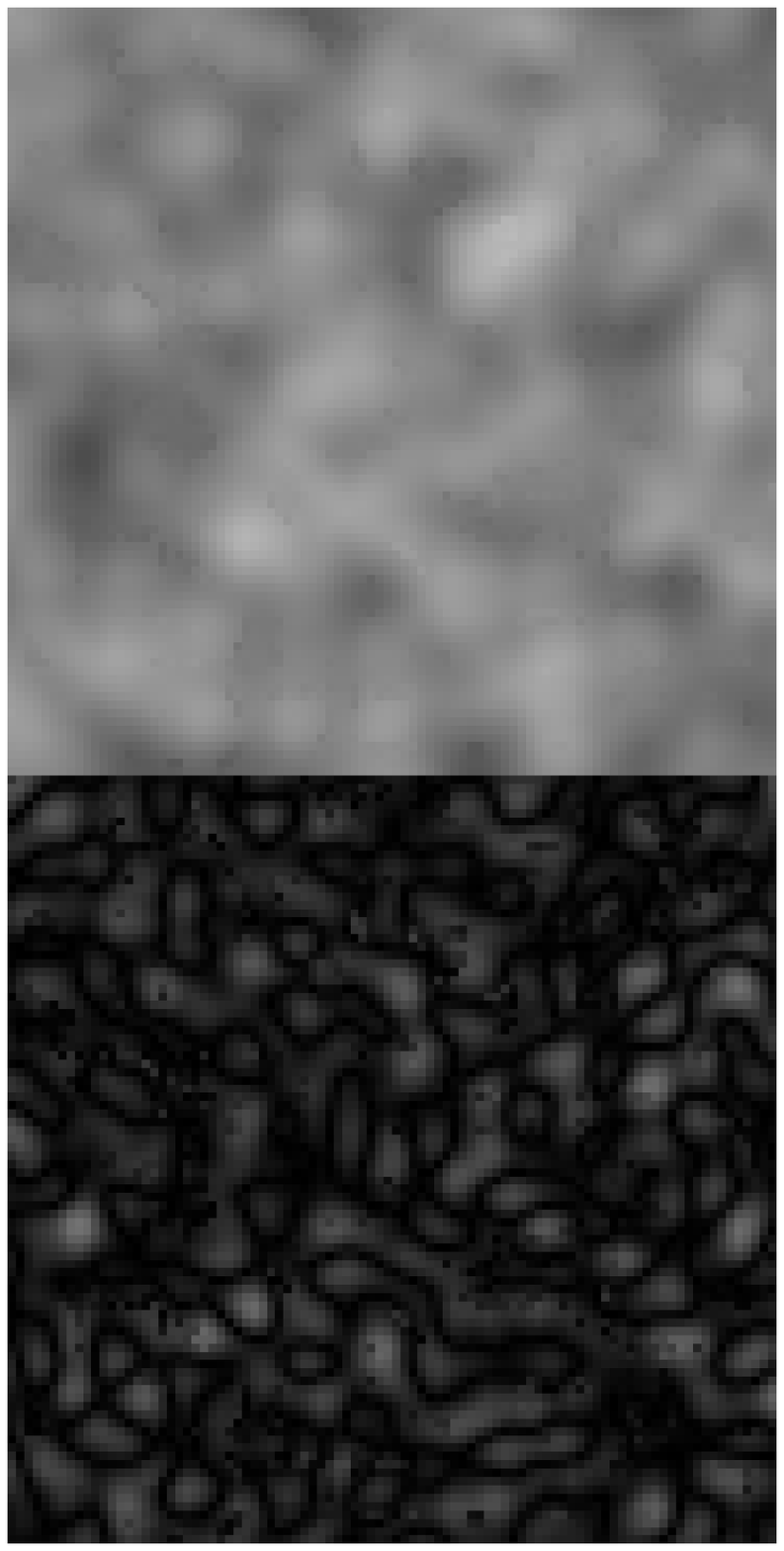}
\caption{\emph{SE}, (7$\times$7)}
\end{subfigure}
\begin{subfigure}[b]{0.24\linewidth}
\includegraphics[width=\textwidth]{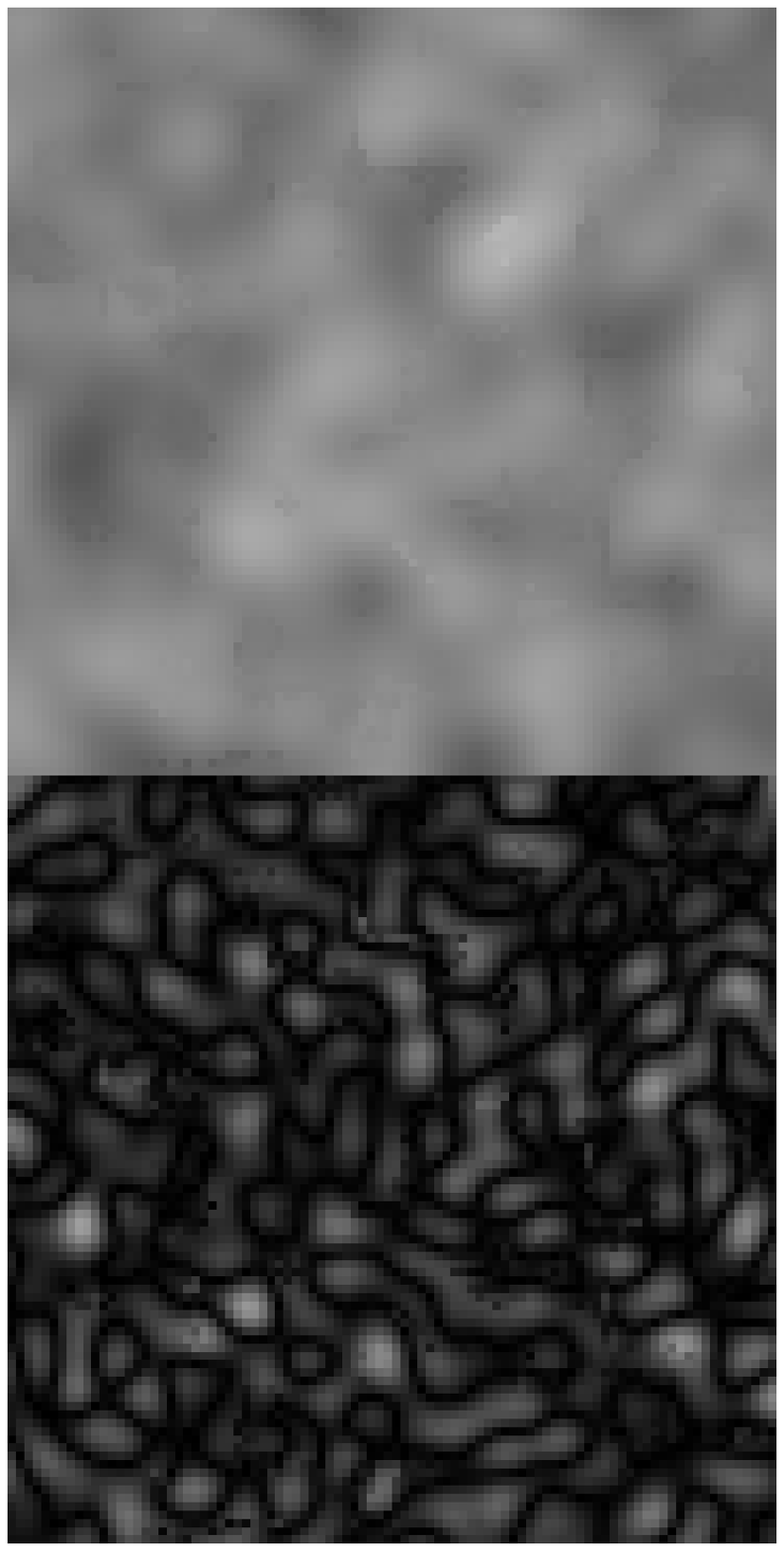}
\caption{\emph{SE}, (9$\times$9)}
\end{subfigure}\vspace{5mm}
\\
\begin{subfigure}[b]{0.24\linewidth}
\includegraphics[width=\textwidth]{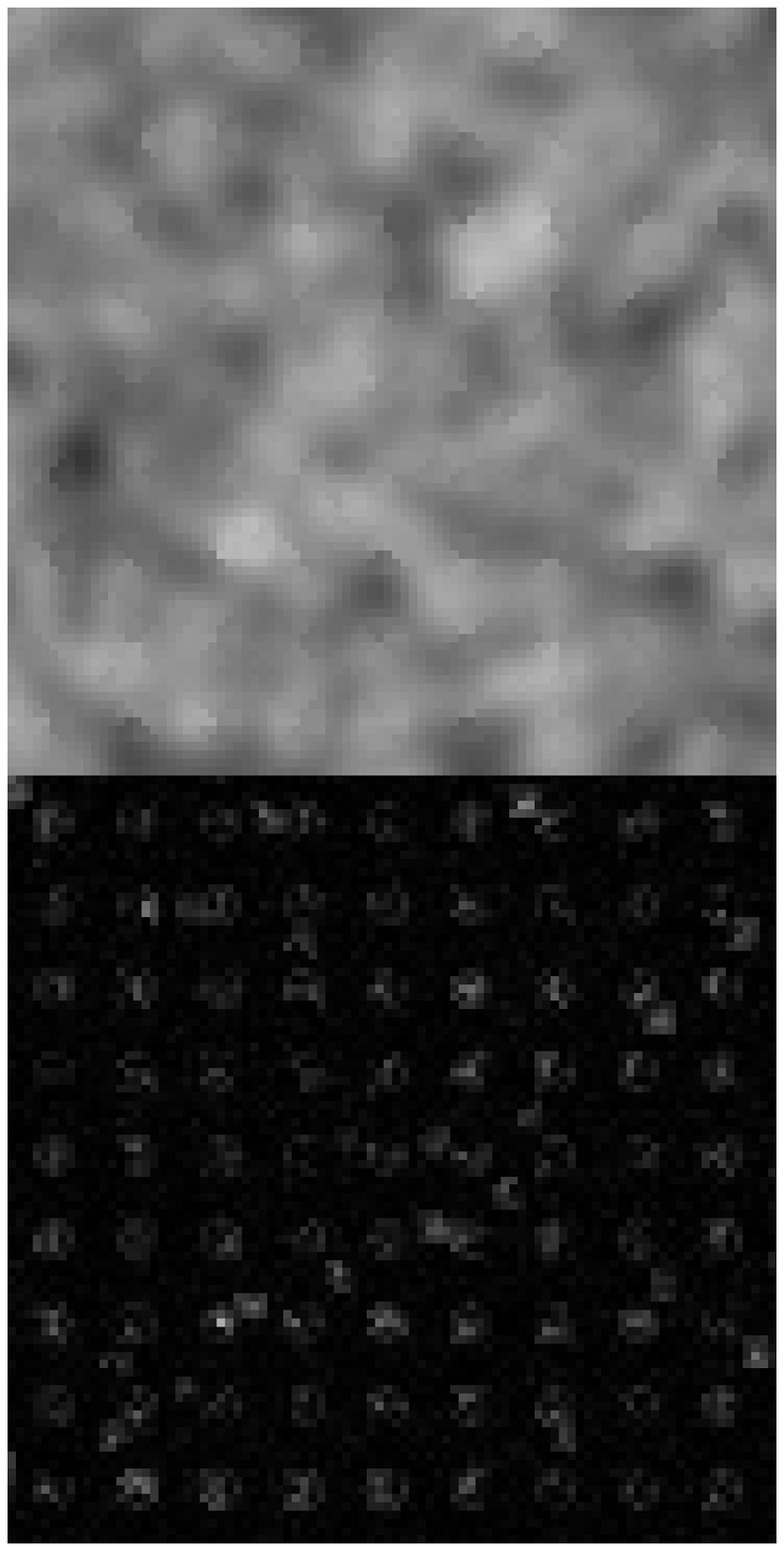}
\caption{Proposed, $k$=1, A}
\end{subfigure}
\begin{subfigure}[b]{0.24\linewidth}
\includegraphics[width=\textwidth]{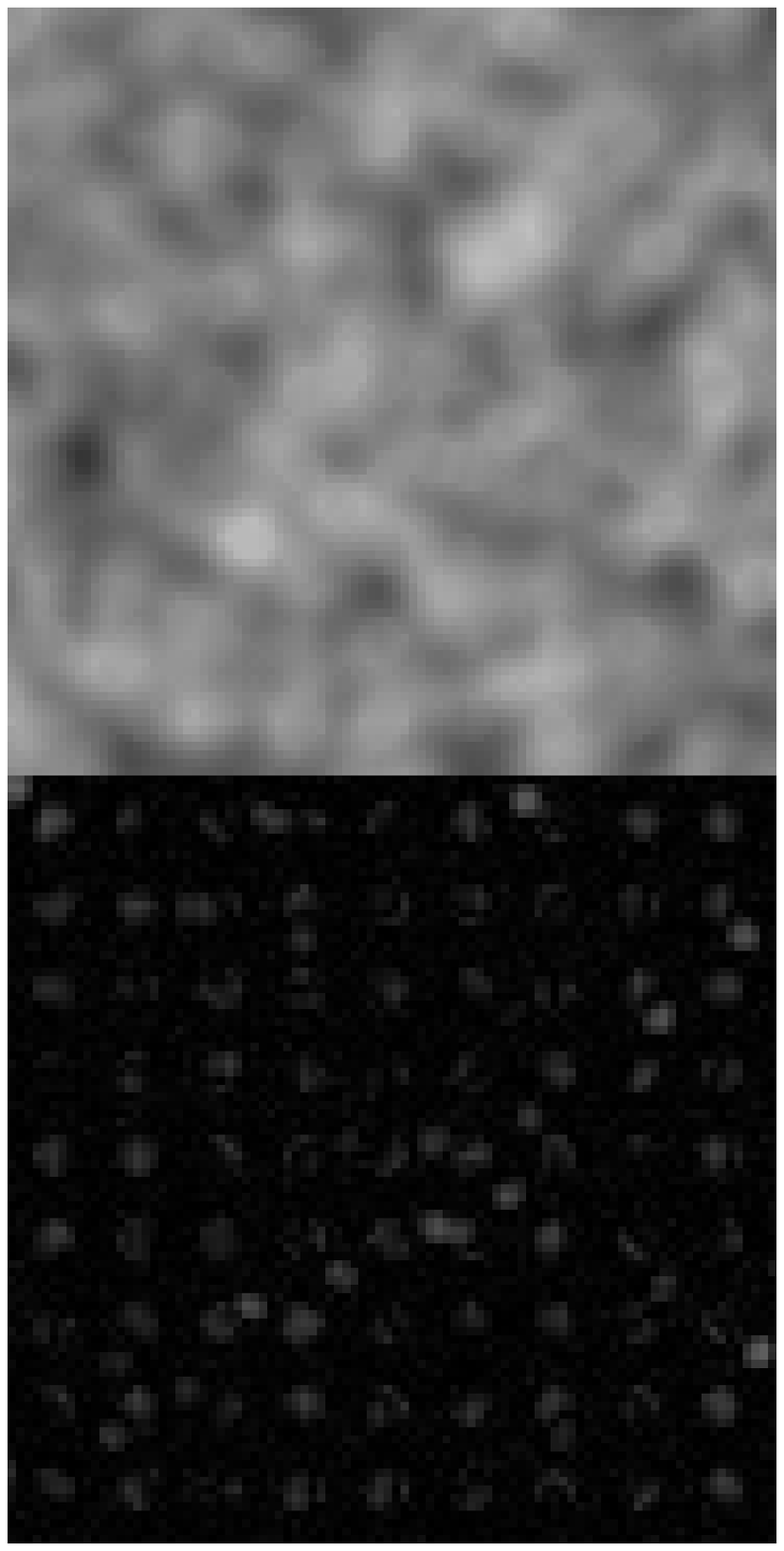}
\caption{Proposed, $k$=1, B}
\end{subfigure}
\begin{subfigure}[b]{0.24\linewidth}
\includegraphics[width=\textwidth]{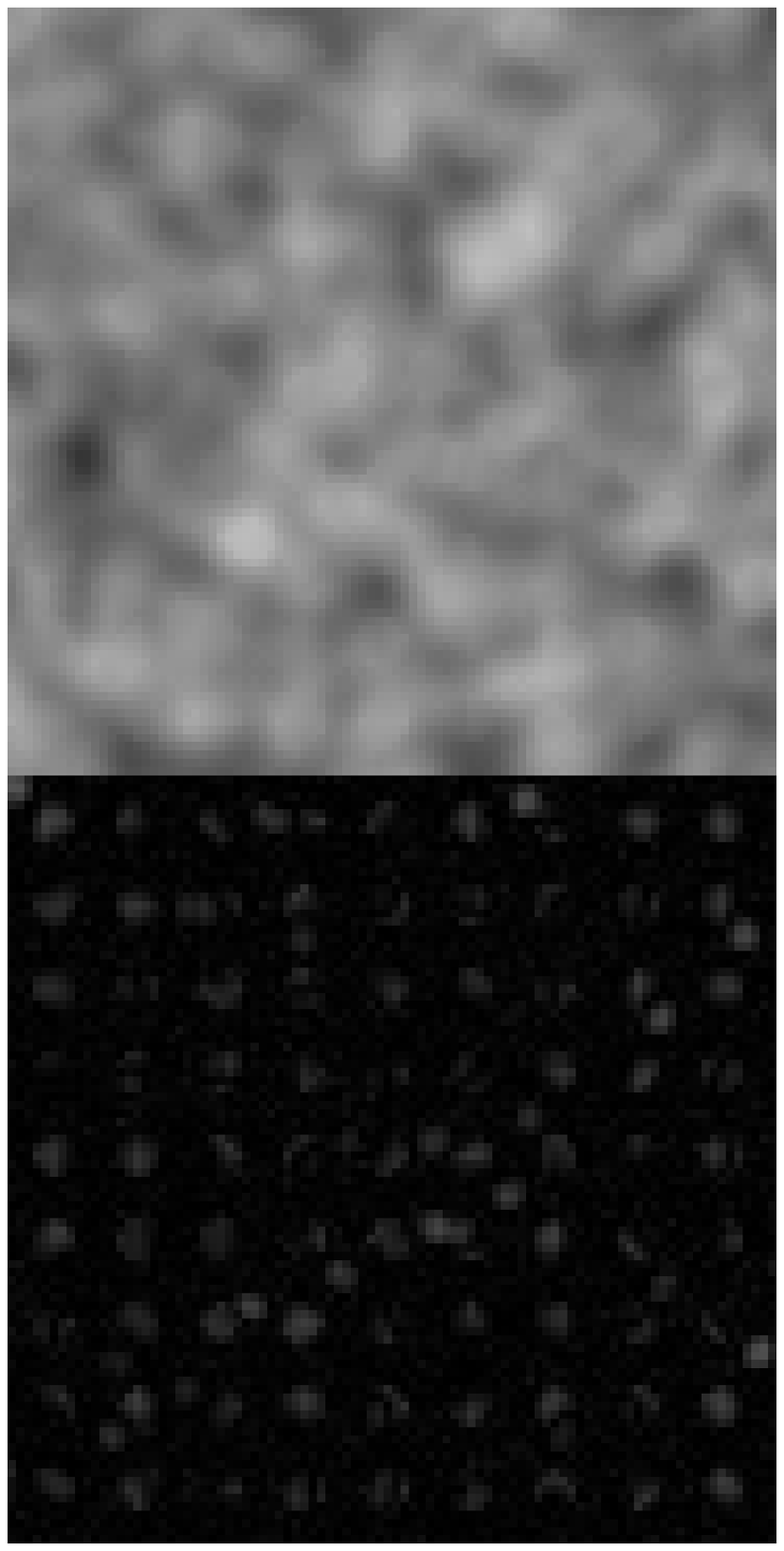}
\caption{Proposed, $k$=1, C}
\end{subfigure}
\begin{subfigure}[b]{0.24\linewidth}
\includegraphics[width=\textwidth]{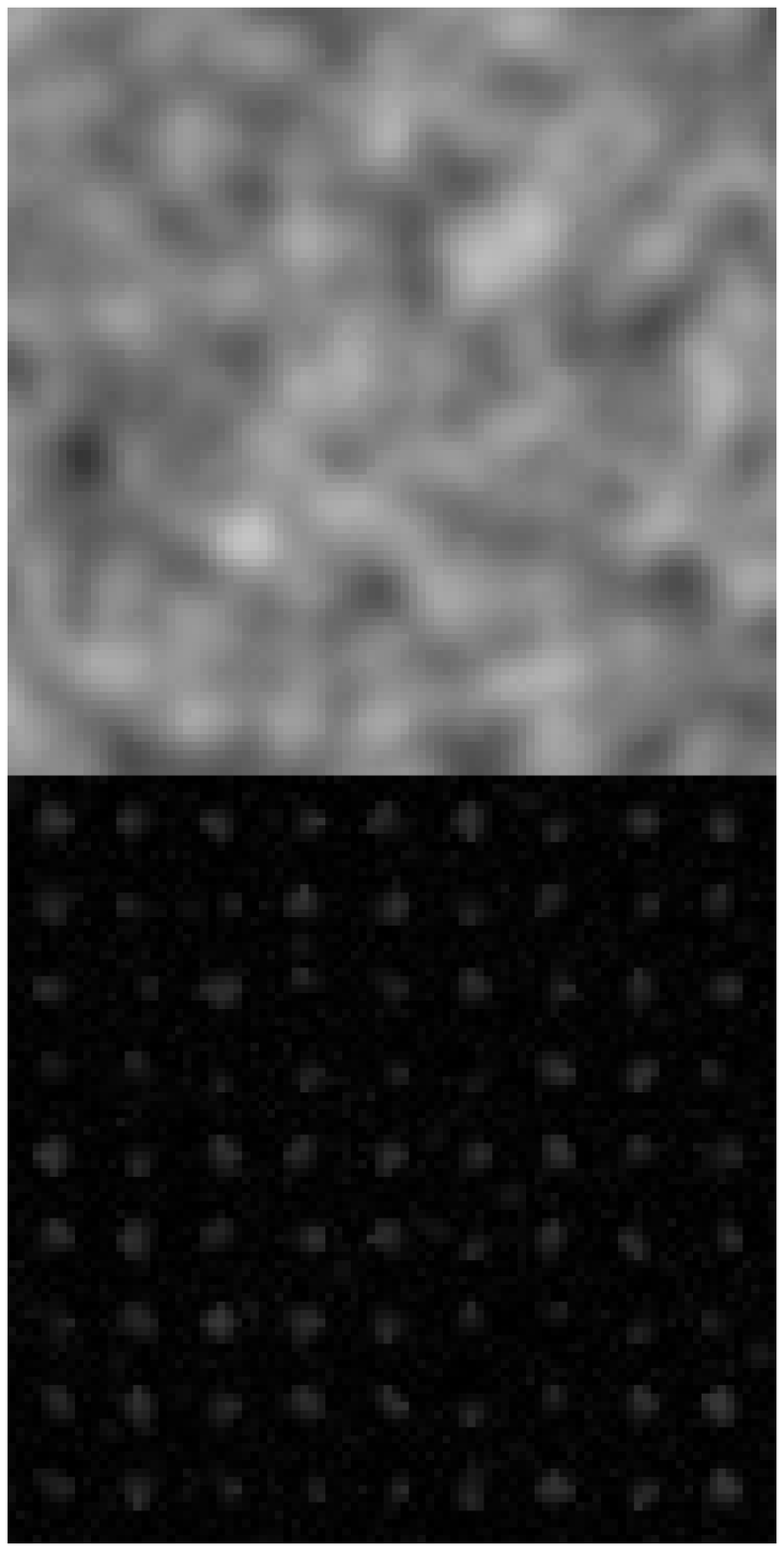}
\caption{Proposed, $k$=1, D}
\end{subfigure}
\caption{Exemplary results of background estimation (upper) with corresponding absolute error map (lower) for compared methods. A, B, C and D refer to various interpolation methods, respectively: nearest neighbor (A), linear (B), cubic (C) and biharmonic (D). The background surface parameters are: $\sigma_s=2000~{[} e^{-} {]}$, $l=8~{[}\textrm{pix}{]}$.}
\label{outcomes}
\end{figure*}

Exemplary outcomes of the competitive algorithms for the surface generated with $\sigma_s=2000~{[} e^{-} {]}$, $l=8~{[}\textrm{pix}{]}$ are presented in Fig. \ref{outcomes}. Each background estimation is accompanied by corresponding error map, which exhibits the absolute difference between the reference background and the estimated one. It should be noted, that for our method, only the small residual errors within the stars are visible. For the other methods, the errors are significantly bigger and, depending on the mask size, they appear either within the objects (like in Fig. \ref{outcomes} a and e) or within the background structures (like in Fig. \ref{outcomes} d and h).

To compare the background estimations provided by the analyzed methods, we employed the most straightforward quality measure - the Root Mean Square Error (RMSE). This metric is calculated using the differences between pixels intensities in reference and transformed image:
\begin{equation}
\textrm{RMSE} = \sqrt{\frac{1}{N}\sum_{j=1}^N (I_j - I_j')^2},
\label{RMSEeq}
\end{equation}
where $N$ is a the number of image pixels, $I_j$ and $I_j'$ is the intensity of pixel $j$ respectively in a filtered and reference image.

For the completeness of our tests, we obtained the results of the photometry performed on included stars after the background removal. We utilized the aperture photometry, where the radius was set at 3 pixels, thus nearly the whole PSF was included in the aperture. The reference number of counts for our artificial star was 30474 [ADU], which is -11.21 instrumental magnitudo. We obtained the average magnitudo error in each analyzed image, employing the formula:
\begin{equation}
\Delta m = \sqrt{\frac{1}{S}\sum_{i=1}^{S} (m_i - m_{ref})^2},
\label{RMSEmag}
\end{equation}
where $S$ is the number of stars in the image ($S=81$), $m_i$ is the magnitudo of $i$-th star in the image and $m_{ref}=-11.21$ [mag] is the reference magnitudo. 

%-------------------------------------------------------------------------------------------------------------------------
\section{Results}

\begin{figure*}
\centering

\begin{subfigure}[b]{0.33\linewidth}
\includegraphics[width=\textwidth]{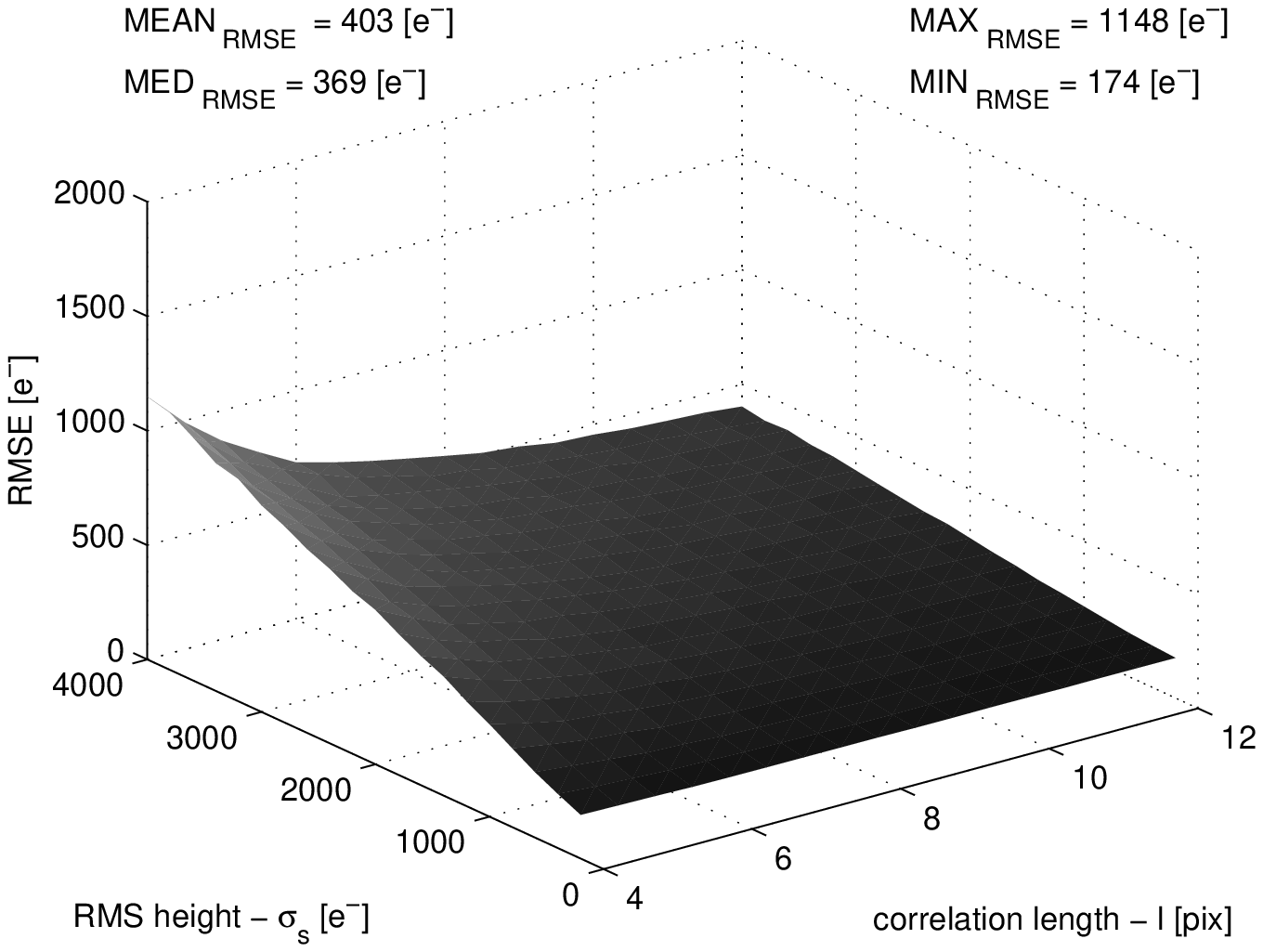}\caption{5$\times$5 mask}
\end{subfigure}
\begin{subfigure}[b]{0.33\linewidth}
\includegraphics[width=\textwidth]{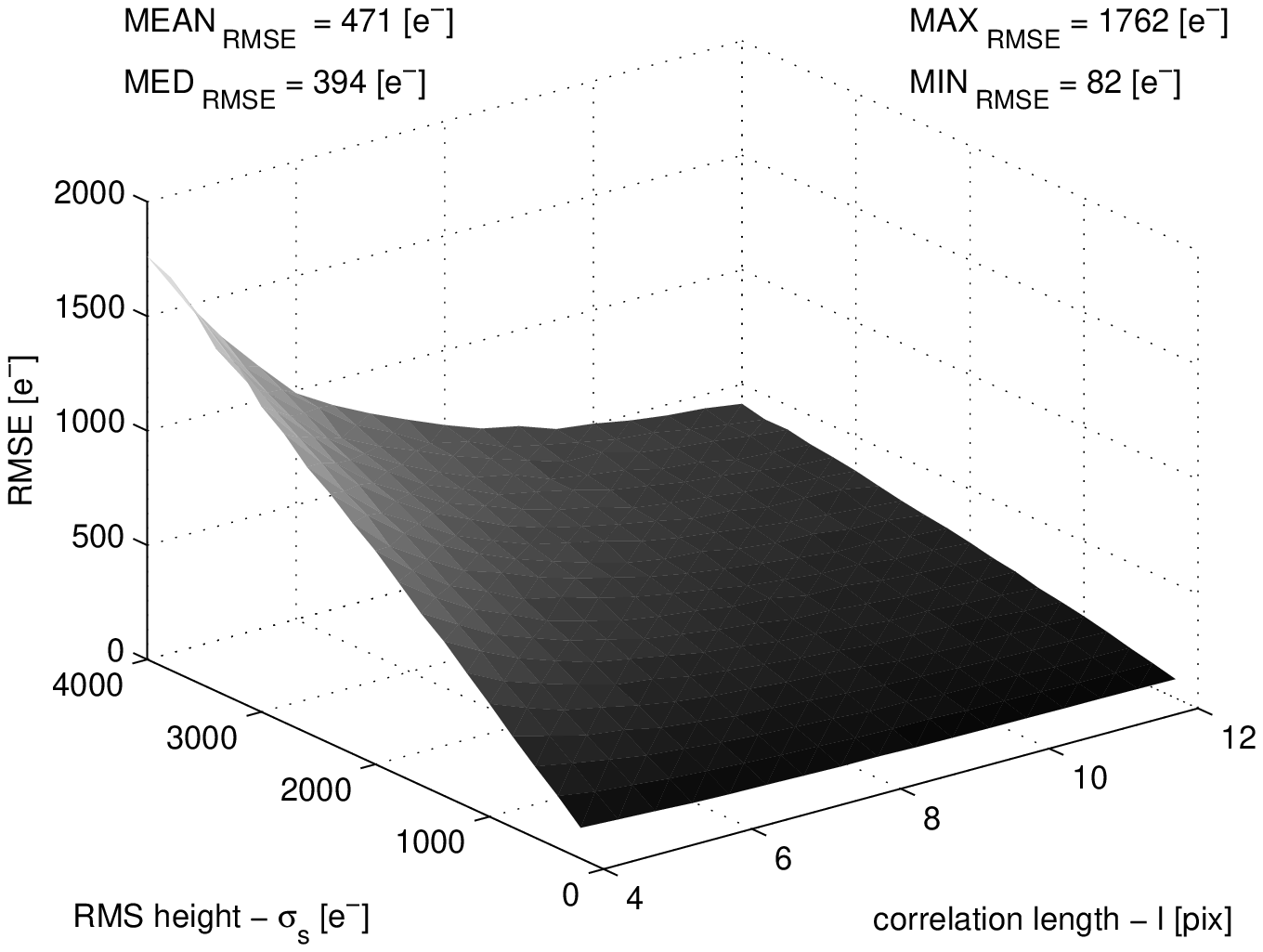}\caption{7$\times$7 mask}
\end{subfigure}
\begin{subfigure}[b]{0.33\linewidth}
\includegraphics[width=\textwidth]{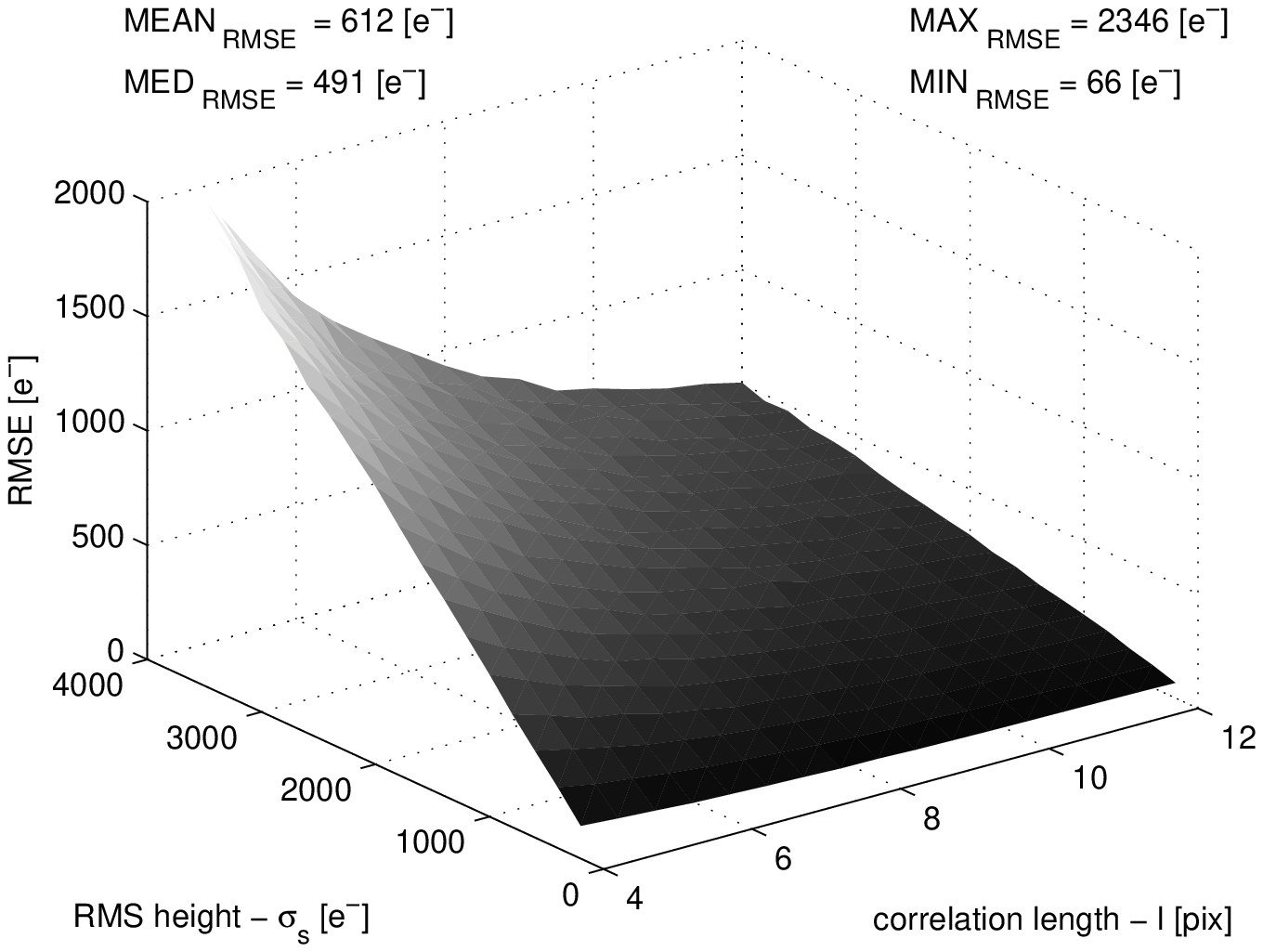}\caption{9$\times$9 mask}
\end{subfigure}

\caption{RMSE of median filtering based background estimation for different mask sizes and all $l$-$\sigma_s$ background parameters.}
\label{res1}
\end{figure*}

\begin{figure*}
\centering
\begin{subfigure}[b]{0.33\linewidth}
\includegraphics[width=\textwidth]{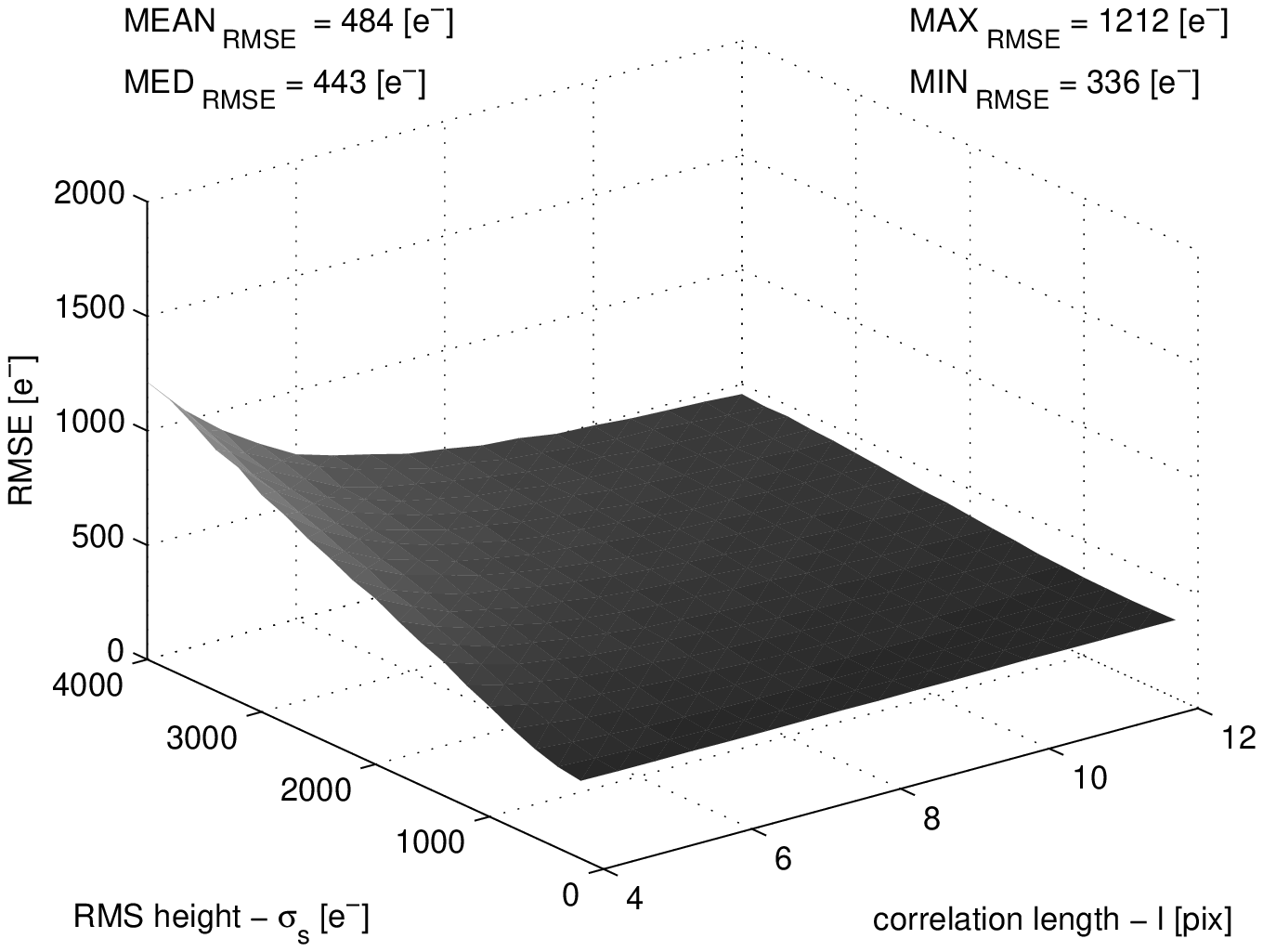}\caption{5$\times$5 mask}
\end{subfigure}
\begin{subfigure}[b]{0.33\linewidth}
\includegraphics[width=\textwidth]{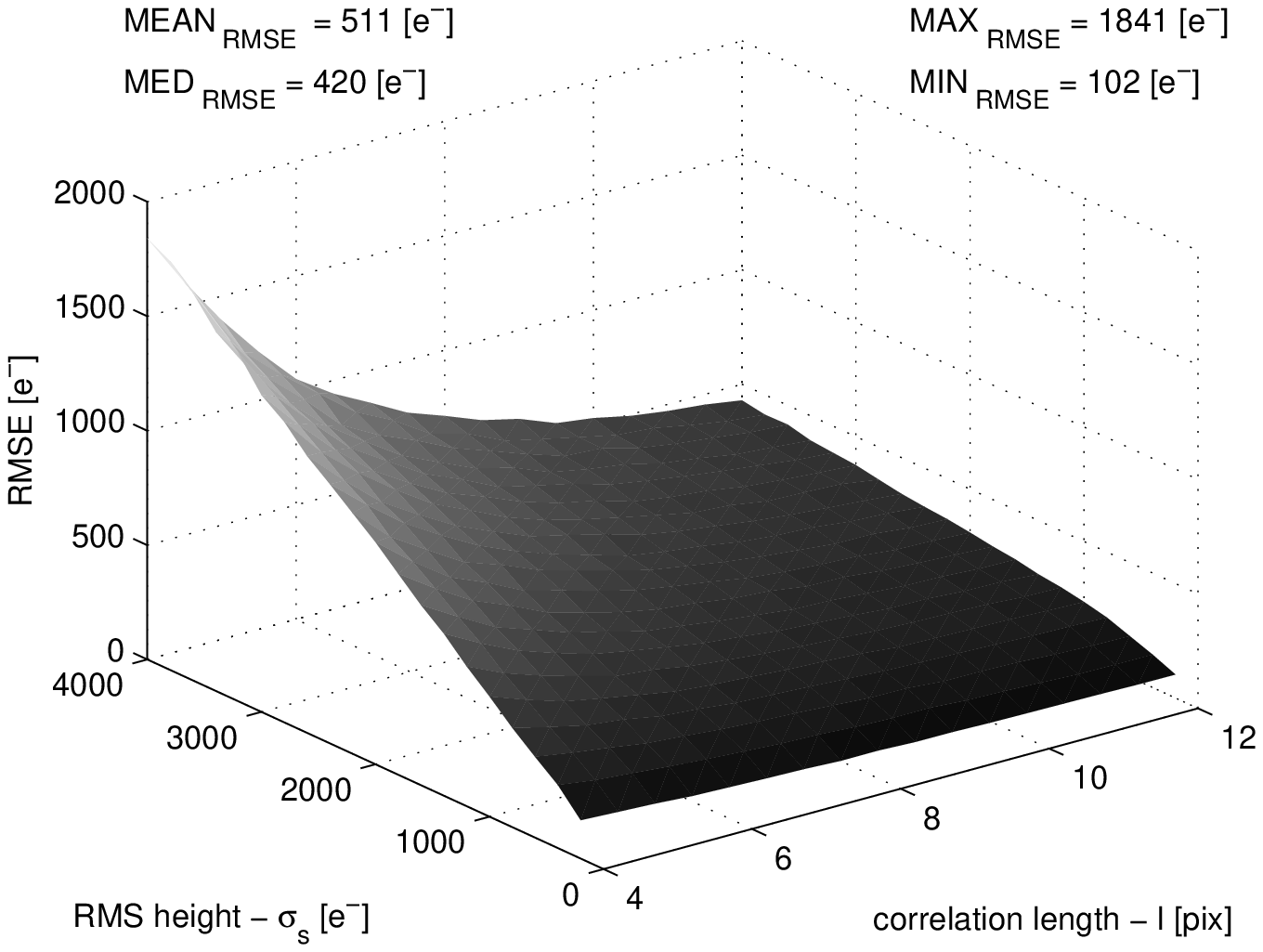}\caption{7$\times$7 mask}
\end{subfigure}
\begin{subfigure}[b]{0.33\linewidth}
\includegraphics[width=\textwidth]{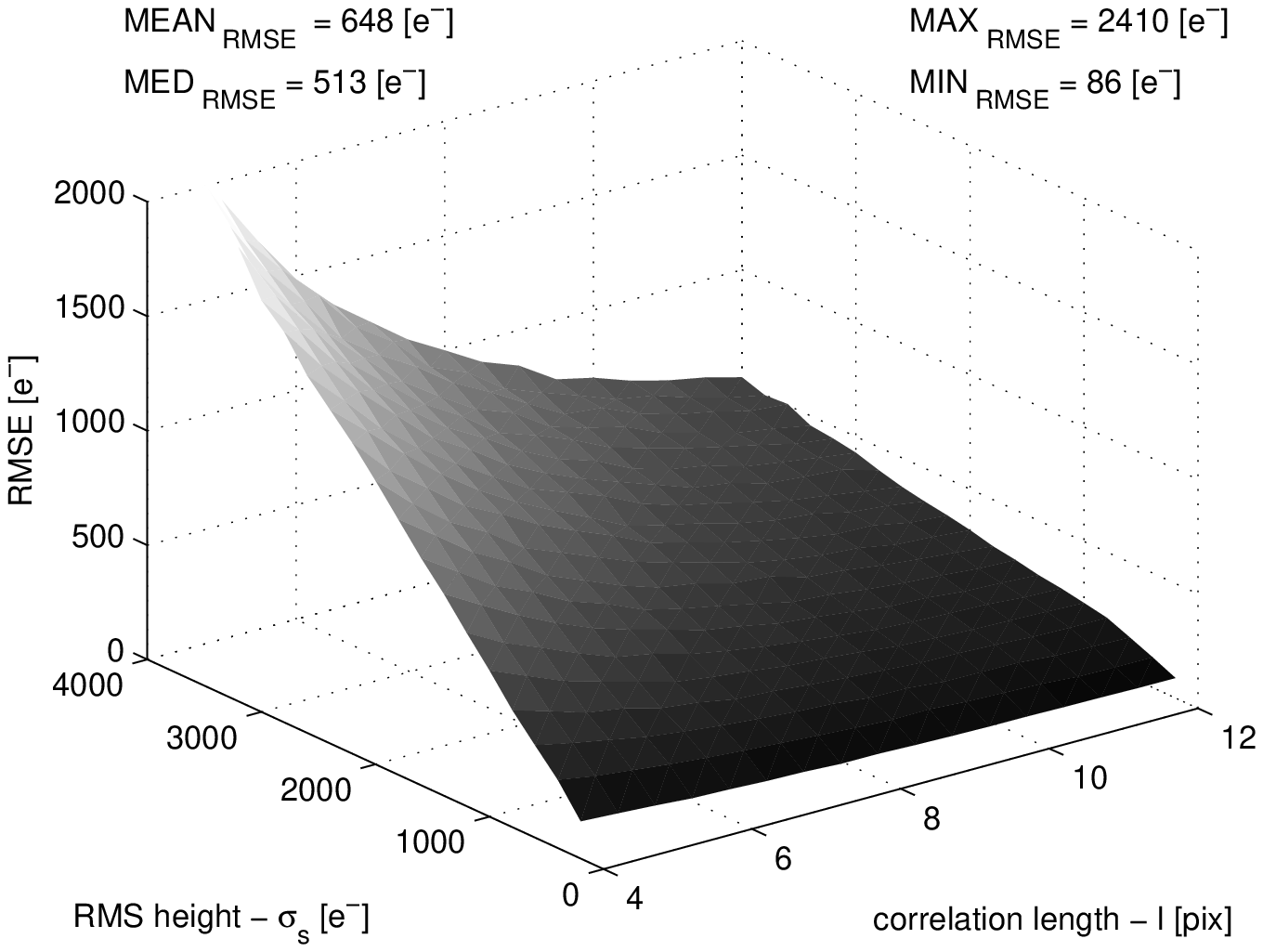}\caption{9$\times$9 mask}
\end{subfigure}
\caption{RMSE of \emph{SExtractor} based background estimation for different mask sizes and all $l$-$\sigma_s$ background parameters.}
\label{res2}
\end{figure*}

\begin{figure*}
\centering
\begin{subfigure}[b]{0.35\linewidth}

\includegraphics[width=\textwidth]{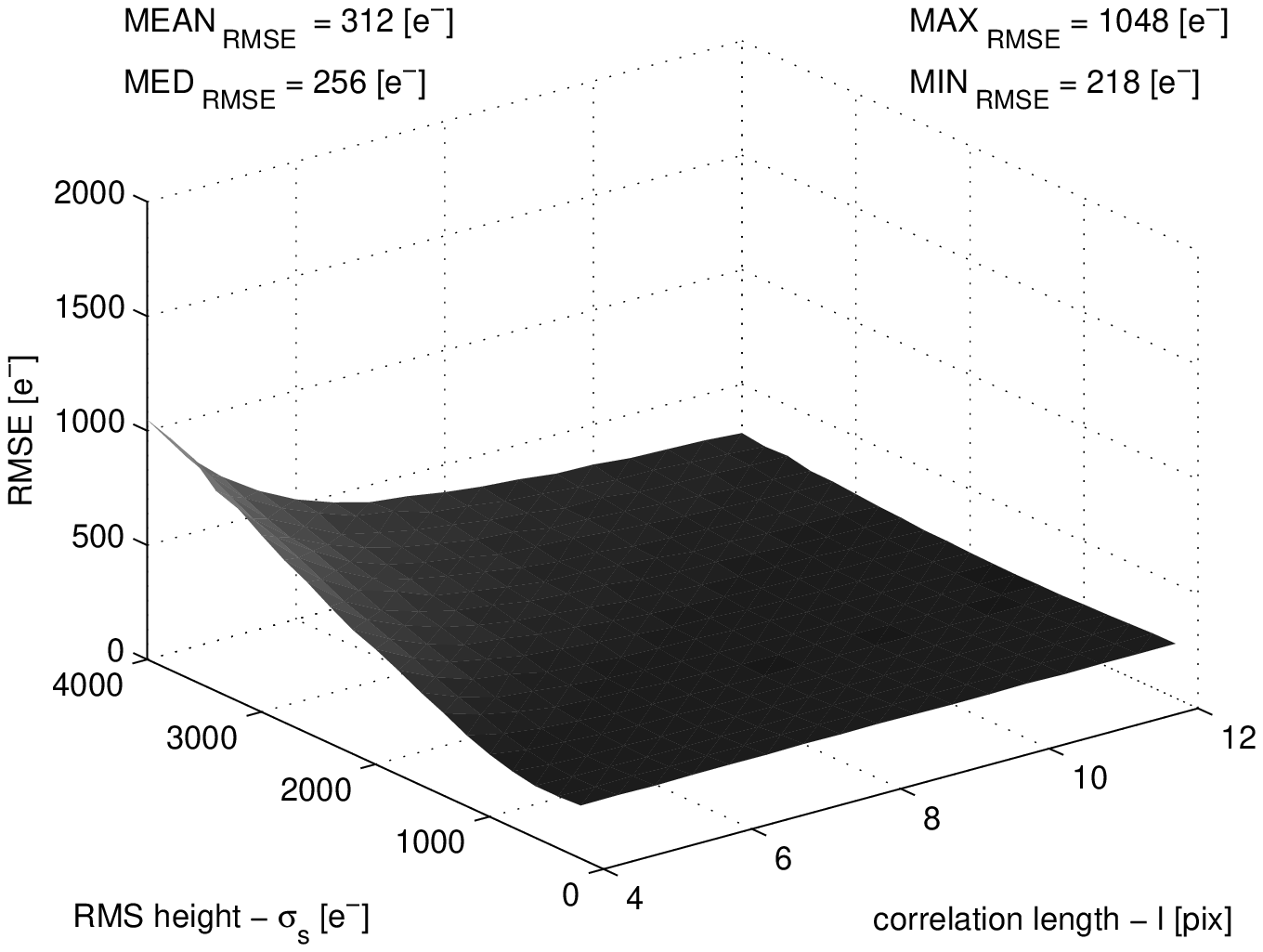}\caption{Interpolation A}
\end{subfigure}
\begin{subfigure}[b]{0.35\linewidth}

\includegraphics[width=\textwidth]{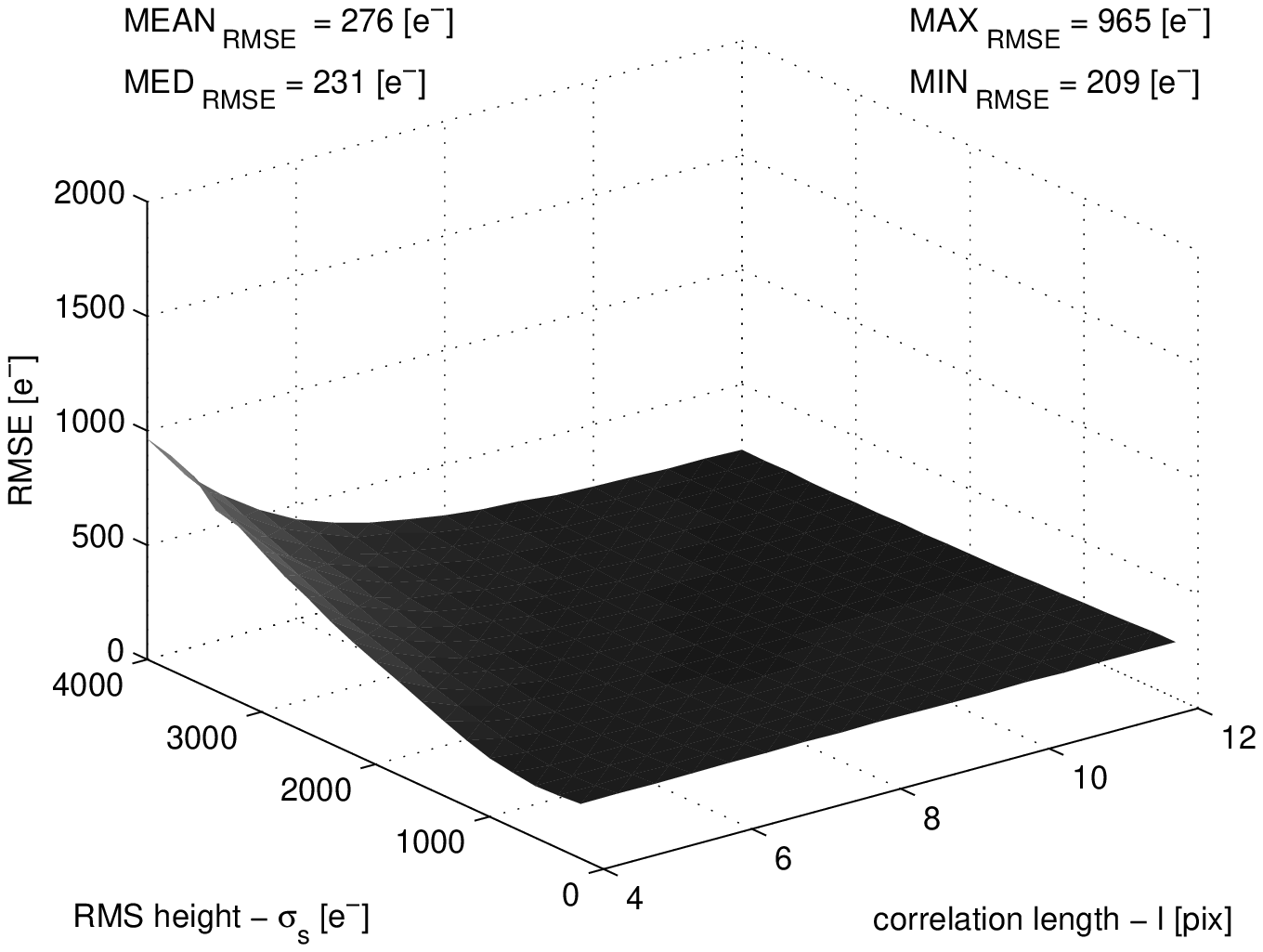}\caption{Interpolation B}
\end{subfigure}
\\
\begin{subfigure}[b]{0.35\linewidth}

	\includegraphics[width=\textwidth]{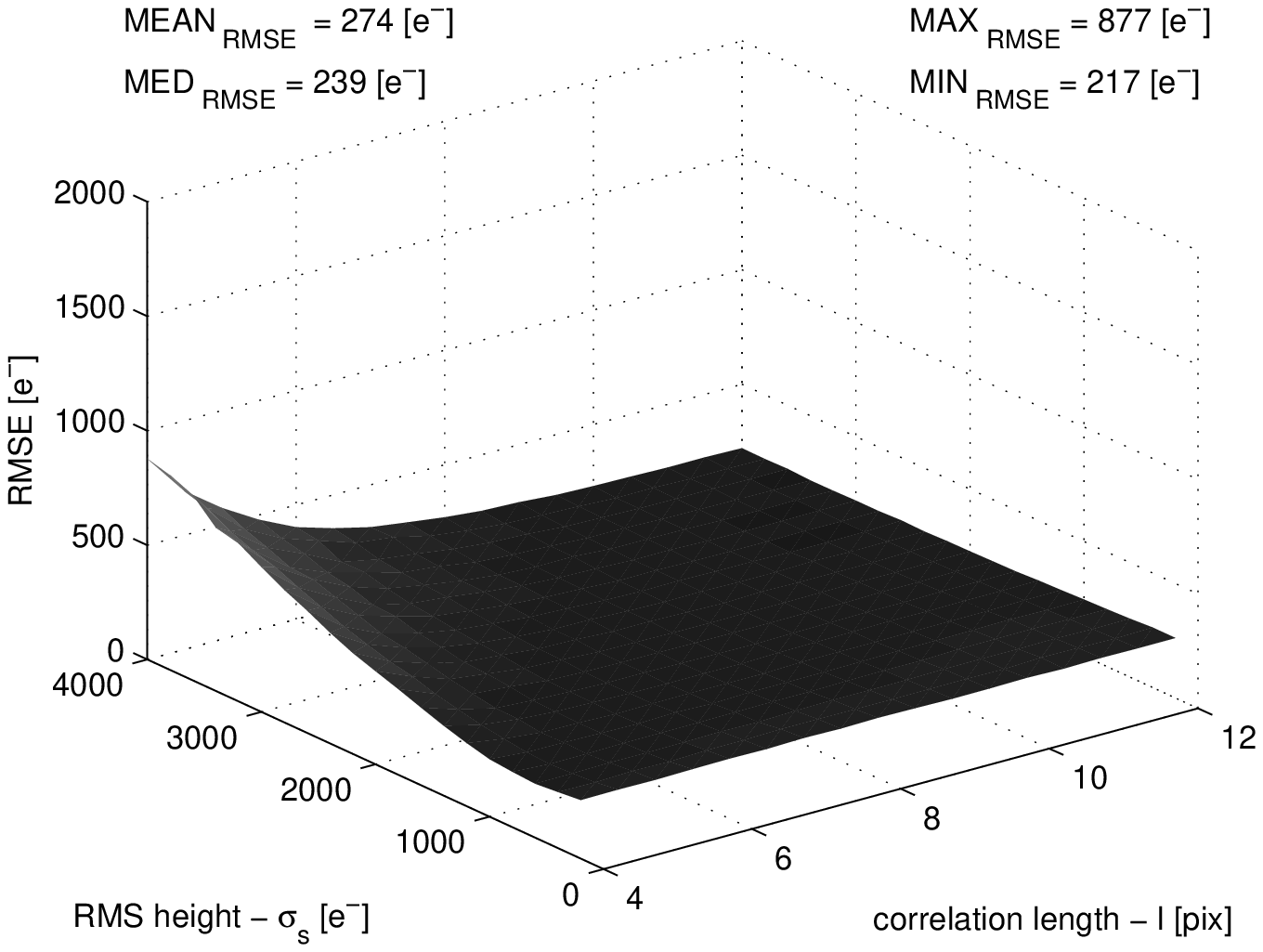}\caption{Interpolation C}
\end{subfigure}
\begin{subfigure}[b]{0.35\linewidth}

	\includegraphics[width=\textwidth]{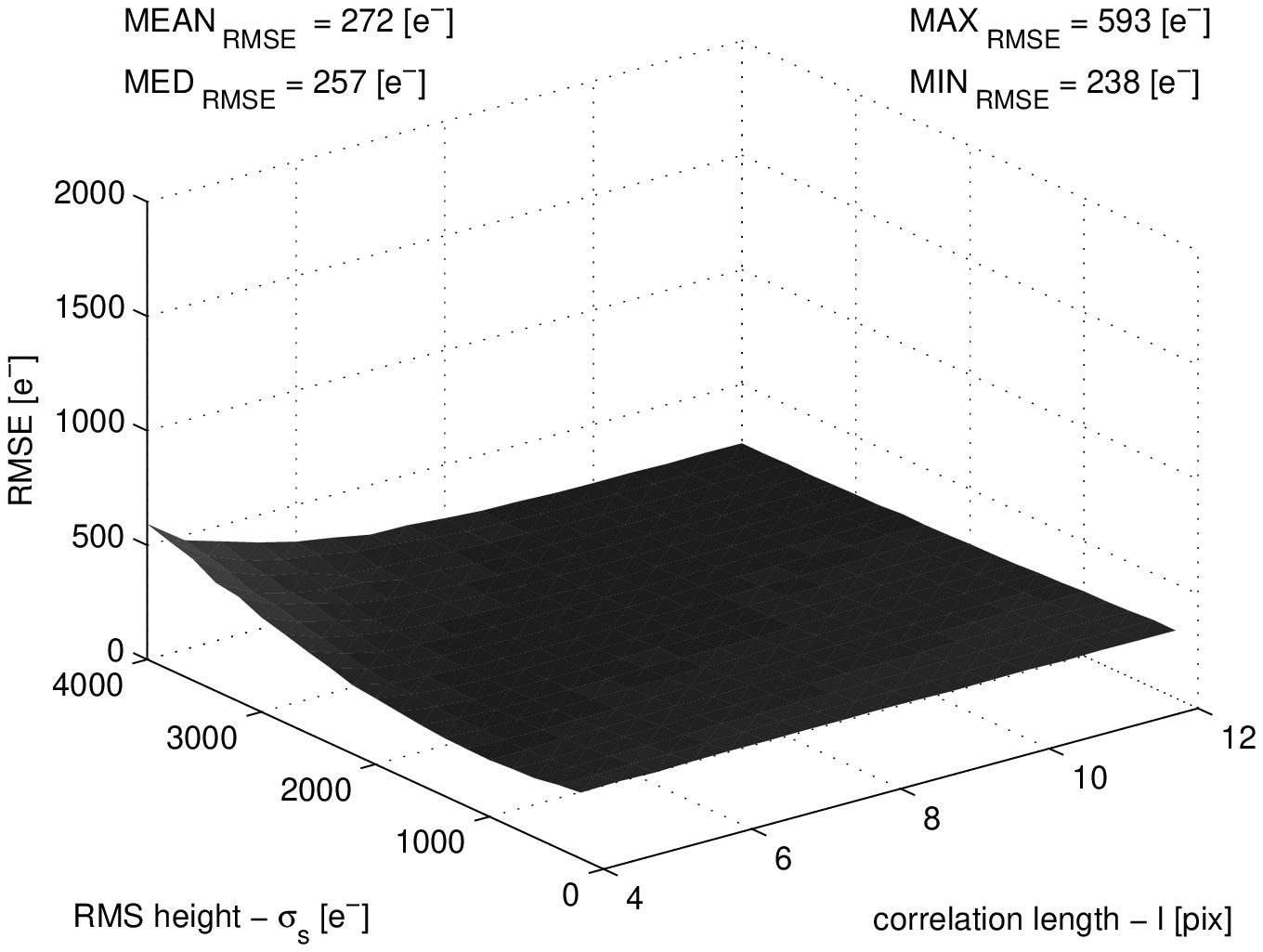}\caption{Interpolation D}
\end{subfigure}
\caption{RMSE of proposed background estimation method for different interpolation methods, sensitivity parameter $k=1$ and all $l$-$\sigma_s$ background parameters.}
\label{res3}
\end{figure*}

\begin{figure*}
\centering
\begin{subfigure}[b]{0.35\linewidth}

\includegraphics[width=\textwidth]{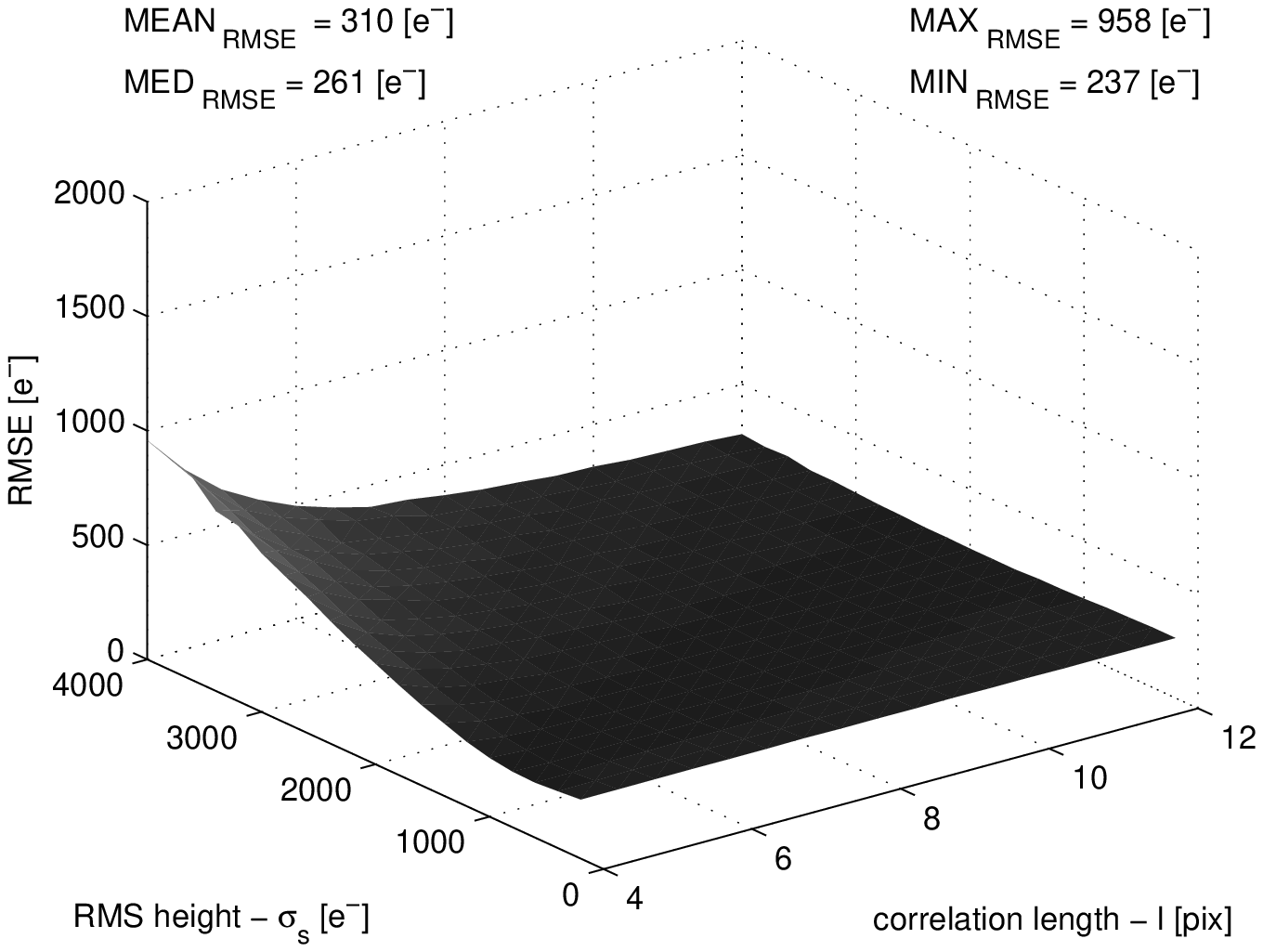}\caption{Interpolation A}
\end{subfigure}
\begin{subfigure}[b]{0.35\linewidth}

\includegraphics[width=\textwidth]{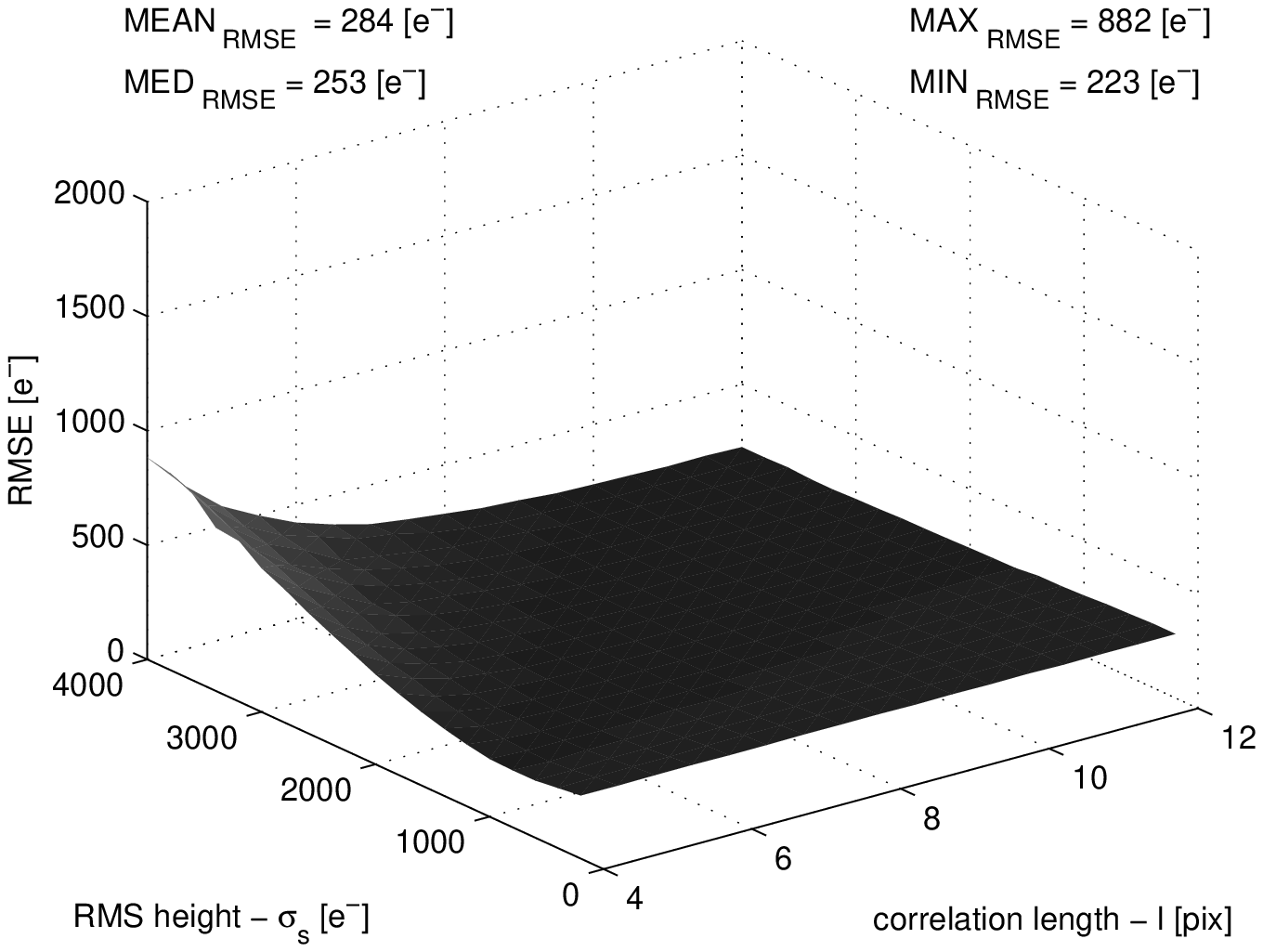}\caption{Interpolation B}
\end{subfigure}
\\
\begin{subfigure}[b]{0.35\linewidth}

	\includegraphics[width=\textwidth]{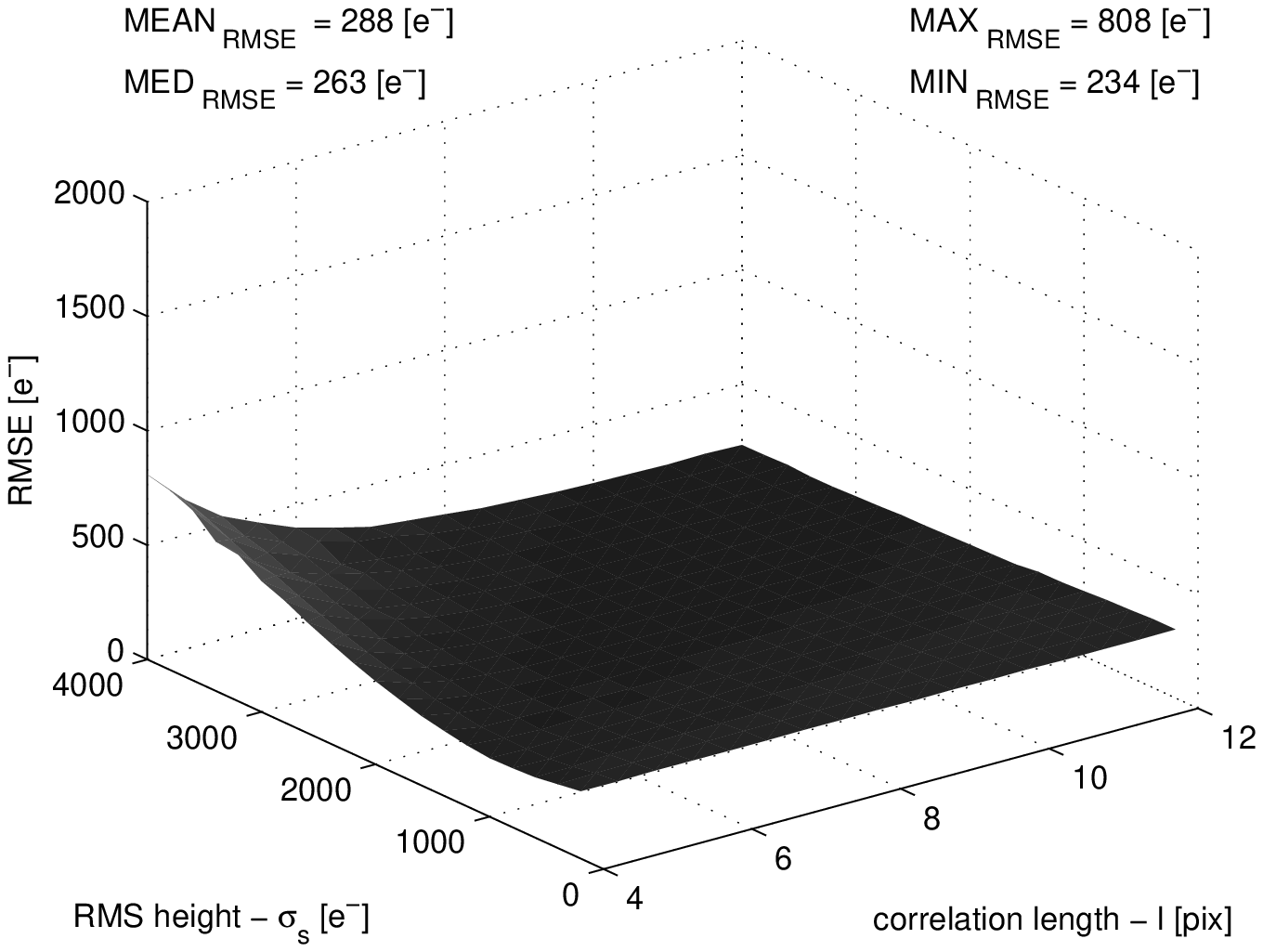}\caption{Interpolation C}
\end{subfigure}
\begin{subfigure}[b]{0.35\linewidth}

	\includegraphics[width=\textwidth]{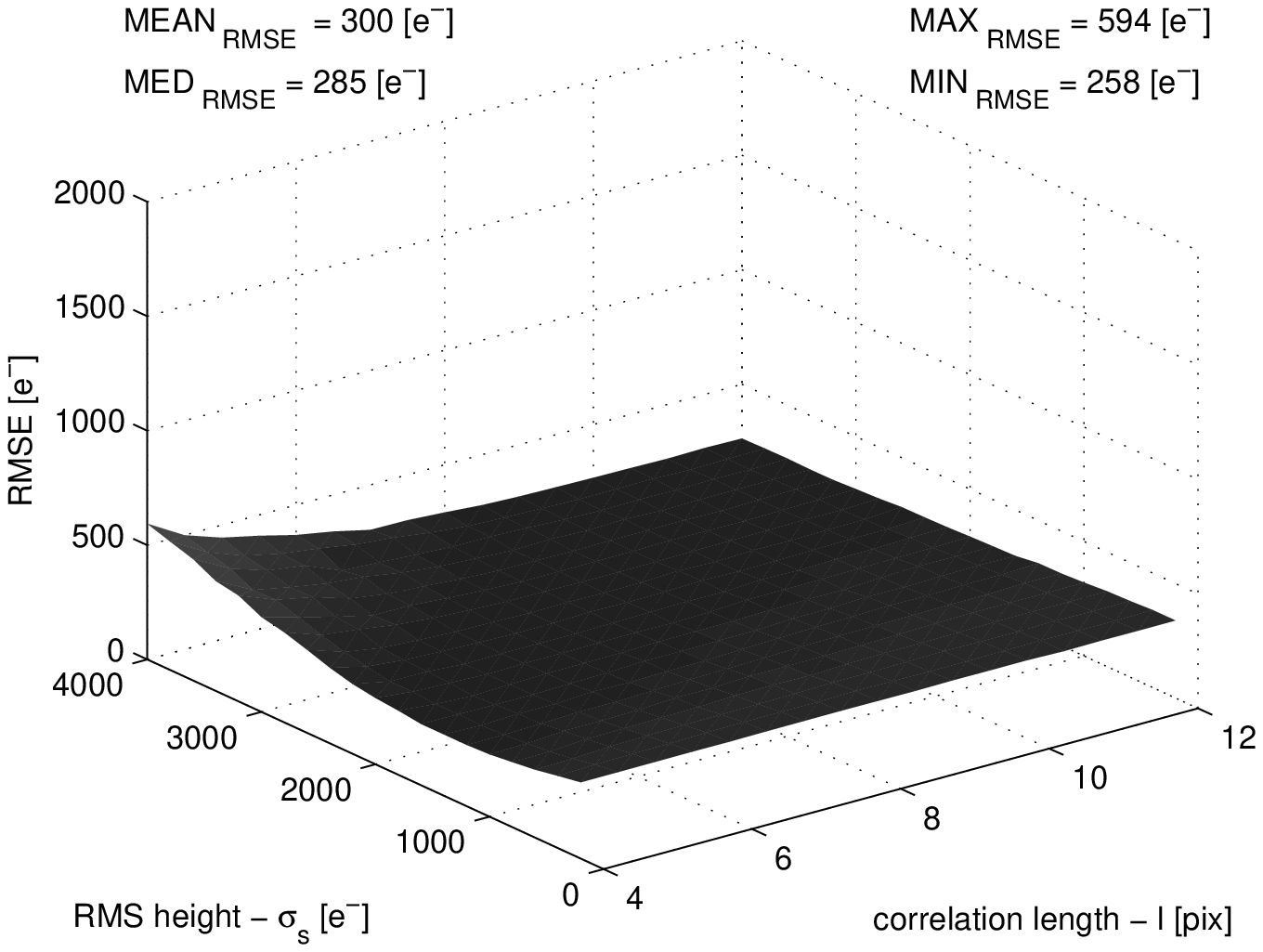}\caption{Interpolation D}
\end{subfigure}
\caption{RMSE of proposed background estimation method for different interpolation methods, sensitivity parameter $k=2$ and all $l$-$\sigma_s$ background parameters.}
\label{res4}
\end{figure*}

\begin{figure*}
\centering
\begin{subfigure}[b]{0.35\linewidth}

\includegraphics[width=\textwidth]{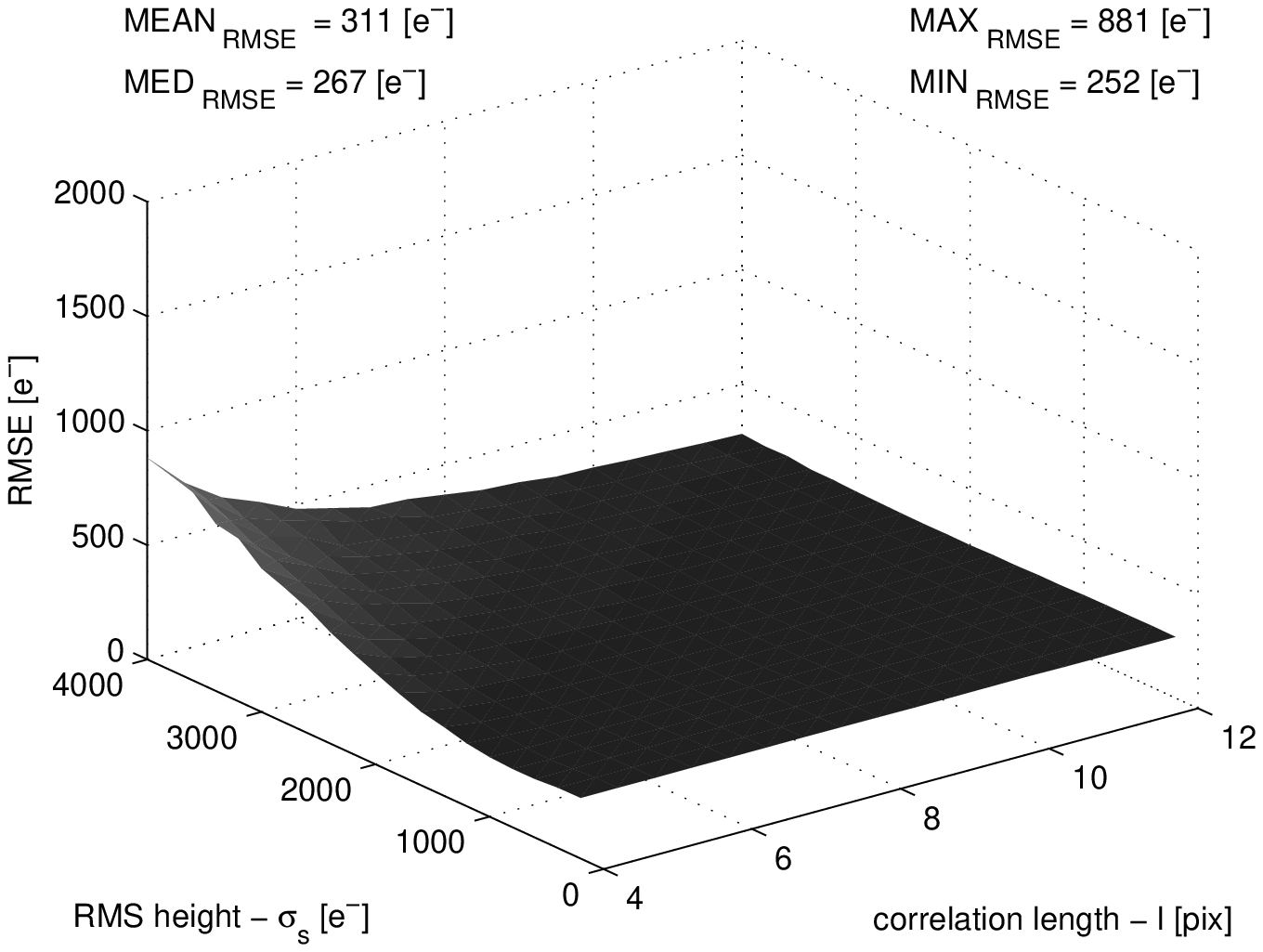}\caption{Interpolation A}
\end{subfigure}
\begin{subfigure}[b]{0.35\linewidth}

\includegraphics[width=\textwidth]{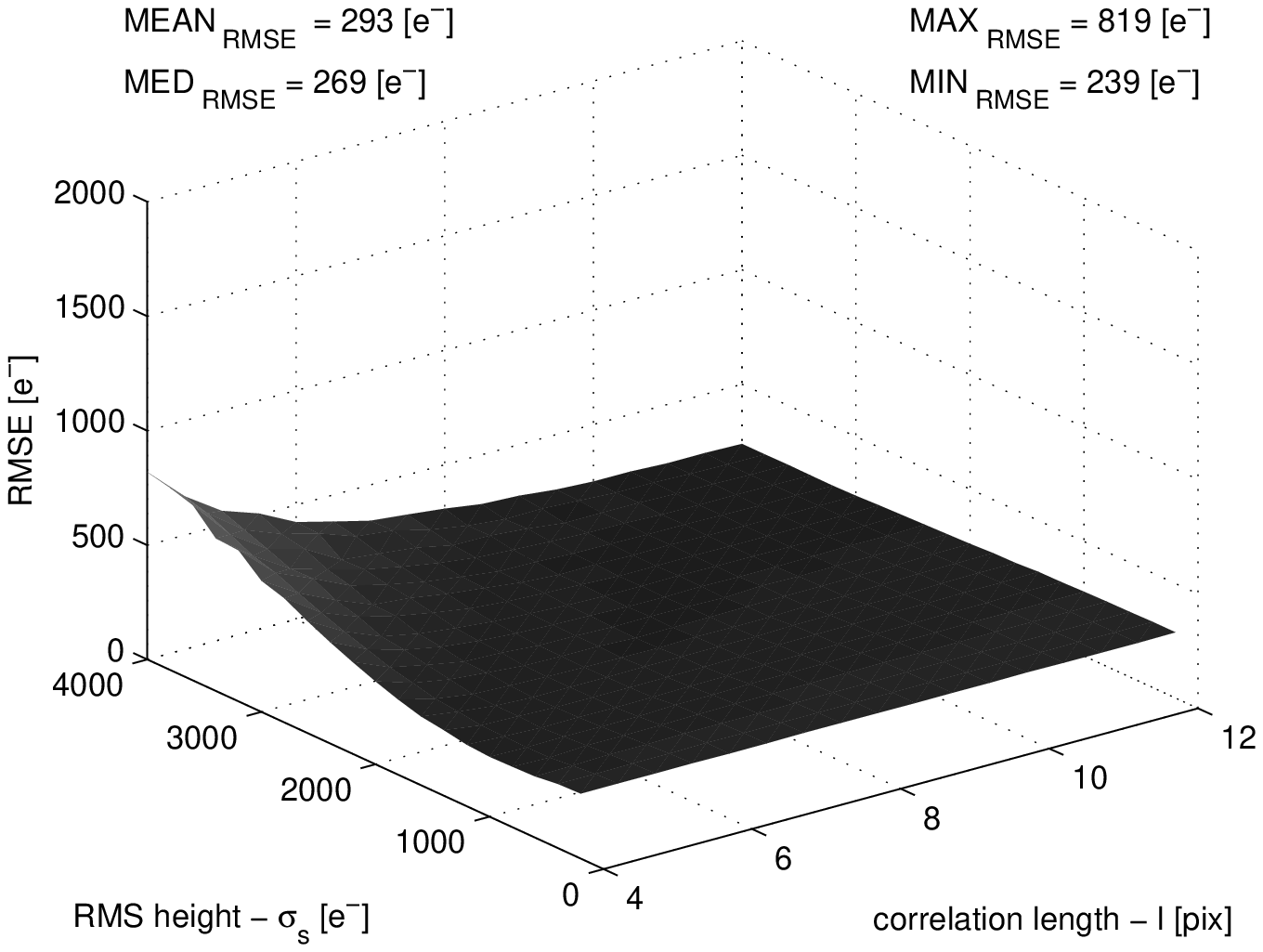}\caption{Interpolation B}
\end{subfigure}
\\
\begin{subfigure}[b]{0.35\linewidth}

	\includegraphics[width=\textwidth]{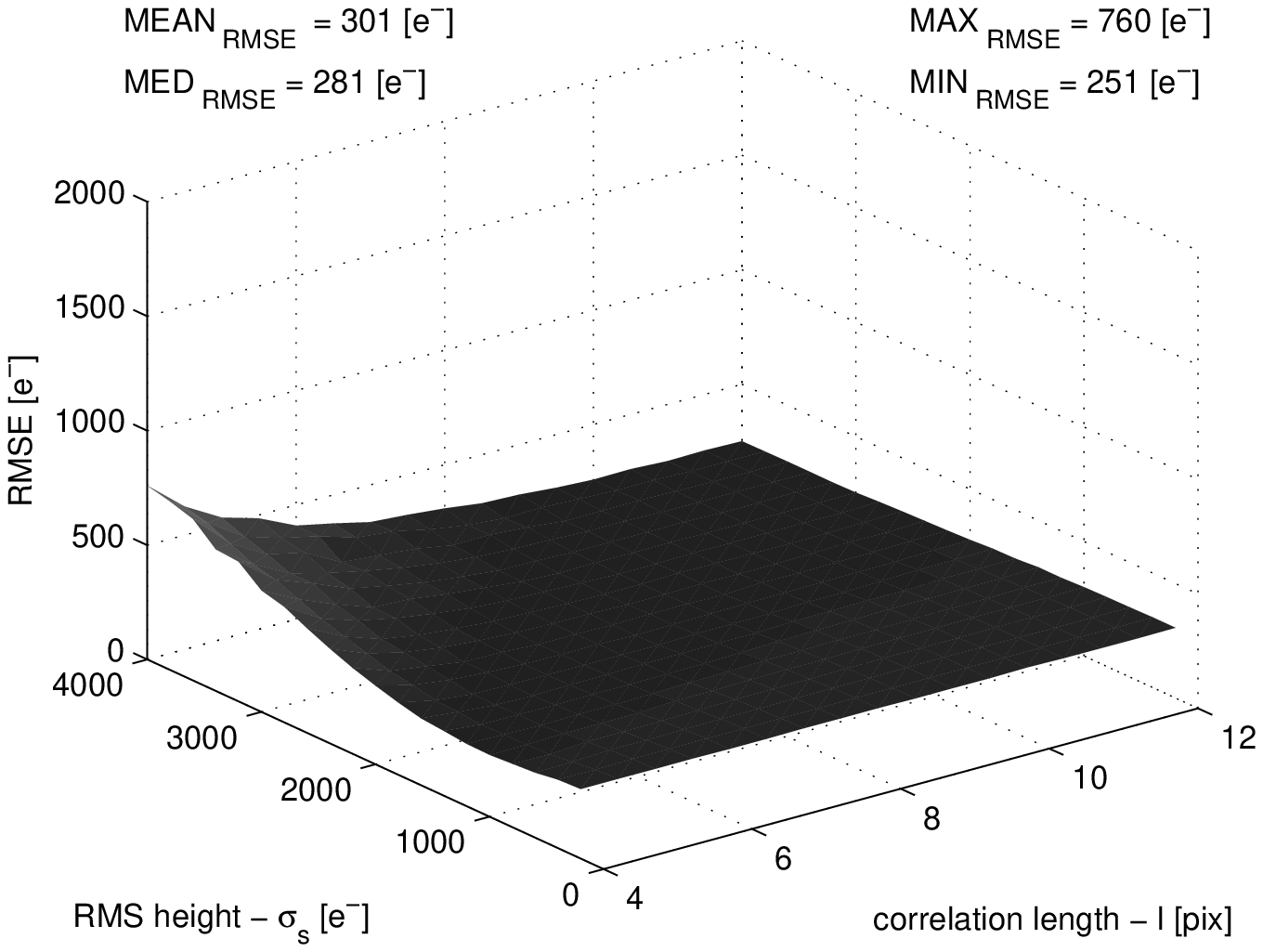}\caption{Interpolation C}
\end{subfigure}
\begin{subfigure}[b]{0.35\linewidth}

	\includegraphics[width=\textwidth]{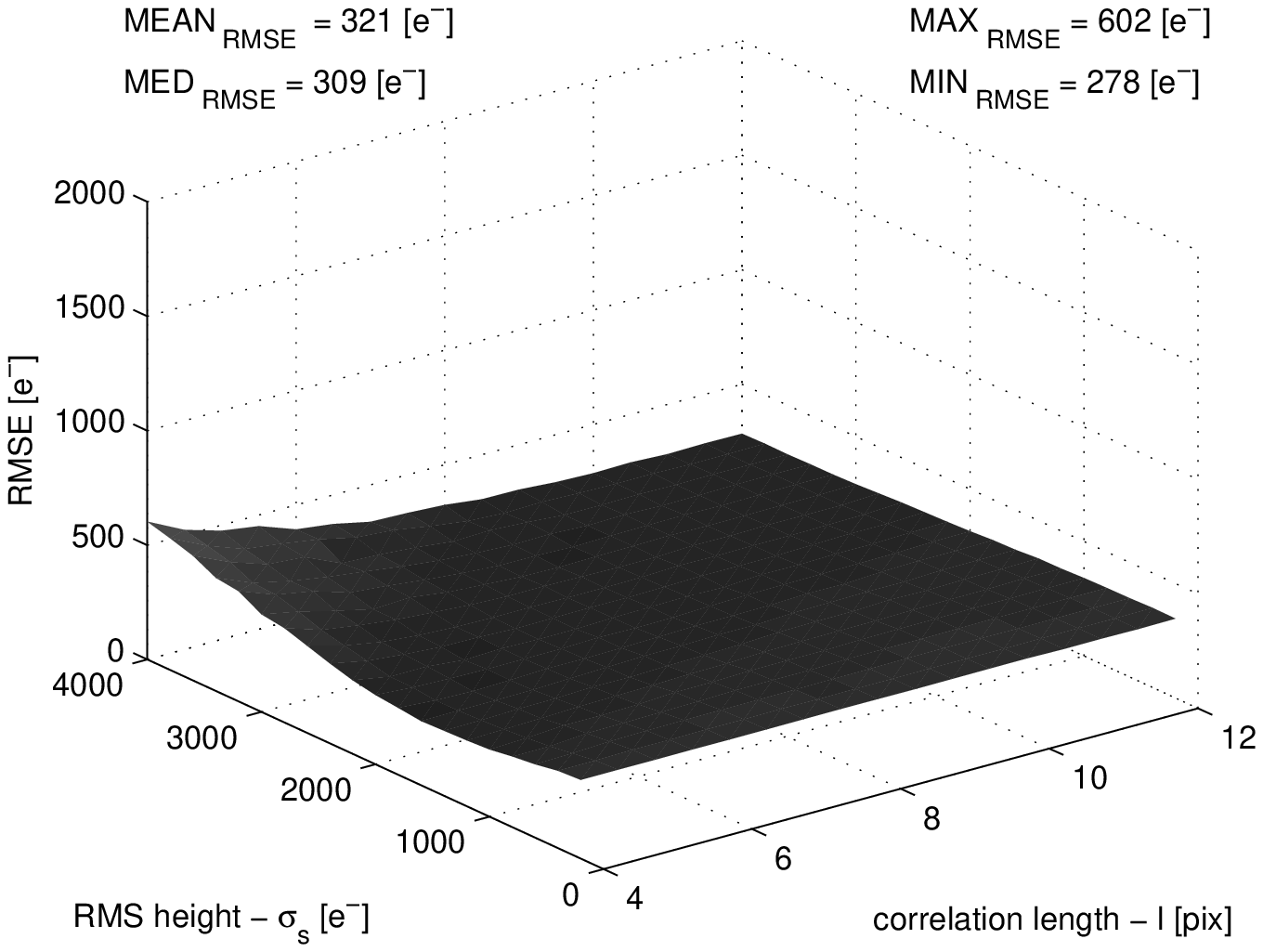}\caption{Interpolation D}
\end{subfigure}
\caption{RMSE of proposed background estimation method for different interpolation methods, sensitivity parameter $k=3$ and all $l$-$\sigma_s$ background parameters.}
\label{res5}
\end{figure*}

The values of RMSE for all the methods are presented in Figs. \ref{res1} - \ref{res5}. The surface plots show the dependencies of RMSE on the parameters of simulated backgrounds (correlation length $l$ and RMS value $\sigma_s$). 

To assess the overall estimation quality, we introduced additional RMSE-based indicators: mean RMSE  ($\textrm{MEAN}_{\textrm{\tiny{RMSE}}}$), median RMSE  ($\textrm{MED}_{\textrm{\tiny{RMSE}}}$), maximal and minimal RMSE ($\textrm{MAX}_{\textrm{\tiny{RMSE}}}$ and $\textrm{MIN}_{\textrm{\tiny{RMSE}}}$). These image quality measures were calculated using all RMSE results for possible $l$-$\sigma_s$ combinations. The results are summarized in Tab. \ref{RMSETab1} and the magnitudo errors are listed in the same way in Tab. \ref{magTab1}.

\begin{figure*}
\centering
\begin{tikzpicture}
[scale=0.3]
%\draw [very thin,top most] (0, 0) grid (20, 17); 
\draw [ultra thick] (0,17) -- (-3.1669,20.1669);
\node at (-3.0002,18.3334) {\Large $l$};
\node at (-1.3334,20.0002) {\Large $\sigma_s$};

\draw [very thick] (20,0) -- (20,17) -- (0,17) -- (0,0) -- (20,0);

\draw [fill=gray!150] (1.4998,-2) rectangle (-0.0002,-1);
\draw [fill=gray!120] (-0.0002,-2.6667) rectangle (1.4998,-3.6667);
\draw [fill=gray!90] (-0.0002,-4.5001) rectangle (1.4998,-5.5001);
\draw [fill=gray!60] (-0.0002,-6.3335) rectangle (1.4998,-7.3335);
\draw [fill=gray!30] (-0.0002,-8.1669) rectangle (1.4998,-9.1669);
\draw [fill=gray!0] (-0.0002,-10.0003) rectangle (1.4998,-11.0003);

\node [right] at (2.4998,-1.5) {Median 7$\times$7};
\node [right] at (2.4998,-3.1667) {Median 9$\times$9};
\node [right] at (2.4998,-5.0001) {Proposed, $k=1$, interpolation A};
\node [right] at (2.4998,-6.8335) {Proposed, $k=1$, interpolation B};
\node [right] at (2.4998,-8.6669) {Proposed, $k=1$, interpolation C};
\node [right] at (2.4998,-10.5003) {Proposed, $k=1$, interpolation D};

\node at (-1.6668,16.5) {\large  4.0};
%\node at (-1,15.5) {\large  4.5};
%\node at (-1,14.5) {\large 5.0};
%\node at (-1,13.5) {\large 5.5};
\node at (-1.6668,12.5) {\large  6.0};
%\node at (-1,11.5) {\large  6.5};
%\node at (-1,10.5) {\large  7.0};
%\node at (-1,9.5) {\large  7.5};
\node at (-1.6668,8.5) {\large  8.0};
%\node at (-1,7.5) {\large  8.5};
%\node at (-1,6.5) {\large  9.0};
%\node at (-1,5.5) {\large  9.5};
\node at (-1.6668,4.5) {\large  10.0};
%\node at (-1,3.5) {\large  10.5};
%\node at (-1,2.5) {\large  11.0};
%\node at (-1,1.5) {\large  11.5};
\node at (-1.6668,0.5) {\large  12.0};

\node  [rotate=90] at (0.5,18.6668) {\large  200};
%\node  [rotate=90] at (1.5,18) {\large  400};
%\node  [rotate=90] at (2.5,18) {\large  600};
%\node  [rotate=90] at (3.5,18) {\large  800};
\node  [rotate=90] at (4.5,18.8335) {\large  1000};
%\node  [rotate=90] at (5.5,18) {\large  1200};
%\node  [rotate=90] at (6.5,18) {\large  1400};
%\node  [rotate=90] at (7.5,18) {\large  1600};
%\node  [rotate=90] at (8.5,18) {\large  1800};
\node  [rotate=90] at (9.5,18.8335) {\large  2000};
%\node  [rotate=90] at (10.5,18) {\large  2200};
%\node  [rotate=90] at (11.5,18) {\large  2400};
%\node  [rotate=90] at (12.5,18) {\large  2600};
%\node  [rotate=90] at (13.5,18) {\large  2800};
\node  [rotate=90] at (14.5,18.8335) {\large  3000};
%\node  [rotate=90] at (15.5,18) {\large  3200};
%\node  [rotate=90] at (16.5,18) {\large  3400};
%\node  [rotate=90] at (17.5,18) {\large  3600};
%\node  [rotate=90] at (18.5,18) {\large  3800};
\node  [rotate=90] at (19.5,18.8335) {\large  4000};

\draw [fill=gray!150,very thick] (0,17) -- (0,13) -- (1,13) -- (1,8) -- (2,8) -- (2,3) -- (3,3) -- (3,0) -- (0,0) -- (0,13);
\draw [fill=gray!120,very thick] (0,17) -- (0,13) -- (1,13) -- (1,8) -- (2,8) -- (2,3) -- (3,3) -- (3,0) -- (7,0) -- (7,2) -- (6,2)-- (6,4) -- (5,4) -- (5,7)  -- (4,7)-- (4,10) -- (3,10)-- (3,13) -- (2,13)-- (2,16) -- (1,16) -- (1,17);
\draw [fill=gray!90,very thick] (0,17) -- (3,17) -- (3,15) -- (4,15) -- (4,12) -- (5,12) -- (5,8) -- (6,8) -- (6,5) -- (7,5) -- (7,2) -- (8,2) -- (8,0)-- (7,0)-- (7,2) -- (6,2)-- (6,4) -- (5,4) -- (5,7)  -- (4,7)-- (4,10) -- (3,10)-- (3,13) -- (2,13)-- (2,16) -- (1,16) -- (1,17);;
\draw [fill=gray!60,very thick] (0,17) -- (4,17) --(4,16) -- (5,16) -- (5,15) -- (6,15) -- (6,14) -- (7,14) -- (7,13) -- (8,13) -- (8,12) -- (9,12) 
--(9,11)--(10,11) -- (10, 10) -- (12,10) -- (12,9) -- (13,9) -- (13,8) -- (15,8) -- (15,7) -- (18,7) -- (18,6) -- (20,6) -- (20,0) -- (8,0) -- (8,2)-- (7,2) -- (7,5) -- (6,5) -- (6,8)-- (5,8)-- (5,12) -- (4,12) -- (4,15) -- (3,15)  -- (3,17)  ; 
\draw  [fill=gray!30,very thick]  (6,15) -- (7,15)--(7,14)--(8,14) -- (8,13) -- (9,13) -- (9,12) -- (11,12)--(11,11) -- (12,11) -- (12,10) -- (14,10) -- (14,9) -- (16,9) -- (16,8) -- (20,8) -- (20,6)
-- (18,6) -- (18,7)-- (15,7) -- (15,8)-- (13,8)-- (13,9) -- (12,9) -- (12,10)-- (10, 10)  --(10,11)--(9,11)-- (9,12) -- (8,12)   -- (8,13) -- (7,13)-- (7,14)-- (6,14)-- (6,15) ;

\foreach \i in {1,...,19}
{
\draw  [ultra thin]  (\i,0) -- (\i,17);
}

\foreach \i in {1,...,16}
{
\draw [ultra thin] (0,\i) -- (20,\i);
}

\end{tikzpicture}
\caption{Visualization of the most accurate method for a given surface complexity. The method with the lowest $\textrm{MEAN}_{\textrm{\tiny{RMSE}}}$ was selected for each combination of the surface parameters $l$ and $\sigma_S$}
\label{best}
\end{figure*}
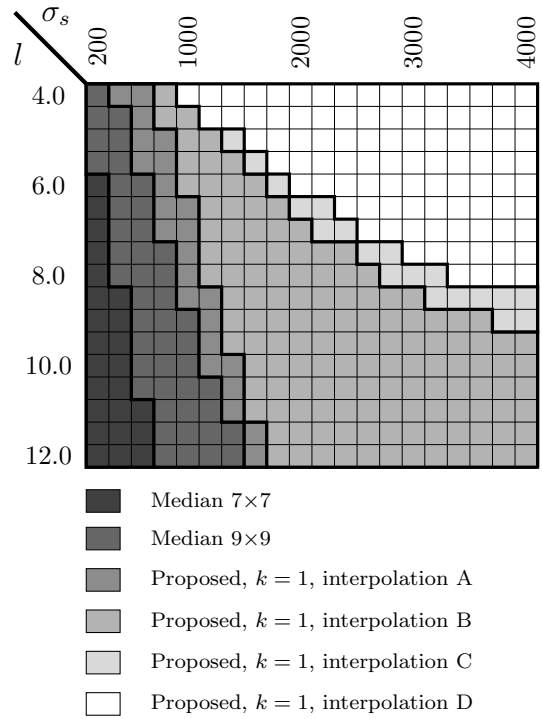

Moreover, in Fig. \ref{best}, we present the visualization of the most accurate method for varying background parameters. For this analysis and a given $l$-$\sigma_s$ combination, we were looking for the lowest RMSE among the competitive methods.

The conclusions from our experiments are as follows:
%\vspace{-0.2cm}
\begin{enumerate}
  \item for the lowest fluctuations of background, we noticed, that the one of the simplest method - the median filtering - is capable to provide very accurate estimations,
  \item the accuracy of \emph{SExtractor} was usually lower, when compared with other competitive approaches,
  \item as the complexity of background increases, the median filtering and \emph{SExtractor} are significantly less accurate than our proposed method,
  \item all interpolation methods employed in our algorithm are useful in different background complexity regimes,
  \item the biharmonic interpolation is the most accurate solution for the most challenging backgrounds,
  \item the best results of our method were achieved for $k=1$, however it is to be noted, that bigger $k$ would be desirable for higher noise levels.
\end{enumerate}

\begin{table*}
\centering
\caption{The summary of RMSE-based measures of overall estimation quality. The smallest mean and median RMSE results are underlined.\vspace{0.2cm}}
\begin{tabular}{l | c | c | c | c | c}  
Method & Parameters & $\textrm{MEAN}_{\textrm{\tiny{RMSE}}}$ & $\textrm{MED}_{\textrm{\tiny{RMSE}}}$ & $\textrm{MAX}_{\textrm{\tiny{RMSE}}}$ & $\textrm{MIN}_{\textrm{\tiny{RMSE}}}$ \\ \hline
Median filter &5$\times$5 & 403 & 369 & 1148 & 174 \\
 &7$\times$7 & 471 & 394 & 1762 & 82 \\
 &9$\times$9 & 612 & 491 & 2346 & 66 \\ \hline

$SExtractor$ & 5$\times$5 & 484 & 443 & 1212 & 336 \\
& 7$\times$7 & 512 & 420 & 1841 & 102 \\
& 9$\times$9 & 648 & 513 & 2410 & 86 \\ \hline

Proposed, &$k=1$ & 312 & 256 & 1048 & 218 \\
interpolation A&$k=2$ & 310 & 261 & 958 & 237 \\
&$k=3$ & 311 & 267 & 881 & 252 \\ \hline

Proposed, &$k=1$ & 276 & \textbf{\underline{231}} & 965 & 209 \\
interpolation B&$k=2$ & 284 & 253 & 882 & 223 \\
 &$k=3$ & 293 & 269 & 819 & 239 \\ \hline

Proposed, &$k=1$ & 274 & 239 & 877 & 217 \\
interpolation C&$k=2$ & 288 & 263 & 808 & 234 \\
 &$k=3$ & 301 & 281 & 780 & 251 \\ \hline

Proposed, &$k=1$ & \textbf{\underline{272}} & 257 & 593 & 238 \\
interpolation D&$k=2$ & 300 & 285 & 594 & 258 \\
 &$k=3$ & 321 & 309 & 602 & 278 \\ \hline
\end{tabular}
\label{RMSETab1}
\end{table*}

\begin{table*}
\centering
\caption{The summary of magnitude errors. The smallest mean and median $\Delta m$ results are underlined.\vspace{0.2cm}}
\begin{tabular}{l | c | c | c | c | c}  
Method & Parameters & $\textrm{MEAN}_{\textrm{\tiny{$\Delta m$}}}$ & $\textrm{MED}_{\textrm{\tiny{$\Delta m$}}}$ & $\textrm{MAX}_{\textrm{\tiny{$\Delta m$}}}$ & $\textrm{MIN}_{\textrm{\tiny{$\Delta m$}}}$ \\ \hline
Median filter &5$\times$5 & 0.61 & 0.60 & 1.21 & 0.29 \\
 &7$\times$7 & 0.47 & 0.45 & 1.14 & 0.12 \\
 &9$\times$9 & 0.47 & 0.43 & 1.29 & 0.07 \\ \hline

$SExtractor$ & 5$\times$5 & 0.64 & 0.69 & 1.08 & 0.20 \\
& 7$\times$7 & 0.48 & 0.46 & 1.12 & 0.13 \\
& 9$\times$9 & 0.46 & 0.40 & 1.21 & 0.08 \\ \hline

Proposed, &$k=1$ & 0.27 & 0.21 & 1.20 & 0.05 \\
interpolation A&$k=2$ & 0.26 & 0.20 & 1.22 & 0.05 \\
&$k=3$ & 0.26 & 0.20 & 1.21 & 0.05 \\ \hline

Proposed, &$k=1$ & 0.27 & 0.20 & 1.18 & 0.06 \\
interpolation B&$k=2$ & 0.27 & 0.20 & 1.21 & 0.06 \\
 &$k=3$ & 0.27 & 0.21 & 1.23 & 0.06 \\ \hline

Proposed, &$k=1$ & 0.27 & 0.20 & 1.16 & 0.07 \\
interpolation C&$k=2$ & 0.27 & 0.19 & 1.18 & 0.08 \\
 &$k=3$ & 0.25 & 0.15 & 1.12 & 0.08 \\ \hline

Proposed, &$k=1$ & \textbf{\underline{0.22}} & \textbf{\underline{0.11}} & 1.19 & 0.08 \\
interpolation D&$k=2$ & 0.24 & 0.12 & 1.11 & 0.08 \\
 &$k=3$ & 0.25 & 0.15 & 1.1 & 0.08 \\ \hline
\end{tabular}
\label{magTab1}
\end{table*}

%---------------------------------------------------------------------
\section{Oversampling and close multiple objects}
Since the proposed method relies on the intensity gradients encountered between pixels, it is very sensitive to the image sampling. In our experiments, we used $\sigma_{PSF}=1$, however it is common, that the stars in many astronomical images have much larger $\sigma_{PSF}$. In such cases, the algorithm may exhibit lower accuracy, because the Poisson and electronic noise remain the same, while the intensity gradients are much smaller. When the oversampling appears, the selected window size has to be larger and also the $k$ parameter has to be decreased to detect less steep slopes of the foreground objects. However, lower $k$ reduces the robustness of the method, thus there are more false positive detections.

\begin{figure*}
        \centering
         \captionsetup{justification=centering}
\begin{subfigure}[b]{0.24\linewidth}
\centering
\includegraphics[width=0.9\textwidth]{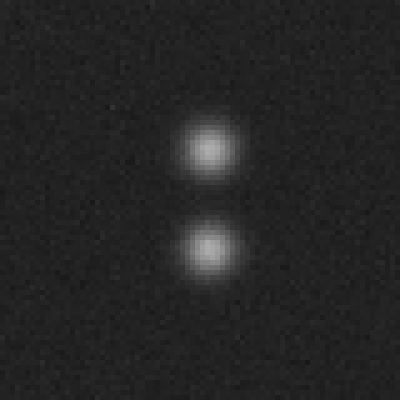}
\caption{Oversampled image \\($\sigma_{PSF}$ = 4).}
\end{subfigure}
\begin{subfigure}[b]{0.24\linewidth}\centering
\includegraphics[width=0.9\textwidth]{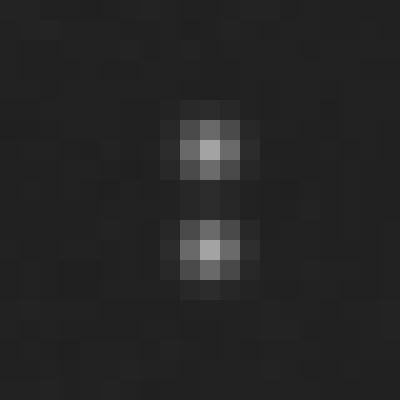}
 \captionsetup{justification=centering}
\caption{The result of 4$\times$4 pixels binning of image (a).}
\end{subfigure}
\begin{subfigure}[b]{0.24\linewidth}\centering
\includegraphics[width=0.9\textwidth]{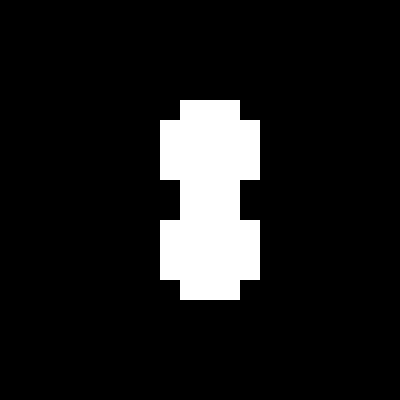}
 \captionsetup{justification=centering}
\caption{Object detection in binned image (b).}
\end{subfigure}
\begin{subfigure}[b]{0.24\linewidth}\centering
\includegraphics[width=0.9\textwidth]{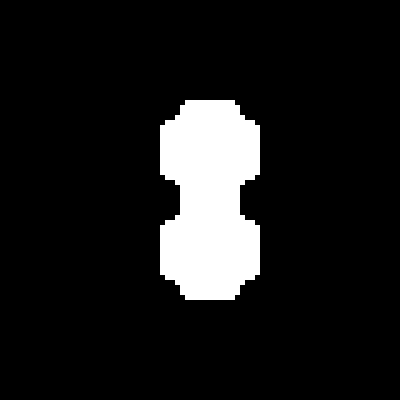}
 \captionsetup{justification=centering}
\caption{Nearest neighbor interpolation of mask in (c).}
\end{subfigure}
\caption{The results of consecutive steps of object detection for oversampled image: (a) exemplary oversampled frame ($\sigma_{PSF}$ = 4), (b) a binned version of the image, ($\sigma_{PSF}$ = 1), (c) detected object pixels in binned image, (d) the corresponding object pixels in the original resolution as obtained from the nearest neighbor interpolation of mask in (c).}
\label{oversample}
\end{figure*}

To easily overcome those problems, the image should be decimated (binned) before performing the object detection. We suggest to choose the binning size, so that in the final image $\sigma_{PSF}$ is close to 1. This way, one may apply the mask sizes and $k$ values close to the ones utilized in our experiments. After the detection phase, the indicated object mask should be resized back to the original image size (employing e.g. the nearest neighbor interpolation). The examples of images obtained during the object detection for significantly oversampled image are presented in Fig. \ref{oversample}. Note also, that when employing robust formula (\ref{formula1}) for binned image, the electronic noise increases according to  $\sqrt{N_{bin}}\sigma_{el}$, where $N_{bin}$ is the number of binned pixels.

\begin{figure*}
        \centering
         \captionsetup{justification=centering}
\begin{subfigure}[b]{\linewidth}
\includegraphics[width=\textwidth]{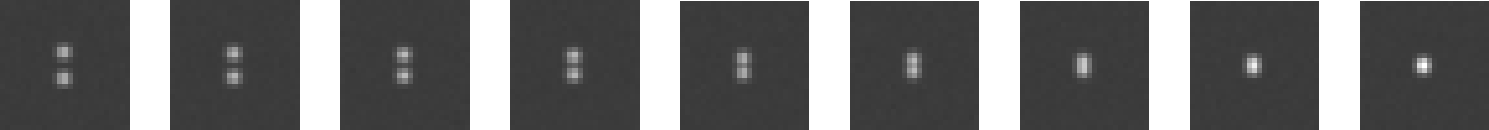}
\caption{A series of close binary objects.\vspace{0.2cm}}
\end{subfigure}
\begin{subfigure}[b]{\linewidth}
\includegraphics[width=\textwidth]{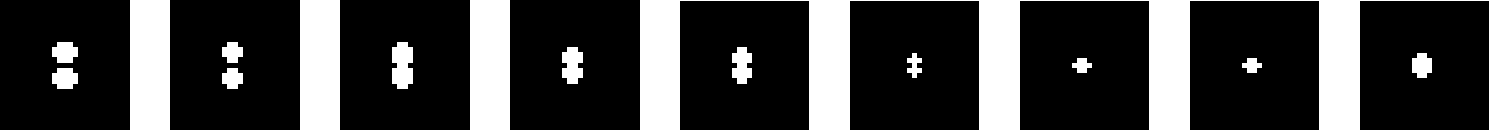}
 \captionsetup{justification=centering}
\caption{Detected object pixels for 5$\times$5 window.\vspace{0.2cm}}
\end{subfigure}
\begin{subfigure}[b]{\linewidth}
\includegraphics[width=\textwidth]{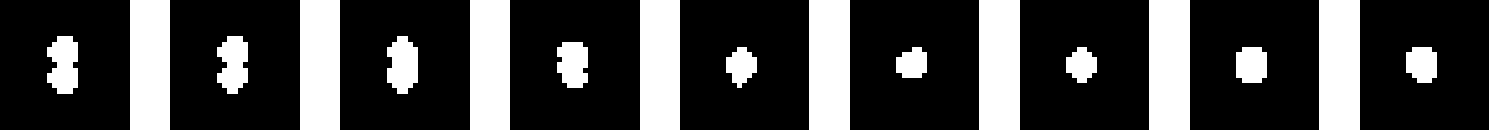}
 \captionsetup{justification=centering}
\caption{Detected object pixels for 9$\times$9 window.}
\end{subfigure}
\caption{The results of object detection for very close objects: (a) a series of approaching star objects, (b) the detected object pixels utilizing 5$\times$5 window and (c) the result of the detection with 9$\times$9 window.}
\label{closeStars}
\end{figure*}

The algorithm may have also reduced accuracy, when the objects are close to each other. In such a situation, the technique may not recognize the object pixels, as there are some ascending routs due to the presence of the close second star. We present some examples of detected objects using 5$\times$5 window in Fig. \ref{closeStars}b. As it can be seen, the detection is performed correctly for well separated  (on the left) and for fully overlapping stars (on the right). For very short distances, the number of detected pixels decreases, and the tails of PSFs are not chosen for the interpolation. To mitigate this problem, we propose to choose slightly larger window size, to include both of such close objects within the window area. In Fig. \ref{closeStars}c we depict the results of pixels classification for a 9$\times$9 window, which appeared to be a reasonable setting. 

%---------------------------------------------------------------------
\section{Applications}
There are plenty of possible applications of the proposed algorithm in astronomical imaging. We provide some of them and include the corresponding examples. 

\begin{figure*}
\centering
\begin{subfigure}[b]{0.49\linewidth}
	\begin{tikzpicture}[scale=0.8,spy using outlines={circle,white,magnification=4,size=3.5cm, connect spies}]
	\node {\pgfimage[width=8cm]{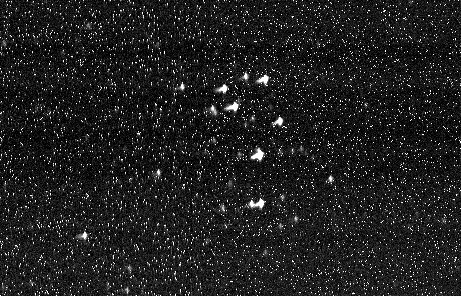}};
	\spy on (-0,0.7) in node [left] at (-1,-1.25);
%\spy on (2.3,-1) in node [left] at (-0.5,-3);
	\end{tikzpicture}
\caption{Original image.}
\end{subfigure}
\begin{subfigure}[b]{0.49\linewidth}
	\begin{tikzpicture}[scale=0.8,spy using outlines={circle,white,magnification=4,size=3.5cm, connect spies}]
	\node {\pgfimage[width=8cm]{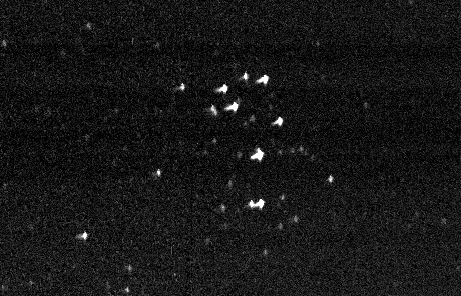}};
	\spy on (-0,0.7) in node [left] at (-1,-1.25);
%\spy on (2.3,-1) in node [left] at (-0.5,-3);
	\end{tikzpicture}
\caption{Extracted background.} 
\end{subfigure}
\caption{An example of utilization of our algorithm for the impulsive noise cancellation in BRITE satellite image depicting Pleyades: (a) a part of original noisy full frame obtained on 28 August 2014 by UniBRITE satellite, (b) the same frame filtered by our background extractor.} 
\label{BRITE}
\end{figure*}

The straightforward way of utilization of the proposed technique is the reduction of impulsive noise, which appears mainly due to the presence of the dark current (in CCDs) and the clock induced noise (in EMCCDs). In such a case, the hot pixels are detected as foreground objects and removed. The first example (see Fig. \ref{BRITE} (a)) depicts the part of the preliminary image obtained by one of the BRITE satellites \footnote{Based on data collected by the BRITE-Constellation satellite mission, built, launched and operated thanks to support from the Austrian Aeronautics and Space Agency and the University of Vienna, the Canadian Space Agency (CSA) and the Foundation for Polish Science and Technology (FNiTP MNiSW) and National Center for Science (NCN).}. The mission \cite{BRITE1,BRITE2} aims at the precise photometry of the brightest stars for the purpose of the astroseismology. However, due to the required small dimensions of the satellite, there was not enough place to include shielding, thus the CCD detectors suffer from the proton induced degradation. We show, that employing our method it is possible to better localize and extract the stars in the field of view. Note, that the stars PSFs are elongated because of the intentional defocussing.

\begin{figure*}
\centering
\begin{subfigure}[b]{0.49\linewidth}
	\begin{tikzpicture}[scale=0.8,spy using outlines={circle,white,magnification=4,size=3.5cm, connect spies}]
	\node {\pgfimage[width=8cm,height=6cm]{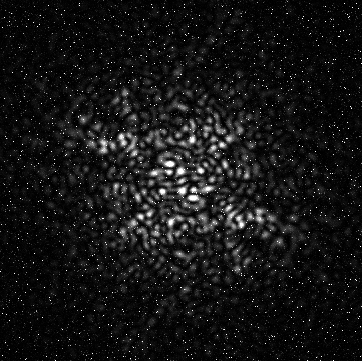}};
	\spy on (1,-0.5) in node [left] at (-0.5,3.25);
%\spy on (2.3,-1) in node [left] at (-0.5,-3);
	\end{tikzpicture}
\caption{Simulated speckle image.}
\end{subfigure}
\begin{subfigure}[b]{0.49\linewidth}
	\begin{tikzpicture}[scale=0.8,spy using outlines={circle,white,magnification=4,size=3.5cm, connect spies}]
	\node {\pgfimage[width=8cm,height=6cm]{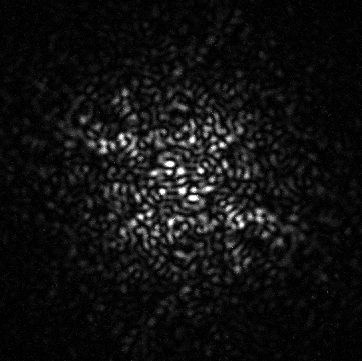}};
	\spy on (1,-0.5) in node [left] at (-0.5,3.25);
%\spy on (2.3,-1) in node [left] at (-0.5,-3);
	\end{tikzpicture}
\caption{Filtered speckle image.} 
\end{subfigure}
\caption{The removal of impulsive noise in speckle imaging: (a) a simulated pattern (2 m telescope, the Fried parameter $r_0$=20 cm) with added EMCCD noise (clock induced noise and dark current spikes), (b) the same frame filtered by our algorithm.} 
\label{spackle}
\end{figure*}

Another example of impulsive noise reduction is presented in Fig. \ref{spackle}, which depicts the utilization of the algorithm in the speckle imaging. In this technique, the camera acquires very short exposures and then, such a series is analyzed to obtain the high resolution outcome. The presented speckle pattern was simulated using the derivations given in \cite{Saha} for 2 m telescope under very good atmospheric conditions (Fried parameter $r_0$= 20 cm). The added impulsive noise consisted of a combination of dark current spikes and clock-induced noise and was obtained from the dark frame of our EMCCD Luca S (Andor).

The proposed method may be also applied for the reduction of the cosmic-rays, which is an important part of the image processing pipelines dedicated to space observations. Most of the spaceborne instruments suffer from the presence of smudges and dots resulting from the collisions of particles with the CCD matrix. To remove such artifacts, the exposures have to be repeated many times to filter out randomly appearing cosmic rays. However, it is a time-consuming process and sometimes it is infeasible due to the variability of observed scene. 

\begin{figure*}
\centering
\begin{subfigure}[b]{0.49\linewidth}
	\begin{tikzpicture}[scale=0.8,spy using outlines={circle,black,magnification=4,size=3.5cm, connect spies}]
	\node {\pgfimage[width=8cm]{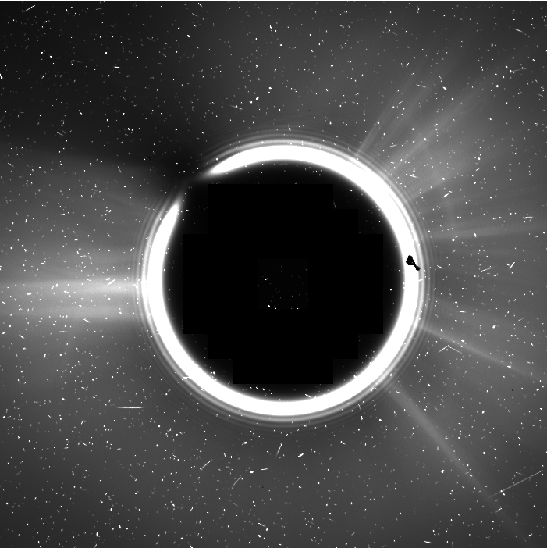}};
	\spy on (-2.6,-0.2) in node [left] at (-1,3.25);
%\spy on (2.3,-1) in node [left] at (-0.5,-3);
	\end{tikzpicture}
\caption{Original SOHO image.}
\end{subfigure}
\begin{subfigure}[b]{0.49\linewidth}
	\begin{tikzpicture}[scale=0.8,spy using outlines={circle,black,magnification=4,size=3.5cm, connect spies}]
	\node {\pgfimage[width=8cm]{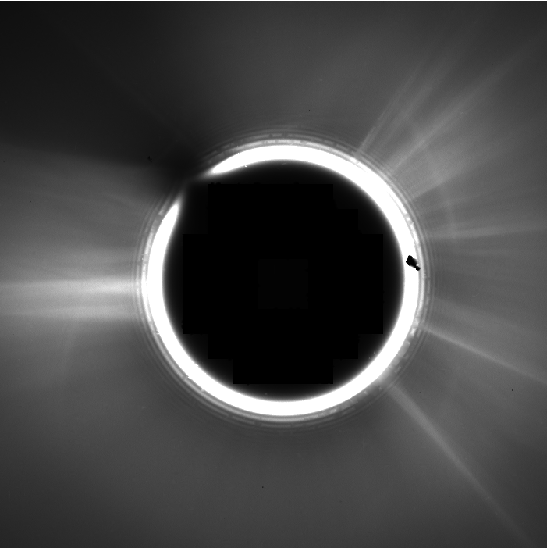}};
	\spy on (-2.6,-0.2) in node [left] at (-1,3.25);
%\spy on (2.3,-1) in node [left] at (-0.5,-3);
	\end{tikzpicture}
\caption{Extracted background.} 
\end{subfigure}
\caption{Removal of cosmic rays in SOHO images: (a) a single frame obtained on 20 April 1998 by LASCO instrument (detector C2, orange filter) two hours after the solar eruption, (b) the same frame filtered by our algorithm. } 
\label{Soho}
\end{figure*}

\begin{figure*}
\centering
\begin{subfigure}[b]{0.49\linewidth}
	\begin{tikzpicture}[scale=0.8,spy using outlines={circle,white,magnification=4,size=3.5cm, connect spies}]
	\node {\pgfimage[width=8cm]{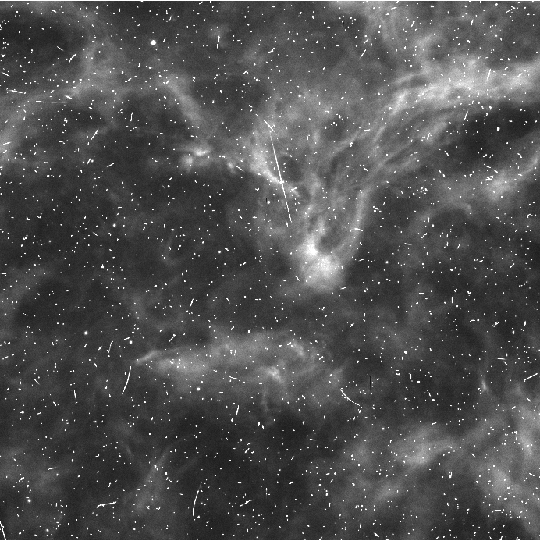}};
	\spy on (0.1,1.2) in node [left] at (-1,-2);
%\spy on (2.3,-1) in node [left] at (-0.5,-3);
	\end{tikzpicture}
\caption{Original HST image.}
\end{subfigure}
\begin{subfigure}[b]{0.49\linewidth}
	\begin{tikzpicture}[scale=0.8,spy using outlines={circle,white,magnification=4,size=3.5cm, connect spies}]
	\node {\pgfimage[width=8cm]{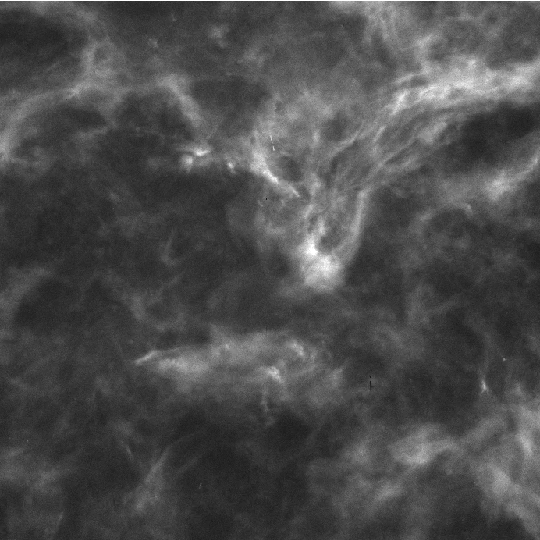}};
	\spy on (0.1,1.2) in node [left] at (-1,-2);
%\spy on (2.3,-1) in node [left] at (-0.5,-3);
	\end{tikzpicture}
\caption{Extracted background.} 
\end{subfigure}
\caption{Removal of cosmic rays in Hubble Space Telescope (HST) image: (a) a single frame obtained on 2 April 2009 by the Wide Field Camera (WFC4) depicting an extended complex nebula, (b) the same frame filtered by our algorithm.} 
\label{HST}
\end{figure*}

In Figs. \ref{Soho} and \ref{HST} we show, that our method is very well suited for such image filtering. We included two examples of different types of space observations: the real-time image obtained from the Solar and Heliospheric Observatory (SOHO) and the long-exposure frame from the Hubble Space Telescope (HST). While for the SOHO mission (\cite{SOHO1}), the images cannot be averaged due to the dynamics of solar corona, the repeating of HST long exposures requires additional time, which may be spent on other observations. Therefore, both of presented examples benefit significantly from the accurate cosmic ray artifacts rejection.

\begin{figure*}
\centering
\begin{subfigure}[b]{0.49\linewidth}
	\begin{tikzpicture}[scale=0.8,spy using outlines={circle,white,magnification=3,size=4cm, connect spies}]
	\node {\pgfimage[width=8cm]{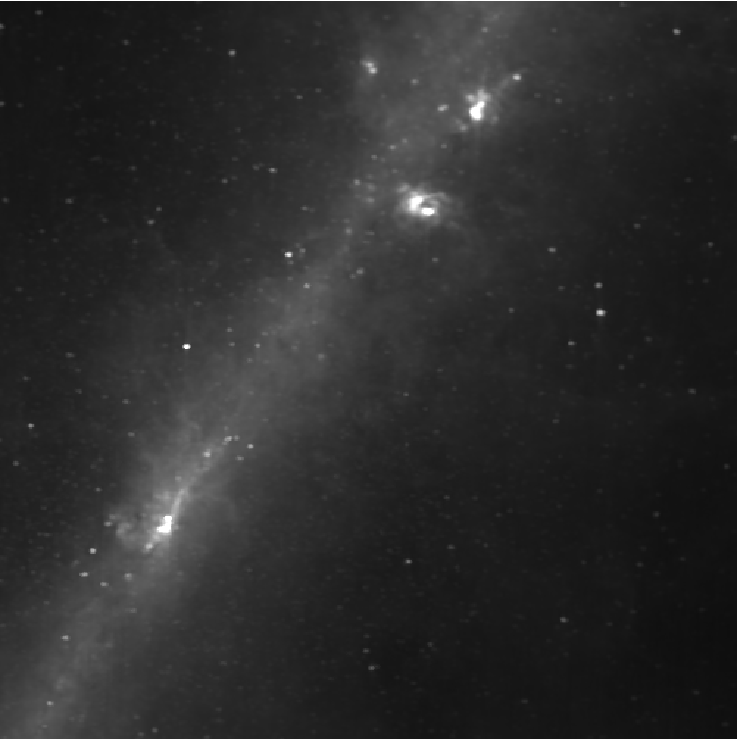}};
	\spy on (0.2,2) in node [left] at (4.9,-2.4);
	\end{tikzpicture}
	\caption{Original IRAS image 1, 12$\mu$m.}
\end{subfigure}
\begin{subfigure}[b]{0.49\linewidth}
	\begin{tikzpicture}[scale=0.8,spy using outlines={circle,white,magnification=3,size=4cm, connect spies}]
	\node {\pgfimage[width=8cm]{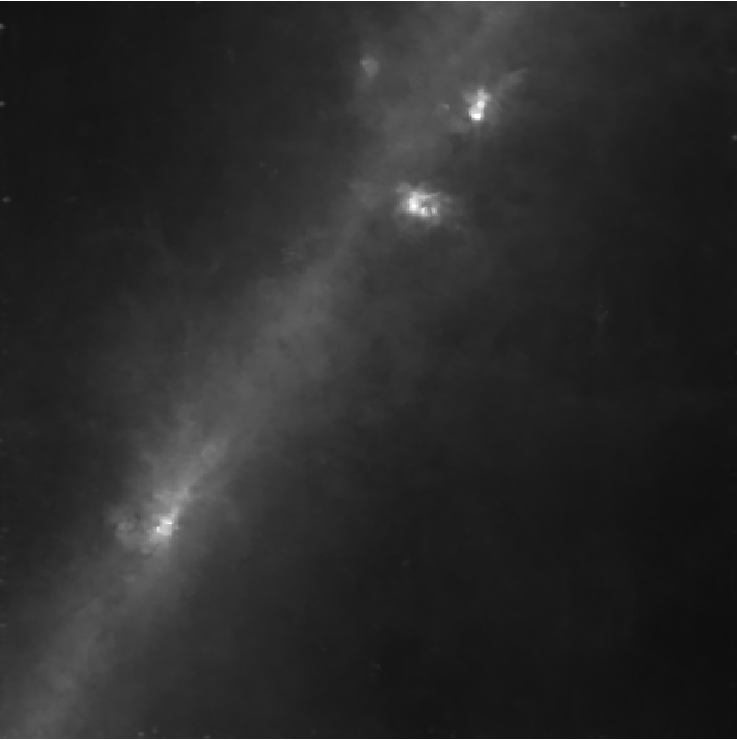}};
	\spy on (0.2,2) in node [left] at (4.9,-2.4);
	\end{tikzpicture}
	\caption{Result of background estimation.}
\end{subfigure}
\caption{Examples of our background estimations for infrared images obtained by the \emph{Infrared Astronomical Satellite} (IRAS, \protect\cite{IRAS}).}
\label{Background_estimations2}
\end{figure*}

The background subtraction performed on exemplary low-resolution image acquired in the infrared is depicted in Fig. \ref{Background_estimations2}. We used the \emph{Infrared Astronomical Satellite} (IRAS \cite{IRAS}) resources stored in the Sky Survay Atlas (available online at \url{http://irsa.ipac.caltech.edu/applications/IRAS/ISSA/}). The image exhibits the extended structures of the Galactic plane. For such application of our method, we assumed no knowledge about the detector noise distribution, thus we did not employ the robust extension of the algorithm and set $k$=0. Even in such simplified version of the algorithm, the outcomes were satisfactory.

\begin{figure*}
\centering
\begin{subfigure}[b]{0.49\linewidth}
	\begin{tikzpicture}[spy using outlines={circle,white,magnification=5,size=3.5cm, connect spies}]
	\node {\pgfimage[width=8cm]{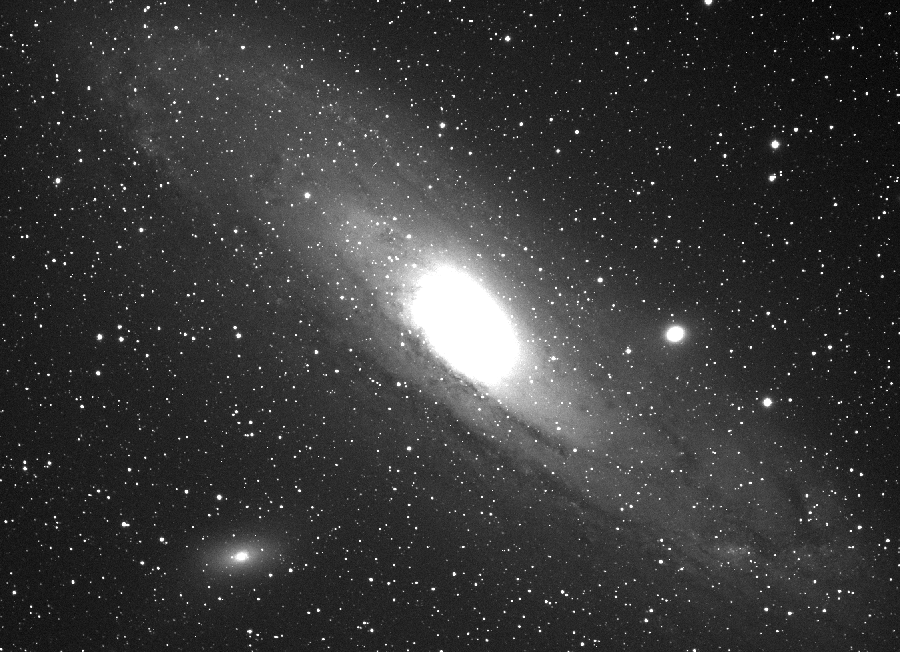}};
	\spy on (-1.85,-2.05) in node [left] at  (-0.5,3.25);
\spy on (2.45,-2) in node [left] at  (4,3.25);
	\end{tikzpicture}
\caption{Original M31 frame.}
\end{subfigure}
\begin{subfigure}[b]{0.49\linewidth}
	\begin{tikzpicture}[spy using outlines={circle,white,magnification=4,size=3.5cm, connect spies}]
	\node {\pgfimage[width=8cm]{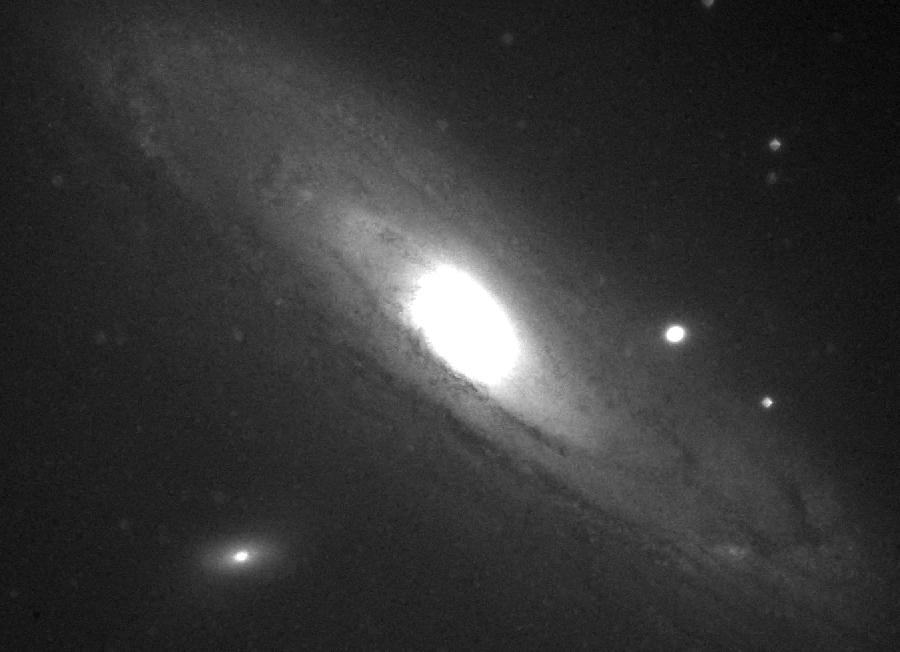}};
	\spy on (-1.85,-2.05) in node [left] at  (-0.5,3.25);
\spy on (2.45,-2) in node [left] at  (4,3.25);
	\end{tikzpicture}
\caption{Result of background estimation.}
\end{subfigure}
\caption{Examples of our background estimations of complex astronomical objects, (equipment: KAI 11002 CCD sensors, apochromatic refractor SkyWatcher ED80 on paralactic mount).}
\label{Background_estimations1}
\end{figure*}

An example of background extraction from real astronomical images obtained by the authors are given in Fig. \ref{Background_estimations1}. We processed single raw frame acquired with the support of Polish Astronomical Society (PTA) and Polish Amateur Astronomical Society (PTMA). The used CCD sensor (Kodak KAI 11000) was well calibrated, as it was previously carefully examined in our dark current investigations (\cite{CCDDarkCurrentPopowicz,AcorrectionalgorithmPopowicz}). The technical parameters of the camera are given in Tab. \ref{CCDtab}. The structures of background Andromeda galaxy were successfully retrieved, while the stars and the impulsive noise in form of hot pixels and cosmic rays, were efficiently rejected.

\begin{figure*}
\centering
\begin{subfigure}[b]{0.49\linewidth}
	\begin{tikzpicture}[scale=0.8,spy using outlines={circle,white,magnification=4,size=3.5cm, connect spies}]
	\node {\pgfimage[width=8cm]{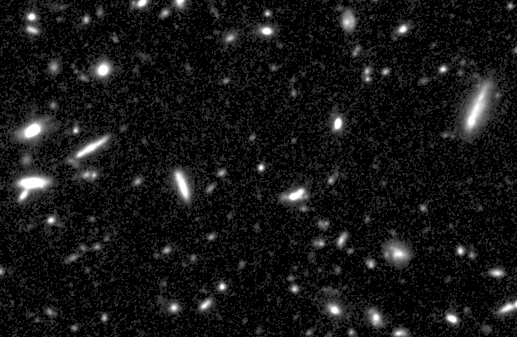}};
	%\spy on (-2.6,2.3) in node [left] at (4,3.25);
%\spy on (2.3,-1) in node [left] at (-0.5,-3);
	\end{tikzpicture}
\caption{Part of HUDF image.}
\end{subfigure}
\begin{subfigure}[b]{0.49\linewidth}
	\begin{tikzpicture}[scale=0.8,spy using outlines={circle,white,magnification=4,size=3.5cm, connect spies}]
	\node {\pgfimage[width=8cm]{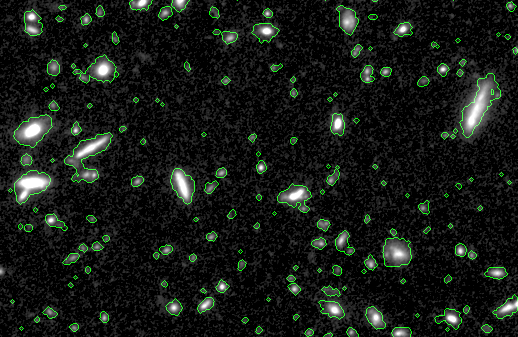}};
	%\spy on (-2.6,2.3) in node [left] at (4,3.25);
%\spy on (2.3,-1) in node [left] at (-0.5,-3);
	\end{tikzpicture}
\caption{The segmentation provided by our algorithm.} 
\end{subfigure}
\caption{The extraction of galaxies in \emph{Hubble Ultra Deep Field} (HUDF) image: (a) original HUDF image obtained on 4 December 2012 by the Wide Field Camera (WFC3), (b) the same frame with indicated contours of detected galaxies.} 
\label{segmentation}
\end{figure*}

In the last example presented in Fig. \ref{segmentation} we prove the capability of our algorithm to detect galaxies in Hubble Ultra Deep Field (HUDF) images. For this purpose, we used only the indication of pixels belonging to the objects and using the morphological basic transformations (erosion and dilation), we obtained the contours of detected galaxies, as depicted in Fig. \ref{segmentation}b. We employed much larger window size (41$\times$41 pixels) to enable the extraction of such extended objects. Therefore, the proposed algorithm may be a useful tool for the segmentation or removal of extended objects.

\begin{table}
\centering
\caption{Characteristics of used CCD cameras. \vspace{0.2cm}}
\begin{tabular}{l | c}
\hline  
Camera&ATIK11000\\ \hline
CCD model&Kodak KAI 11002\\
CCD type&nterline\\ 
Gain&1.07 [ADU/e$^-$]\\ 
Electronic noise $\sigma_{el}$&10.43 [e$^-$]\\ 
Pixel width&9.0 [$\mu$m]\\
Full well capacity&60.0 [ke$^-$]\\ \hline
\end{tabular}
\label{CCDtab}
\end{table}

\section{Conclusions}
In the paper we have presented a novel algorithm for complex background extraction from the astronomical images. Our approach is based on local distance transformation and employs widely used interpolation algorithms: nearest neighbor, linear, cubic and biharmonic. We introduced an adjustable parameter, which controls the algorithm's robustness. We also pointed at the capability of the adaptation of our approach for image processing purposes in other wavelengths, like in the infrared or in radioastronomy.  

We investigated the accuracy of our algorithm and the approach based on $\sigma$-clipping utilized in \emph{SExtractor}. The most straightforward method - the median filtering - was also included in our experiments. For the tests, we generated a range of artificial background images with embedded star objects. To follow the physical characteristics of CCD imaging, we also considered the influence of the electronic and Poisson noise.

The results of our experiments proved high efficiency of the proposed method and a significant superiority over current approaches, which was clearly visible especially for the most challenging backgrounds. According to the quality measurements, the combination of our method with the biharmonic interpolation should be considered as the most accurate and universal tool, when the background complexity is unknown.

We proposed several applications of our algorithm. It may be especially useful in impulsive noise rejection, cosmic rays removal or even in the detection of extended galaxies structures. Both the ground based and the spaceborne observations may benefit from the presented technique.

\section*{Acknowledgments}
Adam Popowicz was supported by Polish National Science Center, grant no. 2013/11/N/ST6/03051: Novel Methods of Impulsive Noise Reduction in Astronomical Images. 
The research was performed using the infrastructure supported by POIG.02.03.01-24-099/13 grant: GeCONiI - Upper Silesian Center for Computational Science and Engineering.

\bibliography{ms}

\end{document}